\newif\ifshortversion
\ifdefined\SHORTVERSION\shortversiontrue\fi
\ifshortversion
  \documentclass[sigconf]{acmart}
\else
  \documentclass[sigconf,nonacm]{acmart}
\fi

\newif\ifshowedits
\showeditstrue

\usepackage{amsmath,amsfonts}
\usepackage{bm}
\usepackage{amsthm}
\usepackage{graphicx}
\usepackage{textcomp}
\usepackage{xcolor}

\usepackage{paralist}
\usepackage[inline]{enumitem}
\newtheorem{theorem}{Theorem}[section]
\newtheorem{corollary}{Corollary}[theorem]
\newtheorem{lemma}{Lemma}
\newtheorem{proposition}{Proposition}
\newtheorem{definition}{Definition}
\newtheorem{remark}{Remark}

\usepackage{xspace}
\usepackage{booktabs}
\usepackage{multirow}
\usepackage{graphicx}
\usepackage{thmtools}	

\usepackage{hyperref}
\usepackage{cleveref}
\usepackage{tikz}
\usetikzlibrary{arrows.meta,positioning,fit,shapes.geometric,calc}

\ifshowedits
  
\else
  
\fi

\newcommand{\extref}[1]{\ifshortversion\citep{anonESP}\else\Cref{#1}\fi}
\newcommand{\proofin}[1]{\ifshortversion\else The proof is in~\Cref{#1}.\fi}

\usepackage{array}
\usepackage{textcomp}
\usepackage{stfloats}
\usepackage{url}
\usepackage{verbatim}
\usepackage{graphicx}

\usepackage{booktabs}           %
\usepackage{multirow}
\usepackage{caption}            %
\usepackage{xspace}
\usepackage{pifont}%

\usepackage{subcaption}
\usepackage{adjustbox}
\usepackage{enumitem}
\usepackage{nicefrac}
\usepackage{xcolor}

\newcommand{\obj}{\ensuremath{F}\xspace}
\newcommand{\tobj}{\ensuremath{\widetilde{F}}\xspace}

\newcommand{\objAppx}{\ensuremath{\widetilde{\obj}}\xspace}

\newcommand{\spara}[1]{\vspace{1mm}\noindent\textbf{#1}}
\newcommand{\xmark}{\ding{55}}%

\newcommand{\pv}{\ensuremath{\bm{p}}\xspace}
\newcommand{\Wset}{\ensuremath{\mathcal{W}}\xspace}
\newcommand{\wed}{\ensuremath{\omega}\xspace}
\newcommand{\Gw}{\ensuremath{G_{\Wset}}\xspace}
\DeclareMathOperator{\Bern}{Bern}
\DeclareMathOperator{\We}{W}
\DeclareMathOperator{\Rf}{H}
\DeclareMathOperator{\hardCost}{C}
\DeclareMathOperator{\V}{V}
\DeclareMathOperator{\edPress}{c}
\newcommand{\bigO}{\ensuremath{\mathcal{O}}\xspace}
\newcommand{\triSet}{\ensuremath{\Delta}\xspace}
\newcommand{\espresso}{\textsc{espresso}\xspace}

\newcommand{\Wmax}{\ensuremath{\mathcal{W_{\text{max}}}}\xspace}
\newcommand{\Hess}{\ensuremath{\bm{H}}\xspace}
\newcommand{\R}{\ensuremath{\mathbb{R}}\xspace}
\newcommand{\Mmat}{\ensuremath{\bm{M}}\xspace}
\newcommand{\Imat}{\ensuremath{\bm{I}}\xspace}
\DeclareMathOperator{\diag}{diag}
\DeclareMathOperator{\inter}{\mathbf{int}}
\DeclareMathOperator{\doma}{\mathbf{dom}}
\newcommand{\qv}{\ensuremath{\bm{q}}\xspace}
\newcommand{\espressoNL}{\textsc{n}-\textsc{espresso}\xspace}
\newcommand{\nolips}{\textsc{NoLips}\xspace}
\newcommand{\agg}{\ensuremath{S}\xspace}
\newcommand{\errAcc}{\ensuremath{\varepsilon_{\mathsf{NL}}}\xspace}
\newcommand{\degen}{\ensuremath{\zeta(G)}\xspace}
\newcommand{\dMax}{\ensuremath{d_{\text{max}}}\xspace}
\newcommand{\espressoNLAppx}{\textsc{n}-\textsc{espresso}-\textsc{s}\xspace}
\newcommand{\espressoR}{\textsc{n}-\textsc{espresso}-\textsc{r}\xspace}
\newcommand{\espressoGD}{\textsc{espresso}-\textsc{gd}\xspace}
\newcommand{\blGreedy}{\textsc{greedy}\xspace}
\newcommand{\blGreedyKDD}{\textsc{greedy}\ensuremath{_\mathsf{D}}\xspace}

\newcommand{\blVertexCover}{\textsc{vertex-cover}\xspace}

\newcommand{\stcden}{\textsc{stc}-\textsc{den}\xspace}
\newcommand{\espressoDEN}{\textsc{espresso}-\textsc{den}\xspace}
\newcommand{\blPeel}{\textsc{stc}-\textsc{peel}\xspace}
\newcommand{\roundExp}{\textsc{exp}\xspace}
\newcommand{\dens}{\ensuremath{q}\xspace}
\newcommand{\lamD}{\ensuremath{\lambda_\mathsf{D}}\xspace}

\newcommand{\grad}{\ensuremath{\nabla}\xspace}
\newcommand{\gradAppx}{\ensuremath{\tilde{\nabla}}\xspace}
\newcommand{\errEst}{\ensuremath{\tilde{\varepsilon}}\xspace}

\newcommand{\appxTri}{\ensuremath{\tilde{t}}\xspace}
\newcommand{\appxTriv}{\ensuremath{\tilde{\bm{t}}}\xspace}
\newcommand{\countTri}{\ensuremath{{t}}\xspace}
\newcommand{\countTriv}{\ensuremath{\bm{t}}\xspace}
\newcommand{\cntv}{\ensuremath{\chi}\xspace}
\newcommand{\cntvv}{\ensuremath{\bm{\chi}}\xspace}
\newcommand{\E}{\ensuremath{\mathbb{E}}\xspace}
\newcommand{\Var}{\ensuremath{\mathrm{Var}}\xspace}
\newcommand{\neigh}{\ensuremath{\mathcal{N}}\xspace}
\newcommand{\Prob}{\ensuremath{\mathbb{P}}\xspace}
\newcommand{\Pcomp}{\ensuremath{\mathbf{P}}\xspace}
\newcommand{\NP}{\ensuremath{\mathbf{NP}}\xspace}

\newcommand{\pokData}{\texttt{PK}\xspace}
\newcommand{\hollData}{\texttt{HW}\xspace}
\newcommand{\bmgData}{\texttt{BMG}\xspace}
\newcommand{\otcData}{\texttt{BIT}\xspace}
\newcommand{\cmData}{\texttt{CM}\xspace}
\newcommand{\euData}{\texttt{EU}\xspace}
\newcommand{\egofbData}{\texttt{EGO}\xspace}
\newcommand{\twitchData}{\texttt{TW}\xspace}
\newcommand{\gemsecData}{\texttt{GS}\xspace}
\newcommand{\wikiData}{\texttt{WT}\xspace}

\newcommand{\famER}{{\textsf{ER}}\xspace}
\newcommand{\famSBM}{{\textsf{SBM}}\xspace}
\newcommand{\famBA}{{\textsf{BA}}\xspace}
\newcommand{\famRGG}{{\textsf{RGG}}\xspace}

\newcommand{\CODE}{\url{https://github.com/iliesarpe/espresso}\xspace}

\crefname{enumi}{}{}
\Crefname{enumi}{}{}
\crefname{equation}{Eq.}{Eqs.}
\crefname{algocf}{Alg.}{Algs.}
\Crefname{algocf}{Algorithm}{Algorithms}
\Crefname{equation}{Eq.}{Eqs.}

\usepackage[ruled,vlined,linesnumbered]{algorithm2e}

\begin{document}

\title{ESPRESSO: Probabilistic Edge Strength under Triadic Pressure}

\author{Lutz Oettershagen}
\authornote{Both authors contributed equally to this research.}
\orcid{0000-0002-2526-8762}
\affiliation{%
	\institution{University of Liverpool}
	\city{Liverpool}
	\country{United Kingdom}}

\author{Ilie Sarpe}
\authornotemark[1]
\orcid{0009-0007-5894-0774}
\affiliation{%
	\institution{KTH Royal Institute of Technology}
	\city{Stockholm}
	\country{Sweden}}

\copyrightyear{2027}
\acmYear{2027}
\setcopyright{cc}
\setcctype{by}
\acmConference[WSDM '27]{Proceedings of the Twentieth ACM International Conference on Web Search and Data Mining}{February 15--19, 2027}{Hong Kong, Hong Kong}
\acmBooktitle{Proceedings of the Twentieth ACM International Conference on Web Search and Data Mining (WSDM '27), February 15--19, 2027, Hong Kong, Hong Kong}
\acmDOI{10.1145/3862322.3863473}
\acmISBN{979-8-4007-3117-4/2027/02}

\begin{abstract}
	Strong Triadic Closure (STC) is a fundamental model of tie strength in social networks, labeling each connection as \emph{strong} or \emph{weak}: if a node has strong ties to two other nodes, then those nodes must be connected.
	Most existing STC methods produce hard binary labelings, which cannot capture the uncertainty and conflicting structural evidence of real networks.
	We propose \espresso, a STC-based model for \emph{probabilistic structural tie strength}.
	{Our model} assigns each edge a probability of being strong, inducing a distribution over binary edge labelings. Each probability represents a structural score induced by the model. 
	The model's probabilities are computed by optimizing an objective function capturing the expected STC cost under this distribution, with entropy regularization.
	Our objective can be optimized by mirror descent, with a closed-form learning rate: the exact algorithm \espressoNL achieves linear convergence, above a given regularization threshold.
	We further develop sampling-based variants with approximation guarantees that avoid costly triangle enumeration over gradient updates and scale to large networks.
		Experiments show {that our objective function has
		predictable parameter behavior, that its probabilities differentiate tie strength on a real social network, and that they improve the discovery of dense subgraphs under STC. Our sampling-based variants scale to large and dense graphs where our exact algorithms exceed the time or memory limits}.
\end{abstract}



\maketitle

\section{Introduction}

Ties in networks can be viewed as strong or weak:
strong ties model trust, interaction, cohesion, and shared social context,
whereas weak ties often act as bridges between distant
subnetworks~\cite{granovetter1973strength}.
Inferring such \emph{tie strength} is a fundamental problem in social-network
analysis~\cite{GilbertK09,lu2010link,stolz2021predicting},
with applications including trust analysis~\cite{massa2007trust,guha2004propagation},
information diffusion~\cite{granovetter1973strength,larson2017weakness,onnela2007structure},
and community and role discovery~\cite{RozenshteinTG2017,wickrama2024dense}.
In this work, we consider tie-strength inference from an observed unweighted
graph using topology alone.

For this problem, a large body of work infers tie strength from metadata, including supervised,
ensemble, neural, and temporal approaches~\cite{GilbertK09,kahanda2009using,
	jones2013inferring,cheng2022comprehensive,zhuang2021modeling,flamino2021ml}.
Such metadata, however, may be noisy, incomplete, private, or unavailable.
Topology-based methods instead infer relational strength solely from the structure of the observed graph, without relying on additional information~\cite{XiangNR10,PhamSL16,liu2020new}.
Graph topology is also exploited in related but distinct tasks, such as link prediction and representation learning~\cite{katz1953new,
	liben2003link,perozzi2014deepwalk,grover2016node2vec,zhang2018link}.
Unlike link prediction, which scores node pairs according to the likelihood of a new edge, we assume a fixed graph and infer structural tie strength for its existing edges.
{Topology-based tie-strength methods can also produce continuous scores~\cite{XiangNR10,PhamSL16,liu2020new}, not based on the 
\emph{Strong Triadic Closure (STC)} principle~\cite{sintos2014using}; in contrast, we derive}
continuous edge strengths directly from the STC.\looseness=-1

\begin{figure*}
	\centering
	\input{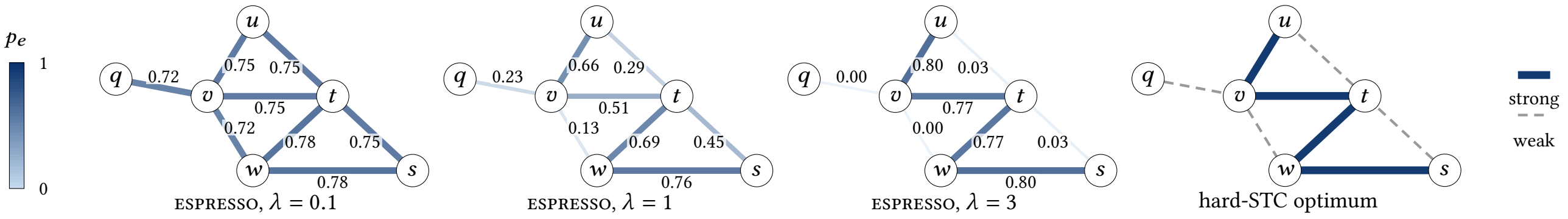}
	\caption{\espresso solutions on the same graph for different values of $\lambda$, with $\alpha=0.7$ and $\gamma=0.5$ fixed. As $\lambda$ increases, the penalty for STC violations becomes stronger and the inferred edge probabilities become more differentiated. Edge width and opacity scale with the strong-edge probability $p_e$. The rightmost panel shows one of the four optimal hard-STC labelings; thresholding the solution for \espresso at $\lambda=3$ at $p_e=0.5$ yields this labeling.}
	\label{fig:example}
\end{figure*}

STC provides a structural approach to tie strength~\cite{sintos2014using}.
It states that if a node has strong ties to two other nodes, then those two
nodes should be connected; otherwise the two strong ties form an
STC \emph{violation}.
This models the observation that strong ties are embedded in locally
cohesive structures.
Classical STC formulations infer a binary strong/weak labeling maximizing
the number of strong edges while avoiding violations, an $\NP$-hard
problem~\cite{sintos2014using}.
Moreover, the corresponding maximization problem is hard to
approximate within a factor of \(n^{1-\varepsilon}\), for any constant
\(\varepsilon>0\), unless \(\Pcomp=\NP\)~\cite{sintos2014using}, where $n$ is the number of nodes.
Subsequent work has studied algorithms, relaxations, constraints, and
applications of the STC formulation~\cite{GruttemeierK20,KonstantinidisNP18,
	RozenshteinTG2017,veldt2022correlation,MatakosG22,wickrama2024dense,
	oettershagen2022inferring,oettershagen2024inferring,
	oettershagen2026consistent,konstantinidis2020maximizing,golovach2020parameterized}.
	The LP relaxations of~\citet{adriaens2020relaxing} admit fractional edge strengths, {which however do not capture nuanced or uncertain probabilistic strength}.
	Thus, existing STC formulations remain limited in their ability to represent
	the uncertainty of real-world graphs.\looseness=-1

We address existing limitations by viewing structural tie strength as
\emph{probabilistic}.
Each edge $e$ has a probability $p_e$ of being strong, inducing a
fully factorized Bernoulli model over edge labelings, which we call
\espresso \emph{(\textbf{E}dge \textbf{s}trength
\textbf{pre}dic\-tion with \textbf{s}tocha\textbf{s}tic STC
\textbf{o}ptimization)}.
{This probabilistic view is motivated by the ambiguity of real networks.
In practice, an edge may lie in several triangles, where a strong label is consistent with STC,
while also forming \emph{open wedges}, i.e., pairs of edges sharing a node whose other endpoints are not connected, where a strong label introduces STC violations.}
A deterministic assignment cannot represent such conflicting structural
evidence.
Figure~\ref{fig:example} illustrates the solutions of \espresso on a small
graph, for increasing values of $\lambda$, the parameter controlling the
penalty for STC violations.
The probabilities capture both uncertain edges, with probability
close to $1/2$, and edges with certain probabilities
close to $0$ or $1$, according to the structure of the graph.
Furthermore, for a sufficiently large $\lambda$, thresholding these
probabilities recovers a hard-STC solution.
To infer the probabilities of \espresso, we introduce an entropy-regularized
objective balancing
\begin{inparaenum}
	\item the expected number of weak edges;
	\item the expected number of STC violations; and
	\item the entropy of the induced edge probabilities.
\end{inparaenum}
The expectation is over labelings drawn from a fully factorized Bernoulli model.
This yields probabilistic, uncertainty-aware tie strengths retaining
the interpretability of triadic closure.

We establish structural and optimization properties of our objective.
Wedges induce an interaction graph, also known as the \emph{Gallai graph}~\cite{gallai1967transitiv,le1996gallai}, and we show
that its negative spectrum determines the boundary of global convexity for our objective.
Furthermore, our {analysis establishes the} relative smoothness and relative
strong convexity of {our new objective with respect to the negative entropy}.
This allows us to design \espressoNL, a mirror-descent
algorithm with a closed-form learning rate and global convergence guarantees in polynomial time.
We further improve scalability by using sampling techniques.
\textbf{Our contributions are:}
\begin{compactenum}[1.]
	\item We introduce \espresso, a framework for \emph{probabilistic structural
		tie strength} that assigns each edge a probability of being strong according
		to the STC principle. 
	
		\item We characterize the properties of our objective function. 
		We connect the negative spectrum of the interaction graph with the global convexity of our objective.
	We use such properties to design \espressoNL and the scalable sampling variants \espressoNLAppx (and \textsc{-r}), which optimize our objective through mirror descent. 
	
	\item We connect solutions of \espresso to hard-STC labeling and to the
	STC-densest-subgraph of~\citet{wickrama2024dense}.
	
	\item We provide an extensive experimental evaluation our new approach. 
	We show that its probabilities can capture nuanced tie strength compared to existing LPs, supporting STC-hard applications, and that \espressoR scales to large graphs.
\end{compactenum}

\ifshortversion
\smallskip\noindent
Missing proofs, and additional
experiments are in our extended version~\citep{anonESP},
together with the source code.
\fi

\section{Preliminaries}
\label{sec:preliminaries}
Let $G=(V,E)$ be a simple undirected graph, where $V$ is the set of $n=|V|$ nodes, and
$E\subseteq\{\{u,v\}\mid u,v\in V,\ u\neq v\}$ is the set of $m=|E|$ edges (ties).
An \emph{induced wedge} \( \omega = (u, v, w) \) (or simply $(uvw)$) is a path of length two formed by two edges incident to a common node: in other words, if \( \{u,v\}, \{v,w\} \in E \), \( \{u,w\} \notin E \), and \( u \ne w \), we also say that \( \omega \) is \emph{centered} at \( v \).
Let \( \Wset \) be the set of induced wedges of \( G \). 
Similarly, we let $ \Wset_e$ be the set of induced wedges containing edge $e\in E$.
The \emph{Strong Triadic Closure} (STC) principle~\cite{EasleyK2010,sintos2014using} states that in a graph with edges labeled either \emph{strong} or \emph{weak}, 
if a node \( v \) has strong ties to two distinct nodes \( u \) and \( w \), 
then there should be an edge between \( u \) and \( w \) (either strong or weak). 
An \emph{STC violation} occurs when \( \{u,v\} \) and \( \{v,w\} \) are strong, and \( \{u,w\} \notin E \).
\textsc{MaxSTC} captures this principle~\cite{sintos2014using}, it requires labelling each edge as strong or weak, maximizing the number of strong labels while introducing no STC violations.
For a node $u\in V$ we let $\neigh(u)=\{v:\{u,v\}\in E\}$ be the \emph{neighborhood} of $u$, 
and $d_u=|\neigh(u)|$ the \emph{degree} of $u\in V$. 
We also let
$\dMax\doteq\max_{u\in V}d_u$
be the maximum degree of $G$.
A triangle is a triplet of distinct nodes $\{u,v,z\}\subseteq V$ ($\{uvz\}$ for short) where all three edges $\{u,v\},\{u,z\}$ and $\{z,v\}$ exist in $E$; 
with a slight abuse of notation, we use $e\in t$ when $e\in E$ is formed by two nodes of a triangle $t$.
We let $\triSet(G)$ denote the \emph{set} of triangles in $G$, and $\triSet_e\subseteq \triSet(G)$ denote the set of triangles containing $e\in E$. 
We use $\cntv_e = |\triSet_e|$ and $\cntvv(G)\in \mathbb{N}_0^m$ (or simply $\cntvv$) for the vector collecting such counts. 
For any vector $\pv\in(0,1)^m$ and an edge $e=\{u,v\}$ we use
$\countTri_e(\pv)=\sum_{\tau=\{uvz\}\in\triSet_e}(p_{uz}+p_{zv})$,
and we denote with $\countTriv(\pv)\in\mathbb{R}_{\ge0}^m$ the vector collecting such values.

\section{The \espresso model}
\label{sec:espresso_model}

We now introduce \espresso, 
a probabilistic framework modeling structural strengths for all edges $e\in E$ of a graph $G = (V, E)$. 
We associate each edge $e \in E$ with a latent binary variable $Z_e \in \{0,1\}$ indicating whether the tie is \emph{strong} ($Z_e = 1$) or \emph{weak} ($Z_e = 0$). 
Rather than committing to a fixed assignment $\bm{z} \in \{0,1\}^m$, we consider a distribution parameterized by a vector $\pv\in(0,1)^m$ of marginal probabilities, using the fully factorized family (e.g.,~\cite{jordan1999introduction})
\begin{equation}\label{eq:factorization}
	q(\bm{z}; \bm{p}) \doteq \prod_{e \in E} \Bern(Z_e = z_e \mid p_e),
	\, \text{and} \, Z_e \sim \Bern(p_e).
\end{equation}
Our model is motivated by a tractability choice: it controls the exponential space $\{0,1\}^{|E|}$ with a single parameter per edge.

Our goal is to identify an optimal value $\pv \in (0,1)^m$ for an objective function $\obj(\pv)$ capturing the STC principle under our factorized family.  
That is, each $p_e = \Prob[Z_e=1]$ is the \emph{structural tie strength} of $e\in E$,
a continuous value obtained from the topology of $G$ uniquely.
While the $Z_e$ are mutually independent under $q(\cdot;\pv)$, the objective $\obj(\pv)$ couples the parameters of neighboring edges through the wedges in $\Wset$, in accordance with the STC principle.

\textbf{Objective function.} 
We now present the properties of an optimal $\pv$, capturing probabilistic edge strength.
First we introduce our objective function $\obj(\pv; \alpha, \lambda, \gamma)$ for $\alpha, \lambda, \gamma>0$, or simply $\obj(\pv)$ for short, defined as
\begin{align}
	\obj(\bm{p}; \alpha, \lambda, \gamma)
	&\doteq
	\alpha \We(\bm{p})
	\;+\;
	\lambda\,\V(\bm{p})
	\;-\;
	\gamma\Rf(\bm{p}),\label{eq:full_objective}
\end{align}
where the terms $\We(\bm{p}), \V(\bm{p}) $ and $\Rf(\bm{p})$ correspond to
\begin{align}
	\We(\bm{p}) &\doteq \sum_{e\in E} (1-p_e)	,\label{eq:weak_edges}\\
	\V(\bm{p}) &\doteq \sum_{\omega=(u,v,w)\in \Wset} p_{uv}\,p_{vw},\label{eq:stc_violations}\\
	\Rf(\pv) &=-\sum_{e\in E}\bigl[p_e\log p_e + (1-p_e)\log(1-p_e) \bigr].\label{eq:entropy_regularizer}
\end{align}
The objective for \espresso is to find $\pv^* \doteq \arg\min_{\pv\in(0,1)^m}\allowbreak \obj(\pv)$.

\textbf{Connection with hard STC and intuition.} %
In the classical hard STC view each edge has a
binary label $\bm{z}\in \{0,1\}^m$.
A hard labeling should satisfy two competing requirements: 
\begin{inparaenum}[($i$)]
	\item minimize the number of weak edges, 
	i.e., assigning only weak labels trivially satisfies STC
	but is uninformative;
	\item  avoid assigning strong labels
	to both edges of a wedge in $\Wset$, introducing an STC violation.
\end{inparaenum}
This leads to the penalized \emph{hard} STC cost
\begin{equation}
	\hardCost(\bm z; \alpha, \lambda)
	\doteq
	\alpha
	\sum_{e\in E}
	(1-z_e)
	+
	\lambda
	\sum_{\omega=(uvw)\in\Wset}
	z_{uv}z_{vw},
	\label{eq:hard_stc_cost}
\end{equation}
where \(\alpha>0\) penalizes weak edges and
\(\lambda>0\) is the penalty for STC violations. That is, the first term counts
weak edges scaled by $\alpha$, while the second term counts wedges in $\Wset$ (i.e., violations) whose two incident edges are both strong, scaled by $\lambda$. 
Under \espresso by~\Cref{eq:factorization}, an edge \(e\in E\) is strong with probability \(p_e\), that is
$\Prob[z_e=1]=p_e$. 
Hence, the penalized \emph{expected soft} STC cost is
\begin{equation}
	\E_{q(\bm{z}; \bm{p})}[\hardCost(\bm z)]
	=
	\alpha \We(\bm p)
	+
	\lambda \V(\bm p),
	\label{eq:expected_hard_cost}
\end{equation}
since $\E[z_e]=p_e$, and by independence $\E[z_ez_{e'}]=p_ep_{e'}$ using \Cref{eq:factorization}. 
Therefore, $\obj(\pv) = \E[\hardCost(\bm{z})] - \gamma \Rf(\pv)$ where $\Rf(\pv)$ corresponds to an entropy regularizer of the probabilities $\pv$.

Equivalently, $\obj$ is the entropic regularization of the expected hard STC cost at temperature $\gamma$. 
Consider the Gibbs distribution $\pi_\gamma(\bm z)\propto\exp(-\hardCost(\bm z)/\gamma)$ over strong/weak labels, $\obj(\pv)$ and $\mathrm{KL}\bigl(q(\cdot;\pv)\,\|\,\pi_\gamma\bigr)$ differ by a constant independent of $\pv$~\cite{wainwright2008graphical}. 
In turn $q(\cdot;\pv^*)$ is the distribution in the family of \Cref{eq:factorization} minimizing $\mathrm{KL}(\cdot\,\|\,\pi_\gamma)$; see~\extref{app:gibbs-mean-field} for the exact derivation.
Hence $p_e^*$ is the marginal of the distribution $q(\cdot;\pv^*)$, i.e., our definition of the structural tie strength of $e\in E$. 
However, $p_e^*$ is \emph{not} the Gibbs marginal $\Prob_{\pi_\gamma}[Z_e=1]$, since $q(\cdot;\pv)$ does not model dependencies among edge labels.
Thus, $p_e^*$ is a structural score induced by our STC objective, and not a calibrated probability of observed tie strength.
The expectation in~\Cref{eq:expected_hard_cost} is linear in each $p_e$, and minimizing 
$\E[\hardCost(\bm z)]$ yields a hard labeling $\pv \in\{0,1\}^m$. 
However, the objective $\obj(\pv)$ contains the entropy of $q(\cdot;\pv)$ from term $-\gamma \Rf(\pv)$, controlled by $\gamma>0$:
$\Rf(\pv)$ is maximized at $\pv=\tfrac{1}{2}\bm{1}$ and minimized at $\pv\in\{0,1\}^m$, so it discourages unnecessary hard decisions.
The interplay between entropy and the STC terms determines the optimization properties enjoyed by $\obj$, through a graph-dependent threshold that we obtain in~\Cref{lemma:Fprops}.

Importantly, the optimum of $\obj(\pv)$ balances three components: a global preference for strong ties,
the penalty for STC violations, and uncertainty.
In turn, this gives \espresso a natural \emph{social interpretation}. 
An edge $e\in E$ embedded in a community (e.g., a dense subgraph) can be assigned a high value of $p_e$, since making $e$ strong creates few open wedges in $\Wset$. 
In contrast, an edge $e\in E$ acting as a bridge can be assigned a low probability $p_e$, since
making $e$ strong would violate the requirement of reducing the number of wedges in $\Wset$ with two strong edges. 
Finally, edges $e\in E$ for which the properties of $G$ do not yield a strong topological signal can receive $p_e\approx 0.5$, modeling uncertainty rather than forcing a hard label.%

\subsection{Objective properties}
\label{sec:interpretation}

We now characterize important properties of our objective $\obj(\pv)$. First, we introduce some notation.
Let $e=\{u,v\}\in E$. For each $\omega=(uvw)\in\Wset_e$, let $e'=\{v,w\}$ denote the second edge of $\wed=\{e,e'\}\in\Wset_e$.
Fix $\pv\in(0,1)^m$, we let $\edPress_e(\pv)\doteq\sum_{\omega\in\Wset_e}p_{e'}$ be the \emph{induced wedge pressure} on $e\in E$.
This quantity measures the aggregate probability assigned to neighboring edges $e'$ that can form an induced strong wedge in $\Wset$ with $e$.
Finally, we let $\sigma(z)\doteq(1+\exp(-z))^{-1}$, for $z\in\R$, be the sigmoid function.

We now obtain a closed form of the gradient of $\obj(\pv)$ and characterize the structure of any local minimizer of $\obj(\pv)$.

\begin{proposition}
	\label{prop:local_stationary_form}
	Consider $\obj(\pv;\alpha,\lambda,\gamma)$ for \(\alpha,\lambda,\gamma>0\). 
	The objective function $\obj(\pv): [0,1]^m \to \mathbb{R}$ is differentiable over the open hypercube $(0,1)^m$ with gradient
	\begin{equation}
		\nabla_e\obj(\pv) = \frac{\partial \obj(\pv)}{\partial p_{e}} =
		-\alpha + 
		\lambda \edPress_e(\pv) - 
		\gamma \,\log\Bigl(\tfrac{1-p_{e}}{p_{e}}\Bigr).
		\label{eq:gradient_entropy_open}
	\end{equation}
	Furthermore, at any
	local minimizer \(\pv^* \in (0,1)^m\), the edge probabilities satisfy
	\[
	p_e^*
	=
	\sigma(
	\gamma^{-1}(\alpha-\lambda \edPress_e(\pv^*))
	).
	\]
\end{proposition}
\proofin{app:proofsObjective}

\textbf{Structure of a minimizer.} Proposition~\ref{prop:local_stationary_form} shows that a local minimizer for \espresso has a
simple local interpretation. 
The value
$
\alpha-\lambda \edPress_e(\pv^*)
$
provides a structural score for edge \(e\):
\begin{inparaenum}[($i$)]
	\item the parameter \(\alpha\)
	is a reward for edge $e$ being strong;
	\item the term
	\(\lambda \edPress_e(\pv^*)\) is the local STC pressure against making \(e\) strong:
	it is large when \(e\) has a large number of wedges $\wed=\{e,e'\}\in \Wset_e$ that are themselves likely
	to be strong, i.e., most edges $e'$ have high value of $p_{e'}^*$;
	\item the entropy parameter \(\gamma\) acts as a temperature. 
	That is,
	a larger \(\gamma\) rewards uncertainty, while a smaller \(\gamma\) encourages
	probabilities closer to hard STC assignments.
\end{inparaenum}

Note that if an edge participates in no open
wedges, i.e., $\Wset_e=\varnothing$, then \(\edPress_e(\pv^*)=0\), and at any local minimizer $\pv^*\in(0,1)^m$ we have
$
p_e^*=\sigma(\alpha/\gamma).
$
That is, when $\Wset_e=\varnothing$ an edge $e
\in E$ has no STC penalty and its value of $p_e^*$ depends only on the parameters $\alpha$ and $\gamma$. 
Conversely, if \(\edPress_e(\pv^*)\) is
\emph{large}, then the term $\lambda\edPress_e(\pv^*)$ dominates $p_e^*$
forcing a low $p_e^*$.
The threshold \(\edPress_e(\pv^*)=\alpha/\lambda\)
corresponds to \(p_e^*=1/2\): when the value of $\edPress_e(\pv^*)>\alpha/\lambda$, the local STC pressure outweighs
the global preference for strong edges controlled by $\alpha$.

By \Cref{prop:local_stationary_form}, $p_e^*$ accounts for the values $p^*_{e'}$ of the edges $e'$ with $\{e,e'\}\in\Wset$, i.e., tie strength is aligned with graph structure.

\subsection{Optimization}\label{sec:optimization}

We now introduce our main algorithm (\espressoNL) to optimize the objective function $\obj$ from Eq.~\eqref{eq:full_objective}.

\textbf{Properties for optimization.} We show that under a specific setting of the regularization parameter $\gamma$, our objective function is \emph{convex}.
With this insight, a first approach to design an efficient method would be to apply gradient descent (GD)~\cite[Section 9.3]{Boyd2004CO} with a fixed learning rate to optimize $\obj$. 
However, since function $\obj$ is \emph{not} $L$-smooth, we cannot obtain any optimality guarantees.\footnote{A differentiable function $f:X\mapsto Y$ is $L$-smooth if there exists $L>0$ such that $||\nabla f(\bm{x}) - \nabla f(\bm{y})|| \le L ||\bm{x} - \bm{y}||$ for any $\bm{x},\bm{y}\in X$.}
Furthermore, since the optimization of $\obj$ is over $(0,1)^m$, an approach based on GD requires maintaining the iterates in the interior of the domain. 
Note that due to the the entropy contribution the gradient diverges in magnitude near the boundary. 
That is, for each coordinate, $\nabla_e\obj(\pv)\to-\infty$ as $p_e\to0^+$ and $\nabla_e\obj(\pv)\to+\infty$ as $p_e\to1^-$. 
A projected GD method therefore requires clipping the iterates to an interval $(\varphi,1-\varphi)^m$ for some small $\varphi>0$.
To overcome such challenges, we show that
function $\obj$ is \emph{$L$-smooth and $\mu$-strongly convex relative} to the Fermi-Dirac entropy-type function $h(\bm{p}) \doteq -\Rf(\bm{p}) = \sum_{e\in E} p_e\log p_e + (1-p_e) \log (1-p_e)$. 
This property allows us to use the \nolips optimization framework from~\cite{Bauschke2017NoLips,Lu2018NoLipsConvex}. 
The resulting approach is a mirror descent procedure called~\espressoNL\ (\underline{N}\textsc{oLips} optimization for \underline{\espresso}). \espressoNL computes a solution arbitrarily close to the global optimum of $\obj$, controlled by a parameter $\errAcc>0$, in \emph{polynomial time}.

\ifshortversion\else
In the rest of the manuscript for symmetric matrices $A, B \in \mathbb{R}^{m \times m}$ we use $A \succeq B$ to denote that $A - B$ is positive semidefinite, i.e., $\mathbf{x}^\top A \mathbf{x} \geq \mathbf{x}^\top B \mathbf{x}$
for all $\mathbf{x} \in \mathbb{R}^m$ ($A \succeq 0$ means $A$ is positive semidefinite).
\fi
For a function $h:X\mapsto Y$ that is strictly convex, continuously differentiable, and defined over a convex set $X$, and $\pv,\qv \in X$ we let
$
D_h(\bm{q},\bm{p}) \doteq h(\bm{q}) - h(\bm{p}) - \langle\nabla h(\bm{p}), \bm{q}-\bm{p}\rangle ,
$
be the \emph{Bregman divergence} between $\pv$ and $\qv$, note that 
$D_h(\pv,\qv)$ is not a distance since it is asymmetric in its arguments.
{For our choice of $h$, $D_h(\bm{q},\bm{p})$ is the sum over the edges of the KL divergences between Bernoulli distributions $\bm{q}$ and $\bm{p}$ (see \extref{lemma:BregmanExpr}).}
Given a function $f$, and a set $X$ we use textbook notation: $\doma f$, $\inter X$, to denote respectively the domain of a function $f$, the interior of a set $X$. 
We also let $|\Wmax| \doteq \max_{e\in E} |\Wset_e|$ denote the largest number of induced wedges incident to any single edge.
Furthermore, let $\Mmat\in\{0,1\}^{m\times m}$ be the adjacency matrix of the wedge-interaction graph, also known as the \emph{Gallai graph}~\cite{gallai1967transitiv,le1996gallai}: $\Mmat_{e,e'}=1$ iff $\{e,e'\}\in\Wset$. 
For an edge $e=\{u,v\}$, $|\Wset_e|=d_u+d_v-2-2\cntv_e$ corresponds to its degree in the interaction graph (recall that $\cntvv $ collects the number of triangles of each edge).
Hence, the Gallai graph is regular iff
$|\Wset_e|$ is constant over $e\in E$.
Since $\Mmat$ is symmetric its eigenvalues are real, and we denote by $\lambda_{\min}(\Mmat)$ and $\lambda_{\max}(\Mmat)$ 
its smallest and largest eigenvalue, respectively.\footnote{Note that these terms are not related to the parameter $\lambda$ of $\obj$.}
We then define
\[
\xi_-\doteq-\lambda_{\min}(\Mmat),\qquad
\xi_+\doteq\lambda_{\max}(\Mmat),\qquad
\gamma_c\doteq\frac{\lambda\xi_-}{4}.
\]
Note that if  $\Wset\neq\emptyset$ then both $\xi_-,\xi_+>0$.
In fact, the eigenvalues of $\Mmat$ sum to its trace, which is $0$, while $\Mmat\neq\bm{0}$, hence $\Mmat$ has both a positive and a negative eigenvalue.
We now characterize $\obj$ using the spectrum of $\Mmat$.

\begin{lemma}\label{lemma:Fprops}
	For $\obj:(0,1)^m\mapsto\R$ and $\alpha,\lambda>0$:
	\begin{inparaenum}[($i$)]
		\item $\obj$ is convex iff $\gamma\ge\gamma_c$;
		\item $\obj$ is $L$-smooth relative to $h=-\Rf$ for $L=\gamma+\lambda\xi_+/4$;
		\item if $\gamma>\gamma_c$, then $\obj$ is $\mu$-strongly convex relative to $h=-\Rf$ for $\mu=\gamma-\gamma_c$.
	\end{inparaenum}
\end{lemma}
\proofin{app:proofsNL}
Computing $\xi_-,\xi_+$ requires a spectral decomposition of $\Mmat$, which is unfeasible in practice. 
{Instead, by Gershgorin's circle theorem~\cite[Corollary~6.1.5]{horn2012matrix}, each eigenvalue of $\Mmat$ has absolute value at most the largest row sum of $\Mmat$, that is, $\xi_-,\xi_+\le|\Wmax|$.
This bound is, in general loose, however $|\Wmax|$ can be computed efficiently.
Thus, we will use \Cref{lemma:Fprops} for our algorithm \espressoNL by replacing $\xi_-$ and $\xi_+$ with $|\Wmax|$. 
That is, we will set $\gamma>\lambda|\Wmax|/4$, for which $\obj$ is $L$-smooth and $\mu$-strongly convex relative to $h$ with $L=\gamma+\lambda|\Wmax|/4$ and $\mu=\gamma-\lambda|\Wmax|/4$.}

\begin{remark}\label{remark:EigenDecomp}
	The exact convexity threshold depends on $\xi_-$, not on the spectral radius $\xi_+$. The latter controls relative smoothness. %
\end{remark}

The threshold $\gamma_c$ has a direct interpretation at
$\bar\pv\doteq\tfrac12\bm{1}$, when every edge is maximally uncertain.
The Hessian of
$\obj$ at $\bar\pv$ is $\lambda\Mmat+4\gamma\Imat$, with smallest eigenvalue
$4(\gamma-\gamma_c)$.
Hence, $\obj$ is convex in a neighborhood of
$\bar\pv$ if and only if $\gamma\ge\gamma_c$.

\begin{proposition}\label{prop:balancedTransition}
	Consider a graph $G$ with a $d$-regular Gallai graph, $d>0$, and let
	$\alpha=\lambda d/2$. Then $\bar\pv=\tfrac12\bm{1}$ is a stationary point of $\obj$ for every $\gamma>0$, and
	\begin{inparaenum}[($i$)]
		\item if $\gamma\ge\gamma_c$, $\bar\pv$ is the unique global minimizer of $\obj$ over $(0,1)^m$;
		\item if $0<\gamma<\gamma_c$, $\bar\pv$ is not a local minimizer of $\obj$.
	\end{inparaenum}
	Moreover, let $0<\gamma<\gamma_c$, let $a_\gamma$ be the unique solution in $(0,1)$ of
	$a=\tanh(\gamma_c a/\gamma)$, and assume that $\Mmat\bm{q}=-\xi_-\bm{q}$ for some $\bm{q}\in\{-1,1\}^m$.
	Then $\pv^{\pm}=\tfrac12(\bm{1}\pm a_\gamma\bm{q})$ are global minimizers of $\obj$,
	and they are the only ones if the eigenspace of $\lambda_{\min}(\Mmat)$ is one-dimensional.
	Finally, $a_\gamma\to1$ as $\gamma\to0^+$.
\end{proposition}
\proofin{app:proofsNL}
For example, the Gallai graph of an even cycle is the cycle itself, $d=2$ and $\xi_-=2$, yielding $\gamma_c=\lambda/2$ and $\alpha=\lambda$. That is, above $\gamma_c$ all probabilities equal $1/2$, while below $\gamma_c$ they alternate between two values and approach a hard STC assignment as $\gamma\to0^+$.

\begin{algorithm}[t]
	\caption{\espressoNL: \nolips for \espresso}\label{alg:NoLips}%
	\setcounter{AlgoLine}{0}
	\PrintSemicolon
	\KwIn{%
		Graph $G=(V,E)$,
		parameters $\alpha>0$, $\lambda>0$, $\gamma>0$,
		accuracy $\errAcc>0$, max iterations $K\in \mathbb{N}$.%
	}
	\KwOut{Edge-strength probabilities $\pv^K\in(0,1)^m$.}
	$L \gets \dfrac{\lambda\,|\Wmax|}{4} + \gamma$\label{alglineNL:params}\;
	$\pv^{0} \gets (1/2,\dots,1/2)\in\R^m$\label{alglineNL:p0}\;
	\lFor{$u\in V$}{$\agg_u^{0} \gets d_u/2$\label{alglineNL:sums}}
	\For{$k \gets 0$ \KwTo $K-1$\label{alglineNL:forIts}}{
		\For{$e = (u,v) \in E$}{
			$\nabla_e \obj(\pv^{k}) \gets \mathtt{Comp\nabla}(G, \pv^{k}, (\agg_u^{k})_{u\in V},e)$\label{alglineNL:gradient}\;
			$p_e^{k+1} \gets
			\sigma\left(
			\log\dfrac{p_e^{k}}{1-p_e^{k}}
			\;-\;
			\dfrac{\nabla_e \obj(\pv^{k})}{L}
			\right)$\label{alglineNL:updateP}\;
		}
		\If{$k < K-1$}{
			\lFor{$u \in V$}{
				$\agg_u^{k+1} \gets
				\agg_u^{k}
				+ \sum_{e \in E : u\in e}
				(p_e^{k+1} - p_e^{k})$\label{alglineNL:updatesums}
			}
		}
	}
	\KwRet{$\pv^{K}$\label{alglineNL:return}}
\end{algorithm}

\textbf{\espressoNL.}
We now describe our algorithm~\espressoNL (Alg.~\ref{alg:NoLips}), which adapts the framework of the~\nolips algorithm proposed in~\cite{Bauschke2017NoLips} to our objective $\obj$.
First, \espressoNL computes the relative-smoothness constant $L$ (using the upper bound on $\xi_+$ and \Cref{lemma:Fprops}), and initializes
$\pv^{0}=\tfrac{1}{2}\bm{1}$, where $\bm{1}\in\R^m$.
Next \espressoNL computes for each node $u\in V$ the aggregates $\agg_u^k$ defined over $k\in [K-1]$ as
\begin{equation}\label{eq:aggCounts}
	\agg_u^k \doteq \agg_u(\pv^k) \doteq
	\sum_{e = \{u,v\}\in E} p_{e}^k ,
\end{equation}
and under $\pv^0$ it holds $\agg_u^0= d_u/2$.
Next, \espressoNL performs the $K$ iterations of the \nolips approach (Line~\ref{alglineNL:forIts}). At each iteration $k\in[0,K-1]$, for each edge $e\in E$, \espressoNL first computes the gradient $\nabla_e\obj(\pv^{k})$ through function $\mathtt{Comp}\nabla$, whose pseudocode is in~\extref{alg:computeGradient}, which evaluates Eq.~\eqref{eq:gradient_entropy_open} according to $\pv^{k}$ (Line~\ref{alglineNL:gradient}). 
We discuss the gradient computation in a later section.
{\espressoNL then performs the mirror-descent step with rate $1/L$, $p_e^{k+1}=(\nabla_e h)^{-1}(\nabla_e h(\pv^{k}) - \tfrac{1}{L} \nabla_e \obj(\pv^{k}))$ (Line~\ref{alglineNL:updateP}), {obtaining $\pv^{k+1}$ (denoted as \emph{iterate})}.
Since $(\nabla_e h)^{-1}=\sigma$, the iterates remain in $\doma \obj$ without any clipping.}
At each iteration $k\ge 0$, after updating all entries of $\pv^{k+1}$, all node aggregate sums $\agg_u^{k+1}, u\in V$ are updated for the next iteration (Line~\ref{alglineNL:updatesums}).
Finally, \espressoNL outputs the solution $\pv^{K}$, which under the correct setting of $K$ guarantees that $\obj(\pv^{K}) - \obj(\pv^*) \le \errAcc$, for any $\errAcc>0$. 
We formalize the guarantees of \espressoNL in the next result, using a result from~\cite{Lu2018NoLipsConvex}.

\begin{theorem}\label{theo:accNL}
	Let $\lambda > 0$, {$\gamma>\gamma_c$, $\mu=\gamma-\gamma_c$, $L=\gamma+\lambda|\Wmax|/4$}, and fix an accuracy parameter $\errAcc>0$. Let $\pv^*$ be a minimizer of $\obj$ in $\doma \obj = (0,1)^m$. By the choice
	\begin{equation}
		{K \ge \tfrac{L}{\mu} \log \left(\errAcc^{-1} \, L m\log 2\right),}
	\end{equation}
	Algorithm~\ref{alg:NoLips} outputs a solution $\pv^{K}$ such that
	$
	\obj(\pv^{K}) - \obj(\pv^*) \le \errAcc.
	$
	{In particular, for $\gamma=\lambda|\Wmax|/4+\delta$ with $\delta>0$ it holds}
	\begin{equation}
		K \ge \frac{\lambda |\Wmax| + 2\delta}{2\delta} \log \left(\frac{(\lambda |\Wmax|+2\delta)m\log 2}{2\errAcc}\right).
	\end{equation}
\end{theorem}
\proofin{app:proofsNL}

{In practice, to further reduce the number of iterations $K$ we consider early stopping conditions described in~\extref{appsec:earlyStoppingNL}.}

\textbf{Gradient computation.} We now describe the details of the subroutine $\mathtt{Comp\nabla}(G, \pv^{k}, (\agg_u^{k})_{u\in V},e)$ (Line~\ref{alglineNL:gradient}). 
First, observe that a computation of the gradient of Eq.~\eqref{eq:gradient_entropy_open} from its definition requires iterating over the wedge set $\Wset$, with size
$
|\Wset|
=
\sum_{v\in V}\binom{d_v}{2}
-
3|\triSet(G)|
\le
\sum_{v\in V}\binom{d_v}{2}.
$
{The number of induced wedges} can be superlinear in $m$ and a naive gradient evaluation has cost
$\bigO(|\Wset|)$ per iteration.
To avoid this bottleneck, we exploit an alternative way of evaluating the gradient based on the set of triangles of the graph $G$.

Recall that $\Mmat$ is the adjacency matrix of the Gallai graph of $G$. It can be used as a linear operator such that for any $\pv\in(0,1)^m$ it holds
$
(\bm{M}\bm{p})_e
\;=\;
\sum_{\omega=\{e,e'\}\in\Wset_e} p_{e'} = \edPress_e(\pv) .
$
Note that instead of enumerating wedges explicitly, we express
$
\bm{M} = \bm{L} - \bm{T},
$
where $\bm{L} \in \{0,1\}^{m\times m}$ is the line-graph adjacency matrix with entries $\bm{L}_{e,e'} =1$ if edges $e,e'\in E, e\neq e'$ are such that $e\cap e' \neq \varnothing$, and 0 otherwise. While $\bm{T} \in \{0,1\}^{m\times m}$ has $\bm{T}_{e,e'} = 1$ if and only if edges $e,e'\in E, e\neq e'$ are in the same triangle in $\triSet(G)$.

For any edge $e=(u,v)\in E$ and any $\pv\in\R^m$, using the line graph matrix $\bm{L}$ we obtain
$
(\bm{L}\bm{p})_{e=(u,v)}
=
\agg_u(\pv)
+
\agg_v(\pv)
-
2 p_{uv},
$
where $\agg_w, w\in V$ is defined in Eq.~\eqref{eq:aggCounts}.
Given $\pv\in(0,1)^m$, the node aggregates $\agg_v(\pv)$ of all nodes $v\in V$ can be thus computed in a single pass over the edges of $E$.

Finally to obtain $(\Mmat \pv)_e$ we need to account for the term introduced by matrix $\bm{T}$, which corresponds to
\[
(\bm{T}\bm{p})_{e=(u,v)}
=
\sum_{z \in \neigh(u)\cap \neigh(v)}
\bigl(
p_{uz} + p_{vz}
\bigr)  =\countTri_e(\pv),
\]
which depends uniquely on the distribution of the triangles in $G$. 
Consequently, we have that for any $e\in E$ it holds
\begin{equation}\label{eq:gradAggSumTrick}
	\edPress_e(\pv) = \agg_u(\pv) + \agg_v(\pv) - 2 p_e - \countTri_e(\pv).
\end{equation}
Therefore, we can evaluate the gradient at each iteration with complexity
$
\bigO(|E| + |\triSet(G)|),
$
instead of $\bigO(|\Wset|)$.
On sparse real-world networks, this may yield near-linear time complexity in $m$ per iteration (e.g., in bounded degeneracy graphs).\footnote{Recall, the \emph{graph degeneracy} is the largest $\kappa$ such that there exists a non-empty $\kappa$-core.}
{However, such an approach requires storing (or listing) all triangles of $G$, which motivates the design of our sampling methods in~\Cref{subsec:randomNespresso}.}

\textbf{Time complexity.} We now bound the time complexity. %

\begin{lemma}\label{lemma:rtEx}
	Under the setting of \Cref{theo:accNL}{, and $\gamma=\lambda|\Wmax|/4+\delta$}, the time complexity of~\espressoNL is
	$
	\bigO(\degen m \frac{\lambda \dMax}{\delta} \allowbreak\log \frac{\lambda m}{\errAcc} ) ,
	$
	where $\degen$ corresponds to the degeneracy of graph $G$.
\end{lemma}
\proofin{app:proofsNL}

The bound above shows why~\espressoNL scales poorly on graphs with large degeneracy (e.g., dense graphs). 
{Furthermore, computing $|\Wmax|$ requires listing all triangles of $G$.
Note that any $B\ge|\Wmax|$ can be used in place of $|\Wmax|$ for the results above, at the expense of slower convergence (a larger $\gamma$ and a smaller step $1/L$). 
For example,  we will consider the value of $B=2(\dMax-1)$ that depends uniquely on the degrees of $G$ and can be computed in $\bigO(m)$.}

\textbf{Non-convex regime.}
When $\gamma<\gamma_c\doteq\lambda\xi_-/4$, $\obj$ is non-convex
by~\Cref{lemma:Fprops}~($i$), so the relative strong-convexity
condition $\mu>0$ in~\Cref{theo:accNL} fails. The bound
$\lambda|\Wmax|/4$ is a conservative sufficient threshold for
convexity. Even in the non-convex regime, $\obj$ remains
$L$-smooth \emph{adaptable} relative to $h$ for
$L=\lambda|\Wmax|/4+\gamma$. For any $L'>L$, \espressoNL with
step size $1/L'$ produces non-increasing objective values
$\obj(\pv^k)$. 
The smallest Bregman divergence
$D_h(\pv^j,\pv^{j-1})$ between successive iterates over the first
$k$ iterations is $\bigO(1/k)$, with a constant that depends on the
initial objective gap $\obj(\pv^0)-\inf_{\pv}\obj(\pv)$ and on
$L'-L$~\cite{Bolte2018Lips}. 
The divergence between successive
iterates also tends to zero. 
In the strongly convex regime
($\gamma>\gamma_c$), \espressoNL converges linearly to the global
minimizer. 
In contrast, the non-convex regime can help distinguish strong and
weak ties because smaller values of $\gamma$ can yield more
differentiated edge probabilities. 
In this regime, the solution
returned by \espressoNL is not guaranteed to be a global minimizer
of $\obj$ and may depend on the initialization $\pv^0$.
We compare the two regimes and discuss their practical implications
in~\Cref{sec:experiments}.

\subsection{Randomized gradient computation}\label{subsec:randomNespresso}
\espressoNL may not be able to scale on large and dense graphs due to the exact gradient computation over all $K$ iterations (Line~\ref{alglineNL:gradient}), which requires the listing of all triangles in $G$. 
In this section we design a new variant, \espressoNLAppx,\footnote{The ``\textsc{s}'' stands for {\emph{sua}, as in \emph{ca phe sua da}}, the Vietnamese iced milk coffee.} which combines \espressoNL with an approximate gradient computation obtained through a \emph{randomized algorithm}, avoiding the explicit listing of all triangles in the graph $G$ at every iteration.
We show that by leveraging the use of an approximate gradient $\gradAppx \obj(\pv^k)$ such that $||\grad\obj(\pv^k) - \gradAppx\obj(\pv^k)||_1 \le  \bigO(\lambda
m\errEst)$ at each iteration $k\ge0$ we can recover a similar result to \Cref{theo:accNL}, for a properly set estimation error threshold $\errEst> 0$.
Our result is based on a key inequality, proven in~\extref{lemma:appxGradGuar}.

\textbf{The algorithm.} \espressoNLAppx, detailed in~\extref{alg:NoLipsAppx}, follows the same steps of \espressoNL with a key modification: 
it replaces the exact gradient computation of $\grad_e\obj$ for $e=(u,v)$ from Line~\ref{alglineNL:gradient} with the approximate gradient computation 
\begin{align}
	\gradAppx_e\obj(\pv) \doteq -\alpha + \lambda(\agg_u(\pv) + \agg_v(\pv) - 2p_e  - \appxTri_e(\pv)) - \gamma\log \tfrac{1-p_e}{p_e}\label{eq:appxGrad}.
\end{align}
At each iteration $k\ge 0$ and for every edge $e\in E$ it holds that $|\appxTri_e(\pv)- \countTri_e(\pv)|\le \errEst$ with error probability bounded by $\eta\in(0,1)$ over all iterations.
For ease of notation we denote the vector corresponding to such estimates as $\appxTriv(\pv)\in\mathbb{R}^m$. {Then, with probability at least $1-\eta$, $\|\grad\obj - \gradAppx\obj\|_1\le \lambda\errEst m$ at every iteration (see \extref{theo:appxGuarantees})}.
\espressoNLAppx uses the approximate gradient $\gradAppx\obj$ in Line~\ref{alglineNL:updateP} to perform the descent step, which corresponds to selecting, at any $k\ge 0$,
\begin{align}
	\pv^{k+1} = \arg\min_{\pv\in (0,1)^m}\{ \langle \gradAppx \obj(\pv^{k}), \pv-\pv^{k}\rangle%
	+ L D_h(\pv,\pv^{k})\} ,
\end{align}

The next result shows the accuracy guarantees. %
\begin{theorem}\label{theo:appxNL}
	Consider the same setting of \Cref{theo:accNL}, where we run~\espressoNLAppx with estimation accuracy $\errEst\le \frac{\errAcc}{2 \lambda K m}$ and error probability $\eta\in (0,1)$. By setting
	\begin{equation}
		K \ge \tfrac{\lambda |\Wmax| + 2\delta}{2\delta} \log \left(\tfrac{(\lambda |\Wmax|+2\delta)m\log 2}{\errAcc}\right),
	\end{equation}
	\espressoNLAppx outputs a solution $\pv^{K}$ such that with probability at least $1-\eta$ it holds 
	$
	\obj(\pv^{K}) - \obj(\pv^*) \le \errAcc.
	$
\end{theorem}
\proofin{app:proofsAppx}

We now bound the time complexity of~\espressoNLAppx.

\begin{lemma}\label{lemma:rtAppx}
	Under the setting of \Cref{theo:appxNL} the time complexity of~\espressoNLAppx is bounded by
	$
	\bigO(\degen m + ms_{\max} \tfrac{\lambda \dMax}{\delta} \log \tfrac{\lambda m}{\errAcc}) ,
	$
	where $s_{\max}$ is an upper bound to the number of samples collected for each edge $e\in E$ at each of the $K$ iterations.
\end{lemma}
\proofin{app:proofsAppx}

{Similar to \espressoNL we also consider an early stopping condition based on an estimate of the objective $\obj$, see~\extref{appsec:earlyStopping}.}

\subsection{\espressoR: a scalable variant}\label{subsec:ristretto}

While \espressoNLAppx uses an approximate gradient, its complexity in \Cref{lemma:rtAppx} has a $\bigO(\degen m)$ term arising from a \emph{single} listing of all triangles in $G$, hindering its scalability on graphs with large degeneracy $\degen$.
We now introduce \espressoR (the ``\textsc{r}'' stands for \emph{ristretto}), a variant of \espressoNLAppx that removes the listing of all triangles through the following optimizations:
\begin{inparaenum}[($i$)]
	\item to set the parameters $\gamma$ and $L$ we use the upper bound $B = 2(\dMax-1)\ge |\Wmax|$ in place of $|\Wmax|$, since computing $|\Wmax|$ requires the listing of all triangles (as discussed after \Cref{lemma:rtEx});
	\item we retain the triangles of edges $e=(u,v)\in E$ only when $\min\{d_u,d_v\}\le \phi$, keeping at most $\phi$ triangles per edge by intersecting the two (small) neighborhoods of $e$, avoiding global enumeration. 
	The triangles of the remaining edges are processed through sampling as in \espressoNLAppx. 
	This bounds the pre-processing by $\sum_{e=(u,v)\in E} \min\{d_u,d_v\}\, \bm{1}[\min\{d_u,d_v\}\le \phi]= \bigO(\phi m) = \bigO(m)$ for a constant $\phi$, i.e., we build and store no more than a linear number of triangles in $m$~\citep{Seshadhri2014CC,Sarpe2025Triad}.
\end{inparaenum}
We design a data-dependent approach for setting $\phi$ and validate it
in~\extref{appsec:selectingPhi}.

Crucially, our analysis of~\espressoNLAppx still holds,  \Cref{lemma:Fprops} holds by replacing $|\Wmax|$ with $B$ (hence a larger $L$).
Furthermore, our sampling algorithm still retains its accuracy guarantees; see~\extref{theo:appxGuarantees}. 
Hence \espressoR retains the guarantees of \Cref{theo:appxNL}, with $B$ replacing $|\Wmax|$. 
This yields the following corollary,
removing the dependence on the degeneracy $\degen$.

\begin{corollary}\label{cor:rtRistretto}
	Under the setting of \Cref{theo:appxNL}, 
	the time complexity of~\espressoR is bounded by
	$
	\bigO\left(m\, s_{\max}\tfrac{\lambda \dMax}{\delta} \log \tfrac{\lambda m}{\errAcc}\right),
	$
	where $s_{\max}$ is a bound on the largest sample size.
\end{corollary}
\proofin{app:proofsAppx}

The complexity is independent of the degeneracy $\degen$.
For a constant sample size $s_{\max}$, as used empirically, the complexity is $\bigO(\dMax m)$ up to a factor $\bigO(\tfrac{\lambda}{\delta}\log\tfrac{\lambda m}{\errAcc})$. Note that a constant-sample setting might not satisfy the accuracy guarantees of~\Cref{theo:appxNL}.

\subsection{Connection with hard STC}\label{subsec:hardSTC}
We can round a solution $\pv^*\in(0,1)^m$ for~\espresso to a hard labeling with no STC violations controlling the number of weak edges.
We say that a labeling 
$\bm z\in\{0,1\}^m$ is \emph{STC-feasible} 
if it has no STC violations, i.e., $\V(\bm z) = \sum_{\omega=(uvw)\in\Wset} z_{uv}z_{vw} = 0$.
A natural way to round $\pv^*$ is by independent Bernoulli rounding: sample
$\bm z\sim q(\cdot;\pv^*)$, i.e.,
$z_e\sim\Bern(p_e^*)$ independently across $e\in E$, as captured by~\Cref{eq:factorization}.
By Eq.~\eqref{eq:expected_hard_cost}, this rounding yields
\begin{equation}\label{eq:roundingIdentity}
	\E_{\bm z\sim q(\cdot;\pv^*)}[\hardCost(\bm z)] = \alpha\We(\pv^*) + \lambda\V(\pv^*) = \obj(\pv^*) + \gamma\Rf(\pv^*).
\end{equation}
By the same argument it holds that  $\E[\We(\bm z)] = \We(\pv^*)$ and $\E[\V(\bm z)] = \V(\pv^*)$. 
That is, the expected number of weak edges and of STC violations of the rounded solution equal their fractional counterparts. 
We use this insight to obtained our rounded solution as follows.

\begin{proposition}\label{prop:rounding}
	Given $\pv^*\in(0,1)^m$, there is a polynomial-time procedure returning an STC-feasible labeling $\bm z'\in\{0,1\}^m$ such that
	\[
	\We(\bm z') \le \We(\pv^*) + \V(\pv^*).
	\]
\end{proposition}
\proofin{app:proofsHardDense}

Proposition~\ref{prop:rounding} shows that when \espresso yields a small expected number of violations $\V(\pv^*)$, the rounded solution $\bm z'$ has $\We(\bm z')$ within $\V(\pv^*)$ of $\We(\pv^*)$. 
The value of $\lambda$ thus controls such gap and the quality of $\bm z'$ as a solution for the hard STC formulation.

\subsection{Connection with STC-densest subgraph}\label{subsec:stcdensest}
We now show how to use a solution of~\espressoNL for the STC-densest subgraph (\stcden) problem studied in~\cite{wickrama2024dense}.
The goal of \stcden is to identify $U\subseteq V$ and an STC-feasible labeling $L$ (i.e., with no violations) of $E[U]$, the induced edges of $U$, to maximize
\begin{equation}\label{eq:objDense}
	\dens(U,L;\lamD) \doteq \frac{m_s(U,L) + \lamD\, m_w(U,L)}{|U|},\qquad \lamD\in[0,1],
\end{equation}
where $m_s(U,L) =|\{e\in E[U] : L(e)=1\}|$ and $m_w(U,L)=|E[U]| - m_s(U,L)$ count the strong and weak edges of the induced subgraph $G[U]$.
The problem is $\NP$-hard for $\lamD<1$, and the techniques proposed in~\cite{wickrama2024dense} require the processing of a \emph{wedge graph}, expensive to compute.
We first show that the expected value of the objective in~\Cref{eq:objDense}, under a labeling $\bm z$ drawn from a vector $\pv$ as in~\Cref{eq:factorization}, has a simple closed form.

\begin{lemma}\label{lemma:expDense}
	Fix $\lamD\in[0,1]$, given any subset of vertices $U\subseteq V$, values $\pv\in(0,1)^m$, and $\hat{w}_e = \lamD + (1-\lamD)p_e$ then it holds that 
	\begin{equation}\label{eq:stcdenReduction}
		\E_{\bm z\sim q(\cdot;\pv)}\left[\dens(U,\bm z;\lamD)\right] = \frac{1}{|U|}\sum_{e\in E[U]} \hat{w}_e.
	\end{equation}
\end{lemma}
\proofin{app:proofsHardDense}
\Cref{lemma:expDense} shows that maximizing the expected STC-density over the subsets $U\subseteq V$ corresponds to
computing the \emph{weighted densest subgraph} of $G$, with nonnegative edge weights $\hat{w}_e\in[\lamD,1]$.
Such a result holds for every $\lamD\in[0,1]$. Recall that the weighted densest subgraph
can be solved exactly in polynomial time, or approximated with a factor of 2~\cite{charikar2000greedy}.
Maximizing the right-hand side of~\Cref{eq:stcdenReduction} yields a solution $U\subseteq V$,
\emph{without} a labeling of $E[U]$ required by \stcden.
We now show that labeling $E[U]$ corresponds to the hard STC problem in~\Cref{subsec:hardSTC} over the induced subgraph $G[U]$.
Fix $U\subseteq V$, since $m_w(U,L) = |E[U]| - m_s(U,L)$, the objective in~\Cref{eq:objDense} can be expressed as
\begin{equation}\label{eq:densFixedU}
	\dens(U,L;\lamD) = \tfrac{\lamD |E[U]| + (1-\lamD)\, m_s(U,L)}{|U|},
\end{equation}
where the term $\lamD |E[U]|$ does not depend on $L$.
Hence, for every $\lamD<1$ and fixed $U\subseteq V$, identifying an STC-feasible labeling $L$ of $E[U]$ to maximize~\Cref{eq:densFixedU} is a \textsc{MaxSTC} problem~\cite{sintos2014using}.
We call \espressoDEN\footnote{\emph{Ca phe den} is a Vietnamese black coffee.} the method that:
\begin{inparaenum}[($i$)]
	\item computes a solution $\pv^*$ to~\espresso;
	\item obtains $U\subseteq V$ by maximizing the weighted density of~\Cref{eq:stcdenReduction}, exactly or within a factor 2;
	\item labels $E[U]$ by solving \textsc{MaxSTC} on $G[U]$, using the values $\pv^*$.
\end{inparaenum}
As we show in our experiments (see \Cref{subsec:stcden}), the resulting method is scalable and reports high quality solutions for \stcden.
Note that, since \textsc{MaxSTC} is $\NP$-hard~\cite{sintos2014using}, for step ($iii$) we adopt a heuristic detailed in~\extref{appsubsec:baselines}, returning an STC-feasible labeling of $E[U]$ with no approximation guarantee for~\Cref{eq:densFixedU}.
Consistently, our results do not contradict the $\NP$-hardness (and inapproximability) of \stcden for $\lamD<1$. Note that step ($iii$) can also be solved in polynomial time by rounding $\pv^*$ with the procedure of~\Cref{prop:rounding}, 
see~\ifshortversion\citep{anonESP}\else\Cref{lemma:stcdenCertificate} and~\Cref{appsubsec:stcdenRounding}\fi{, where we also compare the two methods to perform step ($iii$)}.

\section{Experiments}\label{sec:experiments}

We consider the following research questions:
\begin{enumerate*}[label=(\textbf{Q\arabic*})]
	\item How do the parameters $\alpha$, $\lambda$, and $\gamma$ influence solutions to \obj, and how does the graph spectrum relate the resulting probabilities to hard-STC labelings?
	\label{RQ:params}
	\item Are the continuous edge strengths inferred for \espresso more decisive compared to existing LP-based STC relaxations, and are they connected to social structure?\label{RQ:informedSTC}
	\item How effective are the probabilities produced by \espressoNL for \textsc{stc}-densest subgraph discovery?
	\label{RQ:stcden}
	\item 
	{Does \espressoR scale to large and dense graphs, and is its estimated gradient accurate?}
	\label{RQ:scalability}
\end{enumerate*}

\textbf{Datasets.} We consider public networks from SNAP: \otcData, \cmData, \euData, and \egofbData, together with larger datasets listed in~\Cref{tab:datastats}.
{\otcData is a trust network, \cmData and \euData are message and e-mail networks, and \egofbData is a network of Facebook contacts; we use the first three graphs for \Cref{RQ:params} and all four for \Cref{RQ:stcden}.
We use \pokData and \wikiData to evaluate the gradient estimate of \espressoR, and \hollData and \bmgData for scalability.}
For~\Cref{RQ:informedSTC}, we additionally use Zachary's Karate Club~\citep{zachary1977information} ($34$ nodes, $78$ edges).
{We further consider synthetic graph families.} We detail all datasets in~\extref{appsubsec:data}.
{Experiments on small (and synthetic) graphs are run on an Apple M2 (16GB RAM), while larger experiments are run on a server with four Intel Xeon Gold 6252N CPUs and 3TB RAM, using Python~3.10 (see \extref{appsubsec:envCode}).}

\begin{table}[t]
	\centering
	\caption{Datasets: nodes ($n$) and edges ($m$), degeneracy ($\degen$), maximum degree ($\dMax$), triangle count ($|\triSet|$), wedge count ($|\Wset|$), $|\Wmax|$, and average edge clustering coefficient ($c$).}
	\resizebox{\columnwidth}{!}{%
	\begin{tabular}{lrrrrrrrr}
		\toprule
		Dataset & $n$ & $m$ & $\degen$ & $\dMax$ & $|\triSet|$ & $|\Wset|$ & $|\Wmax|$ & $c$ \\
		\midrule
		\otcData & 5.9 $K$ & 21.5 $K$ & 21 & 795 & 33.5 $K$ & 1.6 $M$ & 995 & 0.26 \\
		\cmData & 1.9 $K$ & 13.8 $K$ & 20 & 255 & 14.3 $K$ & 712.9 $K$ & 396 & 0.13 \\
		\euData & 986 & 16.1 $K$ & 34 & 345 & 105.5 $K$ & 866.8 $K$ & 367 & 0.47 \\
		\egofbData & 4 $K$ & 88.2 $K$ & 115 & 1 $K$ & 1.6 $M$ & 4.5 $M$ & 1.8 $K$ & 0.70 \\
		\hollData & 1.1 $M$ & 56.3 $M$ & 2.2 $K$ & 11.5 $K$ & 4.9 $B$ & 32.9 $B$ & 16.3 $K$ & 0.58 \\
		\pokData & 1.6 $M$ & 22.3 $M$ & 47 & 14.9 $K$ & 32.6 $M$ & 2 $B$ & 16.1 $K$ & 0.11 \\
		\bmgData & 43.1 $K$ & 14.5 $M$ & 1 $K$ & 8 $K$ & 3.6 $B$ & 15 $B$ & 9.4 $K$ & 0.62 \\
		\ifshortversion\else\twitchData & 168.1 $K$ & 6.8 $M$ & 149 & 35.3 $K$ & 54.1 $M$ & 8.7 $B$ & 43.9 $K$ & 0.16 \\\fi
		\ifshortversion\else\gemsecData & 50.5 $K$ & 819.1 $K$ & 69 & 1.5 $K$ & 2.3 $M$ & 120.7 $M$ & 2.7 $K$ & 0.14 \\\fi
		\wikiData & 2.4 $M$ & 4.7 $M$ & 131 & 100 $K$ & 9.2 $M$ & 12.6 $B$ & 108.2 $K$ & 0.11 \\
		\bottomrule
	\end{tabular}}
	\label{tab:datastats}\vspace{-2mm}
\end{table}

\textbf{Methods.} 
We considered our methods for \espresso, as named in~\Cref{sec:optimization}.
We consider several baselines, \blGreedy and \blVertexCover, the \stcden heuristics of~\citet{wickrama2024dense}, LP2 ($d=1$) with Chebyshev-center post-processing from~\citet{adriaens2020relaxing}, and \espressoGD a projected-gradient baseline; all detailed in~\extref{appsubsec:baselines}.
We use \espressoGD to assess the benefits of our methods using mirror descent\ifshortversion.\else, see~\extref{appsec:solutionComparison}.\fi{}

For small graphs, when feasible, we use the complete set
$\mathcal Z^*$ of optimal hard-STC labelings and let
$
r_e \doteq \frac{1}{|\mathcal Z^*|}
\sum_{\bm z\in\mathcal Z^*} z_e,
$ be the
\emph{hard-optimum marginal}, i.e., the fraction of optimal labelings in $\mathcal{Z}^*$ where $e$ is strong,
and let $\bm r\doteq(r_e)_{e\in E}$.
We will use $\mathcal{Z}^*$ to study if the probabilities for \espresso are decisive 
when the hard optimum is unambiguous, and more uncertain when
optimal hard labelings disagree. 
{We will consider $\mathcal Z^*$ only on small graphs, where it can be enumerated efficiently.}
\ifshortversion
Our code is available online at \CODE.
\else
Our code and datasets are available online at \CODE.
\fi

\subsection{Results}

	\subsubsection{\Cref{RQ:params}: Objective and parameters}
\label{subsec:params}

\begin{figure*}[t]
	\centering
	\begin{subfigure}[b]{0.4\linewidth}
		\centering
		\includegraphics[width=\linewidth]{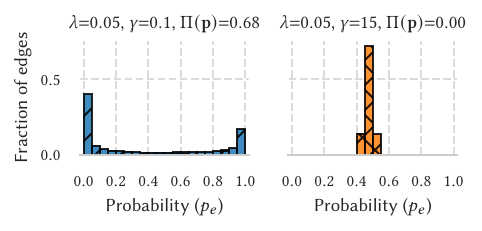}
		\caption{Edge probabilities on \euData}
		\label{fig:eu-probabilities}
	\end{subfigure}\qquad%
	\begin{subfigure}{.25\linewidth}
		\centering
		\includegraphics[width=\linewidth]
		{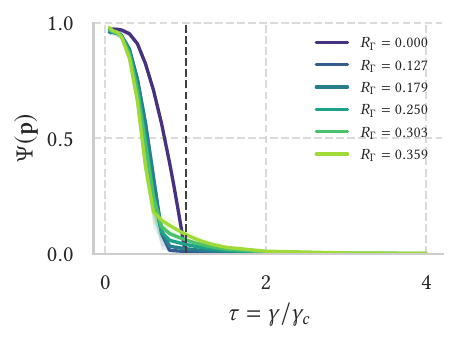}
		\caption{Controlled transition}
		\label{fig:cycles}
	\end{subfigure}%
	\begin{subfigure}{.3\linewidth}
		\centering
		\includegraphics[width=.9\linewidth]
		{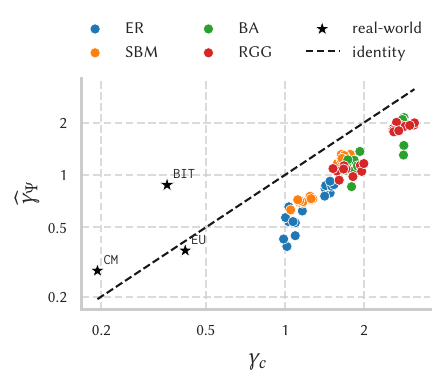}
		\caption{{Predicted vs.\ observed onset}}
		\label{fig:spectral-predicted}
	\end{subfigure}\hfill%
\caption{
	(a) Edge-probability distributions on \euData for $\alpha=1$ and $\lambda=0.05${; $\Pi(\pv)$ is the fraction of edges with $p_e<0.1$ or $p_e\ge0.9$}.
	For small $\gamma=0.1$, the solution is differentiated and bimodal,
	whereas for $\gamma=15$ the probabilities are unimodal near $1/2$.
	(b) For $\alpha=\lambda|\Wset|/m$, the regular Gallai exhibits the transition at $\gamma/\gamma_c=1$, increasing $R_\Gamma$ smooths the transition.
	(c) Theoretical threshold $\gamma_c$ versus the {observed onset}
	$\widehat{\gamma}_{{\Psi}}$ on $80$ synthetic graphs{ ($\lambda=1$)};
	``$\star$'' marks \otcData, \cmData, and \euData{ ($\lambda=0.05$)}.}%
\end{figure*}

By~\Cref{prop:local_stationary_form},
$p_e^*>1/2$ iff $\edPress_e(\pv^*)<\alpha/\lambda$:
$\alpha/\lambda$ therefore determines the threshold for a strong or weak labeling, 
while $\gamma$
controls distance from $1/2$.
On \otcData, \cmData, and \euData we vary
$\alpha\in\{10^{-3},10^{-2},10^{-1},1,10,100\}$ and
$\lambda\in\{5\times10^i:i=-5,\ldots,4\}$ and $\gamma$ such that $\obj$ is convex. 
Results are in \Cref{fig:eu-probabilities}; see~\extref{appsec:additExps} for additional results.
We analyzed the distribution of the probabilities of $\pv^*$ under each setting, and they are  \emph{unimodal} with the distance from $1/2$ controlled by $\lambda$.
In contrast, for $\gamma<\gamma_c$, the distribution of the probabilities optimizing $\obj$ is often \emph{bimodal}.

\textbf{Practical guidance.}
From our results above and \Cref{prop:local_stationary_form},
for applications requiring global optimization guarantees, we set $\gamma$ above the convexity threshold $\gamma_c=\lambda{\xi_-}/4$; leveraging the efficiently computable bound $B=2(\dMax-1)\geq {|\Wmax|}$ if computing ${\xi_-}$ or ${|\Wmax|}$ is impractical, yielding $\gamma>\lambda B/4$.
{Increasing $\alpha/\lambda$ encourages more edges with a high value of $p_e$.
By the choice of $\alpha=\lambda|\Wset|/m$, it holds $p^*_e>1/2$ when $e$ belongs to fewer wedges than the average edge of $E$.}
Small values of $\gamma$ often yield differentiated, and potentially bimodal,
distributions of the probabilities of a solution to $\obj${, at the cost of weaker convergence guarantees}.

\textbf{Spectral transition.}
The exact convexity threshold is
$\gamma_c=\lambda\xi_-/4$.
We fix $\lambda=1$ and
$\alpha=\lambda|\Wset|/m$. 
When the Gallai graph is $d$-regular then
$\alpha=\lambda d/2$ and
$\bar{\pv}=\tfrac12\bm 1$ is a stationary solution (see~\Cref{prop:balancedTransition}).
Since $H(\bar{\pv})=m\log 2$, we refer to $\bar{\pv}$ as the
maximum uncertain solution.
To isolate such effect, we generate $k$-regular graphs with no triangles, $n=200$ and $k\in\{4,6\}$.
The absence of {triangles} forces every edge to participate in exactly $2(k-1)$
wedges, and a regular Gallai graph, hence~\Cref{prop:balancedTransition} holds.
We then apply edge swaps preserving the degrees: the original graph remains
$k$-regular, while the introduced triangles make the Gallai degrees
heterogeneous.
For this controlled experiment, we vary the  parameter
$\tau\doteq\gamma/\gamma_c$ and measure the \emph{Gallai irregularity} by
$
R_\Gamma
\doteq
\frac{2\,\operatorname{std}(\bm d_\Gamma)}{\xi_-}$, with
$
\bm d_\Gamma=\Mmat\bm 1,
$
which vanishes when the Gallai graph is regular.
We also measure the distance from maximum uncertainty by the normalized entropy
deficit
\({\Psi}(\pv)\doteq 1-H(\pv)/(m\log 2)\), which ranges from \(0\) when
\(\pv=\tfrac12\bm 1\) to \(1\) for $\pv\in\{0,1\}^m$.
Our choice of $\alpha$ for this controlled setting forces $\bar{\pv}$ to be the
global minimizer for $\gamma\ge\gamma_c$. Hence, ${\Psi}$ measures the distance
from $\bar \pv$. 
For a general $\alpha$ the convex regime returns an informative
solution and not $\bar{\pv}$.

As shown in~\Cref{fig:cycles}, the regular case exhibits the predicted
transition when $\gamma/\gamma_c=1$.
That is, for $\gamma/\gamma_c\to 0$, ${\Psi}$ approaches 1. 
We also note that increasing $R_\Gamma$
smooths the transition.

We then evaluate the theoretical threshold {$\gamma_c=\lambda\xi_-/4$} on $80$ synthetic
graphs with $n=200$ from Erd\H{o}s--R\'enyi (\famER), stochastic block model
(\famSBM), Barab\'asi--Albert (\famBA), and toroidal random geometric graph (\famRGG)
families, using a mean degree $\bar{k}\in\{8,16\}$, and $10$ seeds per
setting.
{Informally, the \emph{onset} of a graph is the value of $\gamma$ at
which, as $\gamma$ decreases, the edge probabilities move fastest away
from $1/2$ towards $0$ or $1$.}
For each graph, we evaluate $81$ values of
$\gamma\in[0.05,16]$ (logarithmically spaced) and define the {observed onset}
$\widehat{\gamma}_{{\Psi}}$ as
\ifshortversion
the value of $\gamma$ yielding the (absolute) maximum rate of change in ${\Psi}$ (with respect to $\log\gamma$), calculated after a standard smoothing procedure (see~\extref{app:synth_graph}).
\else
{the value of $\gamma$ at which $\Psi$ changes fastest, i.e., the
maximizer of $\left|d\Psi/d\log\gamma\right|$. We compute it from the
smoothed values of $\Psi$ on the grid, and interpolate between grid
points to locate the maximum}
(see~\extref{app:synth_graph}).
\fi
As shown in~\Cref{fig:spectral-predicted}, the theoretical threshold
$\gamma_c$ {predicts the observed onset} $\widehat{\gamma}_{\Psi}$, even if the observed
{onset} is typically below the theoretical value.
Our three real-world datasets are also broadly consistent with
this scale, although \otcData shows a larger deviation from its theoretical behavior.

\textbf{Relation to exact hard STC.}
We next study whether the probabilities computed for the objective $\obj$ relate to optimal hard-STC
labelings on $100$ small graphs for which all solutions can be enumerated
exactly.
We generate $25$ graphs from each of four families:
Erd\H{o}s--R\'enyi, stochastic block model, planted-structure, and
Watts--Strogatz.
The graphs have $7$ to $9$ vertices and $5$ to $17$ edges, we fix
$\alpha=0.7$, $\lambda=1$, and $\gamma=0.5$.
Since $\gamma_c$ is graph-dependent, our fixed value of $\gamma$ does
not force one common regime of $\obj$ across all graphs. 
Furthermore, when $\gamma<\gamma_c$, we use multiple initializations and retain the solution
with smallest objective.
Over our instances, $30/100$ graphs {have} a \emph{unique} hard optimum (see~\extref{tab:hardOptimaValidation}).
When the hard optimum is unique, labeling as strong the edges with $p^*_e>1/2$ (computed with \espressoNL) recovers the
exact optimum on $90\%$ of the $30$ instances.
When several hard optima exist, $\pv^*$ is instead aligned with the
hard-optimum marginals $\bm r$. 
That is, over the $70$ such instances, 
the mean
absolute error between $\pv^*$ and $\bm r$ is $0.121$.
Furthermore, on the $68$ instances where the hard-optimum marginal vector is not
constant, the Pearson and Spearman correlations between $\pv^*$ and $\bm r$ are $0.9$.
For a detailed discussion see~\extref{appsec:hardOptimaValidation}.

Finally, on graphs small enough to enumerate all $2^m$ labelings, $\pv^*$ is also aligned
with the exact marginals of $\pi_\gamma$ (i.e., the Gibbs distribution over all $\{0,1\}$ labelings) in the convex regime, while differing when $\gamma < \gamma_c$, as reported in~\extref{appsec:gibbsMarginals}.

\subsubsection{\Cref{RQ:informedSTC}: Qualitative tie strengths}
\label{subsec:espresso_informed_stc}
We test whether the continuous strengths inferred for \espresso resolve
the structural ambiguity of LP-based STC relaxations.

More in detail, we compare \espressoNL with the LP relaxation of~\citet{adriaens2020relaxing} on Zachary's Karate Club~\citep{zachary1977information}.
Both methods consider as input a simple graph.
{To obtain a solution $\pv^*$ with} \espressoNL we set $\lambda=1$, $\alpha=\lambda|\Wset|/m$, and $\gamma=1.05\gamma_c$, forcing the convex regime.
For the LP we use its second formulation with $d=1$ (denoted with LP2), which reports for each edge a score in $\{0,1/2,1\}$. 
Since the LP optimum is not unique, we report the Chebyshev center of the polytope of its optimal solutions,
avoiding the selection of a specific solution { (see~\extref{appsubsec:lpChebyshev})}.

\begin{figure}[t]
	\centering
	\input{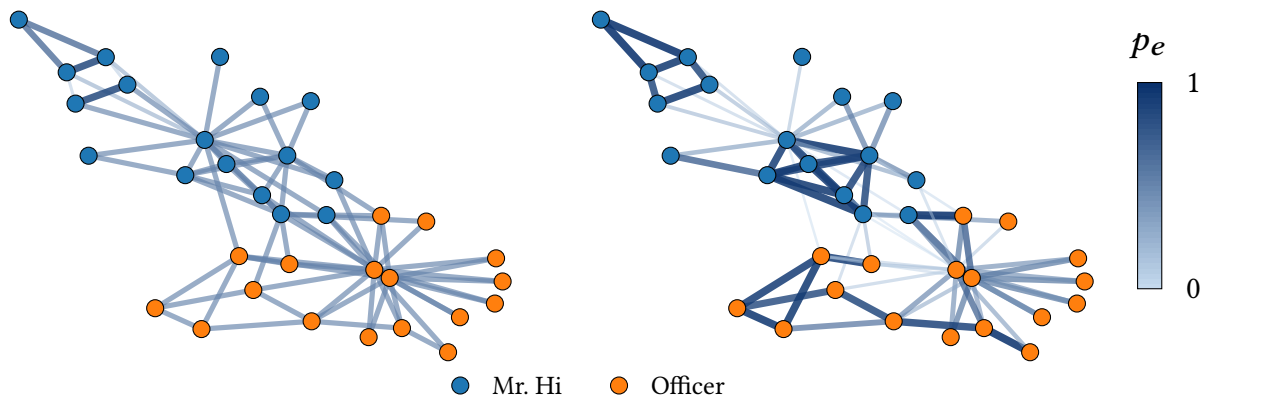}
	\caption{\textbf{Comparison on Zachary's Karate Club.}
	Node colors indicate the two partitions into which the club split, as recorded by~\citet{zachary1977information}.
	Edge width is proportional to the inferred tie strength, ranging in $[0,1]$. }
	\label{fig:karate_comparison}
\end{figure}

{Results are} reported in~\Cref{fig:karate_comparison}.
We first note that the
LP2 method (with Chebyshev center) yields only five distinct values of edge strength, assigning to $70$ out of 78 edges exactly $1/2$.
In contrast, \espressoNL produces $63$ distinct probabilities spanning
$[0.007,0.982]$.
Thus, even on such a small network, \espressoNL substantially reduces the (uninformative) ambiguity of the LP relaxation~\citet{adriaens2020relaxing}, 
providing a fine-grained
ordering of the edges.
{In~\Cref{fig:karate_comparison}, we observe that the values of $\pv^*$ allow to distinguish the two factions recorded by~\citet{zachary1977information}, in contrast
LP2 assigns $1/2$ to every edge of the graph.}
We {also} note that the inferred probabilities have a weak
positive association with pairwise interaction counts, computed from raw data
(Kendall's-$\tau$ for {$\pv^*$} equals $0.244$, and {$0.092$} for LP2).
{Note that interaction counts are a coarse proxy for tie strength, thus on such a small network our result shows that $\pv^*$ matches slightly more closely interaction counts than LP2.
We further relate $\pv^*$ to the edge weights of real data in~\extref{appsec:weightsVsStructure}.}

\subsubsection{\Cref{RQ:stcden}: STC-densest subgraph}
\label{subsec:stcden}

\begin{table}[t]
	\centering
	\caption{\stcden application. We report the objective $\dens(U,L;\lamD)$ for each solution. The runtime of \espressoDEN is the sum of its three steps; in parentheses, we report the runtime of steps $(ii)$ and $(iii)$ (see~\Cref{subsec:stcdensest}). The \emph{speedup} is evaluated against the fastest baseline. ``\xmark'' denotes that the time limit was reached.}
	\label{tab:stcden}
	\resizebox{\columnwidth}{!}{%
	\begin{tabular}{llrrrrrrr}
		\toprule
		& & \multicolumn{3}{c}{Objective $q(U, L; \lamD)$} & \multicolumn{3}{c}{Time (sec)} & \\
		\cmidrule(lr){3-5}\cmidrule(lr){6-8}
		Dataset & $\lamD$ & \espressoDEN & \blGreedyKDD & \blPeel & \espressoDEN & \blGreedyKDD & \blPeel & speedup \\
		\midrule
		\otcData & 0.2 & 3.905 & \textbf{4.000} & 3.726 & 1.46 (0.52) & 680.96 & 25.35 & 17.4$\times$ \\
		 & 0.4 & 6.345 & 7.066 & \textbf{7.098} & 1.43 (0.50) & 846.84 & 25.12 & 17.6$\times$ \\
		 & 0.6 & \textbf{10.905} & 10.411 & 10.444 & 1.79 (0.85) & 854.43 & 24.88 & 13.9$\times$ \\
		 & 0.8 & \textbf{14.016} & 13.760 & 13.782 & 1.72 (0.78) & 855.99 & 25.39 & 14.8$\times$ \\
		\midrule
		\cmData & 0.2 & \textbf{3.731} & 3.489 & 3.465 & 1.93 (1.69) & 154.73 & 13.99 & 7.3$\times$ \\
		 & 0.4 & \textbf{6.924} & 6.778 & 6.773 & 1.59 (1.36) & 179.88 & 14.74 & 9.2$\times$ \\
		 & 0.6 & \textbf{10.263} & 10.067 & 10.055 & 1.13 (0.90) & 186.00 & 14.42 & 12.7$\times$ \\
		 & 0.8 & \textbf{13.479} & 13.353 & 13.348 & 1.31 (1.07) & 183.69 & 14.36 & 11.0$\times$ \\
		\midrule
		\euData & 0.2 & 8.138 & \textbf{8.500} & 6.012 & 1.52 (0.72) & 116.45 & 17.09 & 11.2$\times$ \\
		 & 0.4 & \textbf{12.893} & 11.246 & 11.356 & 1.78 (0.98) & 110.19 & 15.98 & 9.0$\times$ \\
		 & 0.6 & \textbf{17.613} & 16.696 & 16.738 & 2.29 (1.48) & 117.01 & 16.66 & 7.3$\times$ \\
		 & 0.8 & \textbf{22.671} & 22.120 & 22.182 & 2.02 (1.22) & 122.23 & 17.04 & 8.4$\times$ \\
		\midrule
		\egofbData & 0.2 & \textbf{28.218} & \xmark & 19.553 & 34.47 (1.89) & \xmark & 93.33 & 2.7$\times$ \\
		 & 0.4 & \textbf{37.674} & \xmark & 33.849 & 36.51 (3.92) & \xmark & 94.73 & 2.6$\times$ \\
		 & 0.6 & \textbf{51.662} & \xmark & 48.289 & 37.40 (4.81) & \xmark & 95.25 & 2.5$\times$ \\
		 & 0.8 & \textbf{64.598} & \xmark & 62.818 & 35.87 (3.28) & \xmark & 101.00 & 2.8$\times$ \\
		\bottomrule
	\end{tabular}}
\end{table}

We compare \espressoDEN and the two baselines \blGreedyKDD and \blPeel by \citet{wickrama2024dense} on four small datasets for $\lamD\in\{0.2,0.4,0.6,0.8\}$.
The baselines and parameters are described in \extref{appsubsec:baselines}.
For \espressoDEN, we measured the total runtime and also report the time for its steps $(ii)$ and $(iii)$. 
A solution computed for step $(i)$ can be used across multiple values of $\lamD$, requiring only the recomputation of steps $(ii)$ and $(iii)$.
For step $(iii)$ we use \blGreedy to produce an STC-feasible labeling; see~\extref{appsubsec:stcdenRounding}.

\Cref{tab:stcden} shows the results.
First, \espressoDEN is up to $17.6\times$ faster than \blPeel, 
and up to $592\times$ faster than \blGreedyKDD, which does not terminate within 900 seconds on \egofbData.
Our~\espressoDEN is more efficient since it avoids the explicit construction of a wedge graph and of its cover\ifshortversion.\else, detailed in~\extref{appsubsec:baselines}.\fi{}
\espressoDEN also attains the highest $\dens(\cdot;\lamD)$ in $13$ of the $16$ configurations, improving the solution of~\blPeel up to $44.3\%$ on \egofbData for $\lamD=0.2$.

\subsubsection{\Cref{RQ:scalability}: Scalability} 
We evaluate the scalability of \espressoGD, \espressoNL, and \espressoR on our largest graphs \hollData and
\bmgData with a one-hour limit, $K=10$, and $100$GB RAM.
{We set the sample size of \espressoR $s=10$, all methods use $\gamma=\lambda B/4+\delta$, i.e., the convex regime.}
The exact-gradient methods exceed the time limit since the graphs contain
$4.9$B and $3.6$B triangles ($87\times$ and $248\times$ their edge counts),
whereas \espressoR completes in at most $48$ and about $18$ minutes,
respectively, see~\Cref{tab:scalability}.
\begin{table}[t]
	\centering
	\caption{Scalability results. OOM denotes > 100GB RAM.}
	\label{tab:scalability}
	\resizebox{0.8\columnwidth}{!}{%
	\begin{tabular}{llcc}
		\toprule
		Dataset & Method & Outcome & Runtime (s)\\
		\midrule
		\multirow{3}{*}{\hollData} & \espressoGD & OOM & 2\,696.1 \\
		 & \espressoNL & Timeout & 3\,600.4 \\
		 & \espressoR & Success & 2\,873.0 \\
		\midrule
		\multirow{3}{*}{\bmgData} & \espressoGD & Timeout & 3\,600.3 \\
		 & \espressoNL & Timeout & 3\,600.3 \\
		 & \espressoR & Success & 1\,103.6 \\
		\bottomrule
	\end{tabular}}
\end{table}

{We also assess the accuracy of the approximate gradient of \espressoR.
Let $\pv^*$ be the solution of \espressoNL on \pokData and \wikiData (on such datasets \espressoNL terminates within a one hour limit in the convex regime).
We compute the error $\|\grad\obj(\pv^*)-\gradAppx\obj(\pv^*)\|_1=\lambda\|\countTriv(\pv^*)-\appxTriv(\pv^*)\|_1$ for $s\in[2,25]$, over $5$ independent runs for each value of $s$.
For every run the error respects the bound $\lambda\errEst m$ of~\extref{theo:appxGuarantees} for $\errEst=15$, and it decreases as $s$ increases\ifshortversion.\else{} (see~\extref{appsec:gradientAccuracy,appsec:solutionComparison}).\fi}

{Overall, our results show that \espressoR is an efficient and scalable accurate method on large and dense graphs that cannot be processed with exact methods.} 

\balance

\section{Conclusion}
\label{sec:conclusion}
\balance

We introduced \espresso, a probabilistic formulation of STC
that represents structural tie {strength through a fully factorized Bernoulli family and
an entropy-regularized objective $\obj$}.
{The resulting probabilities are a structural score of tie strength induced by the STC principle through $\obj$.}
We characterized the geometry of $\obj$ through the Gallai graph of $G$,
showing that its negative spectrum determines the exact boundary of global
convexity{, and predicts the transition between high-uncertainty and
more differentiated solutions}.
Building on these properties, we developed \espressoNL, an exact
mirror-descent algorithm with convergence guarantees{, and its randomized variants \espressoNLAppx and \espressoR (that avoids triangle enumeration), which provide probabilistic guarantees}.
{Our experiments confirm that the convexity threshold predicts the transition across synthetic graph families, and that the solutions to \espresso are aligned with optimal hard-STC labeling on several instances.}
{Furthermore, \espresso} captures finer-grained edge strengths than {existing LP-relaxations}.
Moreover, we demonstrated the effectiveness of our new algorithms for the downstream task of STC-densest subgraph discovery.

Future work includes {considering models that are not fully factorized, and extending the use of optimal solutions to \espresso for other network-analysis tasks, such as link prediction and clustering}.

\begin{acks}
	This research was partly supported by the ERC Advanced Grant REBOUND (834862),
	and the Wallenberg AI, Autonomous Systems and Software Program (WASP) funded by the Knut and Alice Wallenberg Foundation.
\end{acks}

\section*{Ethical considerations}
\label{sec:ethical-considerations}

\espresso is intended as a structural network-analysis tool for inferring tie strength and supporting downstream graph-analysis tasks. Because its outputs can highlight cohesive relationships and groups, they could potentially be misused for profiling or surveillance. The method relies only on observed network topology, uses no personal attributes, and does not infer missing links. Our experiments use only publicly available and synthetic datasets.

\ifshortversion\else
\section*{Use of generative AI}
\label{sec:ai-usage}

{The authors formulated the problem.
The use of mirror descent to solve it was proposed by Gemini (Google).
The authors proposed the approximate gradient and the approach that avoids
listing all triangles of the graph.
The proofs and key technical details were developed with Claude (Anthropic).
The authors designed the experiments, and the code was written with Claude.
AI tools also assisted with the interpretation of the results, the parsing
of the experimental logs, the literature review, and the writing.
The authors take full responsibility for the content of this paper.}
\fi

\bibliographystyle{ACM-Reference-Format}
\bibliography{literature}

\ifshortversion\else
\appendix

\onecolumn

\section[appendix]{Omitted theoretical results}\label{app:proofs}

\subsection[appendix]{Gibbs and mean-field interpretation}
\label{app:gibbs-mean-field}

We provide the derivation of the Gibbs interpretation stated in the main
text. Let $\hardCost$ be the penalized hard STC cost of
Eq.~\eqref{eq:hard_stc_cost} for fixed $\alpha,\lambda>0$, and let
\[
\pi_\gamma(\bm z)
\doteq
\frac{1}{\mathcal{Z}_\gamma}
\exp\left(-\frac{\hardCost(\bm z)}{\gamma}\right),
\qquad
\mathcal{Z}_\gamma
\doteq
\sum_{\bm z\in\{0,1\}^m}
\exp\left(-\frac{\hardCost(\bm z)}{\gamma}\right).
\]
For any distribution $q(\bm z;\pv)$ in the factorized
family defined in \Cref{eq:factorization}, consider
\begin{align}
\mathrm{KL}\bigl(q(\cdot;\pv)\,\|\,\pi_\gamma\bigr)
&=
\E_q\left[
\log q(\bm z;\pv)
-
\log \pi_\gamma(\bm z)
\right]
\nonumber\\
&=
\E_q[\log q(\bm z;\pv)]
+
\frac{1}{\gamma}\E_q[\hardCost(\bm z)]
+
\log \mathcal{Z}_\gamma.
\end{align}
By computing the entropy of the fully factorized Bernoulli distribution we obtain
\[
-\E_q[\log q(\bm z;\pv)]
=
-\sum_{e\in E}
\left(
p_e\log p_e
+
(1-p_e)\log(1-p_e)
\right)
=
\Rf(\pv).
\]
Therefore, multiplying by $\gamma>0$ and substituting
$\E_q[\log q(\bm z;\pv)]=-\Rf(\pv)$,
\begin{align}
\gamma\,
\mathrm{KL}\bigl(q(\cdot;\pv)\,\|\,\pi_\gamma\bigr)
&=
\E_q[\hardCost(\bm z)]
-
\gamma\Rf(\pv)
+
\gamma\log \mathcal{Z}_\gamma
\nonumber\\
&=
\obj(\pv)
+
\gamma\log \mathcal{Z}_\gamma,
\end{align}
where the second equality follows from
Eq.~\eqref{eq:expected_hard_cost} and the definition of $\obj$.
Hence,
\[
\obj(\pv)
=
\gamma\,
\mathrm{KL}\bigl(q(\cdot;\pv)\,\|\,\pi_\gamma\bigr)
-
\gamma\log \mathcal{Z}_\gamma.
\]
Since $\mathcal{Z}_\gamma$ depends only on $\hardCost$ and $\gamma$, and
not on $\pv$, the minimizers of $\obj$ coincide with
\[
\operatorname{argmin}_{\pv\in(0,1)^m}
\mathrm{KL}\bigl(q(\cdot;\pv)\,\|\,\pi_\gamma\bigr).
\]
Thus, optimizing the objective of \espresso corresponds precisely to the naive mean-field variational
approximation of the Gibbs distribution induced by the penalized hard-STC
cost. If
\[
q^*(\bm z)=q(\bm z;\pv^*),
\]
then
\[
p_e^*=\Prob_{q^*}[Z_e=1].
\]
{That is, $p_e^*$ is the probability that an edge $e\in E$ is strong under $q^*$.
This value, in general, may differ from the probability $\Prob_{\pi_\gamma}[Z_e=1]$ that
$e$ is strong under $\pi_\gamma$.
In fact, the second term of $\hardCost(\bm z)$ (see \Cref{eq:hard_stc_cost}) counts the wedges
$\{e,e'\}\in\Wset$ with $z_e=z_{e'}=1$, i.e., the STC violations, and thus depends on the labels of
the two edges of each wedge. 
Hence, under $\pi_\gamma$ the labels of the edges of a given wedge in $\Wset$ are \emph{not} independent.
In contrast, every distribution in the family of \Cref{eq:factorization} has
independent edge labels, so $q^*$ cannot represent such dependencies.
If $\Wset=\emptyset$ (i.e., every connected component of $G$ is a clique), then $\hardCost$ has no term accounting for the wedges in the graph, and thus $\pi_\gamma$ belongs to the family of \Cref{eq:factorization}, $q^*=\pi_\gamma$, and the two probabilities coincide.}

\subsection[appendix]{Technical preliminaries}\label{app:prelims}

\begin{definition}\label{lemma:PSDConvex}
Consider a function $f:\mathcal{X}\mapsto\mathcal{Y}$ that is twice differentiable, i.e., $\nabla^2 f$ exists at each point in $\mathcal{X}$, which is open. Then $f$ is convex if and only if its domain is convex and its Hessian is positive semidefinite: for all $\bm{x}\in \mathcal{X}$, $\nabla^2f(\bm{x}) \succeq 0 $. 
\end{definition}

\begin{definition}\label{lemma:relativeSmoothandConv}
Consider functions $f,h$ that map $\mathcal{X}\mapsto\mathcal{Y}$ which are twice differentiable at each point over their domain $\mathcal{X}$. We say that
\begin{itemize}
\item $f(\cdot)$ is $L$-smooth relative to $h(\cdot)$ on $\mathcal{X}$ if for any $\bm{p},\bm{q}\in \inter \mathcal{X}$ there exists a scalar $L$ such that
\[
f(\bm{q})\le f(\bm{p}) + \langle\nabla f(\bm{p}),\bm{q}-\bm{p} \rangle + L D_h(\bm{q},\bm{p}),
\]
or alternatively, if twice differentiable, for any $\bm{q}\in \inter \mathcal{X}$ it holds that
\[
\nabla^2f(\bm{p}) \preceq L \nabla^2h(\bm{p}) .
\]
\item $f(\cdot)$ is $\mu$-strongly convex relative to $h(\cdot)$ on $\mathcal{X}$ if for any $\bm{p},\bm{q}\in \inter \mathcal{X}$ there exists a scalar $\mu >0$ such that
\[
f(\bm{q})\ge f(\bm{p}) + \langle\nabla f(\bm{p}),\bm{q}-\bm{p} \rangle + \mu D_h(\bm{q},\bm{p}),
\]
or alternatively, if twice differentiable, for any $\bm{q}\in \inter \mathcal{X}$ it holds that
\[
\nabla^2f(\bm{p}) \succeq \mu \nabla^2h(\bm{p}) .
\]
\end{itemize}
\end{definition}

\begin{definition}[Legendre function, Def. 1 in~\cite{Bauschke2017NoLips}] Let $h:X\to (-\infty, \infty]$ be a lsc proper convex function. It is called:
\begin{enumerate}
\item \emph{essentially smooth}, if $h$ is differentiable on $\inter \doma h$ with moreover $||\nabla h(x^k)||\to \infty$ for every sequence $\{x^k\}_{k\in\mathbb{N}} \subset \inter \doma h$ converging to a boundary point of $\doma h$ as $k\to \infty$;
\item of \emph{Legendre type} if $h$ is essentially smooth and strictly convex on $\inter \doma h$.
\end{enumerate}
\end{definition}

\begin{lemma}[Gershgorin, Theorem 6.1.1~\cite{horn2012matrix}]\label{lemma:GershEigen}
Let $\Hess \in \R^{m\times m}$  be a symmetric matrix. If for each $i\in[m]$ it holds
\[
\Hess_{ii} \ge \sum_{j\neq i}|\Hess_{ij}|,
\]
then every eigenvalue of $\Hess$ is non-negative, that is $\Hess$ is positive semidefinite, i.e., $\Hess \succeq 0$.
\end{lemma}

\begin{lemma}\label{lemma:BregmanExpr}
Under the choice of $h=- \Rf(\pv)$ it holds that for any $\pv,\qv\in \doma h$, the Bregman divergence $D_h(\pv, \qv)  =  h(\pv) - h(\qv) - \langle\nabla h(\qv), \pv-\qv\rangle$ corresponds to 
\begin{equation*}
D_h(\pv, \qv) = \sum_{e\in E}\left[p_e \log\frac{p_e}{q_e} 
+ (1-p_e)\log\frac{1-p_e}{1-q_e}\right] .
\end{equation*}
\end{lemma}
\begin{proof}
We proceed by evaluating the definition of $h$ for $D_h$. That is, consider $\pv,\qv\in\doma h$. Since $\nabla_e h(\qv) = \log(q_e/(1-q_e)), e\in E$, we have that
\begin{align*}
&h(\pv) = \sum_e p_e \log p_e + (1-p_e) \log(1-p_e),\\
&h(\qv) = \sum_e q_e \log q_e + (1-q_e) \log(1-q_e),\\
&\langle \nabla h(\bm{q}) , \pv -\qv\rangle = \sum_e (p_e-q_e)\log\frac{q_e}{1-q_e} .
\end{align*}
substituting into the definition of $D_h$ gives
\begin{align*}
D_h(\pv,\qv) 
&= \sum_e \Bigl[
p_e \log p_e + (1-p_e)\log(1-p_e)
- q_e \log q_e\\
&- (1-q_e)\log(1-q_e) 
- (p_e - q_e)\log\tfrac{q_e}{1-q_e}
\Bigr] \\
&= \sum_e \Bigl[
p_e\log \frac{p_e}{q_e}
- (1-q_e-p_e+q_e)\log(1-q_e) \\
&+ (1-p_e)\log(1-p_e)\Bigr]\\
&= \sum_e \Bigl[
p_e\log \frac{p_e}{q_e}
+ (1-p_e)\log\frac{1-p_e}{1-q_e}
\Bigr] ,
\end{align*}
concluding the proof.
\end{proof}

\subsection[appendix]{Proofs for the objective properties}\label{app:proofsObjective}

\begin{proof}[Proof of~\Cref{prop:local_stationary_form}]
To prove the first part of the claim we start by computing the gradient $\frac{\partial \obj}{\partial p_{e}} $.  
By Eq.~\eqref{eq:full_objective} we can express the gradient as $\frac{\partial \obj(\pv)}{\partial p_{e}} = \alpha \frac{\partial \We(\bm{p})}{\partial p_{e}} + \lambda \frac{\partial \V(\bm{p})}{\partial p_{e}}-\gamma \frac{\partial \Rf(\pv)}{\partial p_{e}}$ where
\begin{align*}
\alpha \frac{\partial \We(\bm{p})}{\partial p_{e}} &= \alpha \frac{\partial}{\partial p_{e}} \sum_{e\in E} (1-p_e) = -\alpha	,\\
\lambda \frac{\partial \V(\bm{p})}{\partial p_{e}} &= \lambda \frac{\partial}{\partial p_{e}} \sum_{\omega=\{e_1,e_2\}\in \mathcal{W}} p_{e_1}\,p_{e_2} = \lambda\sum_{w=\{e,e'\}\in \Wset_e} p_{e'}\\
& =\lambda \edPress_e(\pv),\\
-\gamma \frac{\partial \Rf(\pv)}{\partial p_{e}}  &=-\gamma \frac{\partial}{\partial p_{e}} -\sum_{e\in E}\bigl[p_e\log p_e + (1-p_e)\log(1-p_e) \bigr]\\
&= -\gamma\left(\log (1-p_e) - \log p_e  \right) = -\gamma \log\frac{1-p_e}{p_e},
\end{align*}
yielding the first part of the claim. We now show that any local minimizer of
\(\obj(\bm p)\) lies in the open hypercube \((0,1)^{|E|}\).
Consider \(p_e\in(0,1)\), then by Eq.~\eqref{eq:gradient_entropy_open} when \(p_e\to 0\), it holds that $\log((1-p_e)/(p_e)) \to +\infty$, and since both $\alpha$ and $\lambda \edPress_e(\pv)$ are bounded, the value of $\frac{\partial \obj(\pv)}{\partial p_{e}} $
tends to \(-\infty\). 
Thus increasing \(p_e\) from 0 decreases the objective.
As \(p_e\to 1\), it holds that $\log((1-p_e)/(p_e)) \to -\infty$, so the value of $\frac{\partial \obj(\pv)}{\partial p_{e}} $
tends to \(+\infty\). 
Thus decreasing \(p_e\) from 1 decreases the objective.
The above implies that no local minimizer can occur on the boundary.

Consider any minimizer $\pv^*\in(0,1)^m$, since the derivative must be zero we have that
\[
-\alpha
+
\lambda \edPress_e(\pv^*)
-
\gamma
\log\left(\frac{1-p_e^*}{p_e^*}\right)
=
0.
\]
By reordering the terms we have
\[
\log\frac{p_e^*}{1-p_e^*}
=
\frac{\alpha-\lambda \edPress_e(\pv^*)}{\gamma},
\]
and by the definition of function $\sigma(z)$ it holds $p_e^*
=
\sigma\!(
\frac{\alpha-\lambda \edPress_e(\pv^*)}{\gamma}
)$, concluding the proof.
\end{proof}

\subsection[appendix]{Proofs for the optimization with \espressoNL}\label{app:proofsNL}

\begin{proof}[Proof of~\Cref{lemma:Fprops}]
{We first compute the Hessians of $\obj$ and $h$.
By Eq.~\eqref{eq:gradient_entropy_open} of \Cref{prop:local_stationary_form}, $\nabla_e\obj(\pv)=-\alpha+\lambda\edPress_e(\pv)-\gamma\log\frac{1-p_e}{p_e}$.
Since $\edPress_e(\pv)=\sum_{\omega\in\Wset_e}p_{e'}$, we have $\partial\edPress_e(\pv)/\partial p_{e'}=\Mmat_{e,e'}$ for every $e'\in E$.
Moreover, $\frac{\partial}{\partial p_e}\bigl(-\log\frac{1-p_e}{p_e}\bigr)=\frac{1}{p_e(1-p_e)}$.
Let $\bm{D}(\pv)\doteq\diag\bigl((p_e(1-p_e))^{-1}\bigr)_{e\in E}$.
Then
\[
\nabla^2\obj(\pv)=\lambda\Mmat+\gamma\bm{D}(\pv),
\qquad \nabla^2h(\pv)=\bm{D}(\pv),
\]
where the second equality follows from the same computation applied to $h=-\Rf$.}

{Next, we show that $-\tfrac{\xi_-}{4}\bm{D}(\pv)\preceq\Mmat\preceq\tfrac{\xi_+}{4}\bm{D}(\pv)$.
Since every eigenvalue of $\Mmat$ lies in $[-\xi_-,\xi_+]$, it holds $-\xi_-\Imat\preceq\Mmat\preceq\xi_+\Imat$.
Since $p_e(1-p_e)\le1/4$ for $p_e\in(0,1)$, every diagonal entry of $\bm{D}(\pv)$ is lower bounded by $4$, that is, $\Imat\preceq\tfrac14\bm{D}(\pv)$.
Combining the two inequalities shows that $-\tfrac{\xi_-}{4}\bm{D}(\pv)\preceq\Mmat\preceq\tfrac{\xi_+}{4}\bm{D}(\pv)$ as claimed.
Using such identity we obtain for that every $\pv\in(0,1)^m$ it holds
\begin{equation}\label{eq:hessSandwich}
	(\gamma-\gamma_c)\,\nabla^2h(\pv)
	\ \preceq\ \nabla^2\obj(\pv)
	\ \preceq\ \Bigl(\gamma+\frac{\lambda\xi_+}{4}\Bigr)\nabla^2h(\pv),
\end{equation}
where we multiplied by $\lambda>0$, added $\gamma\bm{D}(\pv)$, and used $\gamma_c=\lambda\xi_-/4$.
We now prove the three claims.
\begin{compactitem}
	\item[$(ii)$] The right inequality of Eq.~\eqref{eq:hessSandwich} corresponds to $\nabla^2\obj(\pv)\preceq L\,\nabla^2h(\pv)$ with $L=\gamma+\lambda\xi_+/4$, hence $\obj$ is $L$-smooth relative to $h$ by \Cref{lemma:relativeSmoothandConv}.
	\item[$(iii)$] If $\gamma>\gamma_c$, the left inequality of Eq.~\eqref{eq:hessSandwich} corresponds to $\nabla^2\obj(\pv)\succeq\mu\,\nabla^2h(\pv)$ with $\mu=\gamma-\gamma_c>0$, hence $\obj$ is $\mu$-strongly convex relative to $h$ by \Cref{lemma:relativeSmoothandConv}.
	\item[$(i)$] If $\gamma\ge\gamma_c$, the left inequality of Eq.~\eqref{eq:hessSandwich} and $\nabla^2h(\pv)=\bm{D}(\pv)\succeq0$ yield $\nabla^2\obj(\pv)\succeq0$ for every $\pv\in(0,1)^m$. Hence $\obj$ is convex by \Cref{lemma:PSDConvex}.
	Conversely, consider $\gamma<\gamma_c$ and $\bar\pv=\tfrac12\bm{1}$.
	Then $\bm{D}(\bar\pv)=4\Imat$, thus $\nabla^2\obj(\bar\pv)=\lambda\Mmat+4\gamma\Imat$, whose smallest eigenvalue corresponds to $4\gamma-\lambda\xi_-=4(\gamma-\gamma_c)<0$.
	Hence $\nabla^2\obj(\bar\pv)$ is not positive semidefinite, and $\obj$ is not convex by \Cref{lemma:PSDConvex}.
\end{compactitem}}
{The above three properties yield the claim, concluding the proof}
\end{proof}

\begin{proof}[Proof of~\Cref{prop:balancedTransition}]
Let $\bm{s}=2\pv-\bm{1}$ and $\tobj(\bm{s})\doteq\obj(\tfrac12(\bm{1}+\bm{s}))$ for $\bm{s}\in(-1,1)^m$.
This transformation is an affine bijection from $(-1,1)^m$ onto $(0,1)^m$ that maps $\bm{0}$ to $\bar\pv$.
Hence our mapping preserves stationary points, local minimizers, and (unique) global minimizers, we thus consider $\tobj$ in our proof.

\textbf{Centered objective.}
Each wedge $\{e,e'\}\in\Wset$ appears two times in the evaluation of $\pv^{\top}\Mmat\pv$, i.e., as $(e,e')$ and as $(e',e)$.
Hence, by Eq.~\eqref{eq:stc_violations}, $\V(\pv)=\tfrac12\pv^{\top}\Mmat\pv$.
Let $\psi(x)\doteq\frac{1+x}{2}\log\frac{1+x}{2}+\frac{1-x}{2}\log\frac{1-x}{2}$ for $x\in(-1,1)$, such that $\Rf(\pv)=-\sum_e\psi(s_e)$.
Substituting $\pv=\tfrac12(\bm{1}+\bm{s})$ over Eq.~\eqref{eq:full_objective} yields
\[
\tobj(\bm{s})=\frac{\alpha m}{2}+\frac{\lambda md}{8}
+\Bigl(\frac{\lambda d}{4}-\frac{\alpha}{2}\Bigr)\bm{1}^{\top}\bm{s}
+\frac{\lambda}{8}\bm{s}^{\top}\Mmat\bm{s}
+\gamma\sum_e\psi(s_e),
\]
where we used $\Mmat\bm{1}=d\bm{1}$, since the Gallai graph is $d$-regular.
Since $\alpha=\lambda d/2$, the coefficient of $\bm{1}^{\top}\bm{s}$ is zero, hence
\begin{equation}\label{eq:centeredObj}
	\tobj(\bm{s})=\frac{\alpha m}{2}+\frac{\lambda md}{8}+\frac{\lambda}{8}\bm{s}^{\top}\Mmat\bm{s}+\gamma\sum_e\psi(s_e),
	\qquad \bm{s}\in(-1,1)^m .
\end{equation}
The first two terms do not depend on $\bm{s}$.
Hence, such terms do not affect the gradient of $\tobj$ or the difference $\tobj(\bm{s})-\tobj(\bm{s}')$ for every $\bm{s},\bm{s}'\in(-1,1)^m$, in the rest of the proof we will consider only gradients or differences for $\tobj$, thus not reporting the first two terms.
Since $\psi'(x)=\tfrac12\log\tfrac{1+x}{1-x}=\operatorname{atanh}(x)$, then
$\nabla\tobj(\bm{s})=\tfrac{\lambda}{4}\Mmat\bm{s}+\gamma(\operatorname{atanh}(s_e))_{e\in E}$, hence $\nabla\tobj(\bm{0})=\bm{0}$.
That is, $\bar\pv$ is a stationary point of $\obj$ for every $\gamma>0$.

\textbf{Lower bound.}
Let $\mathcal{E}_-\doteq\ker(\Mmat+\xi_-\Imat)$ be the eigenspace associated with $\lambda_{\min}(\Mmat)=-\xi_-$.
Since $d>0$, we have $\Wset\neq\emptyset$, hence $\Mmat\neq\bm{0}$.
Since $\Mmat$ is symmetric with zero diagonal, its eigenvalues sum to $\operatorname{tr}\Mmat=0$, hence $\lambda_{\min}(\Mmat)<0$, i.e., $\xi_->0$.
The matrix $\Mmat+\xi_-\Imat$ is thus positive semidefinite, hence
$\bm{s}^{\top}\Mmat\bm{s}\ge-\xi_-\|\bm{s}\|_2^2$, with equality iff $\bm{s}\in\mathcal{E}_-$ (since $\Mmat+\xi_-\Imat\succeq0$, $\bm{s}^{\top}(\Mmat+\xi_-\Imat)\bm{s}=0$ iff $(\Mmat+\xi_-\Imat)\bm{s}=\bm{0}$).
Let
\[
\phi_\gamma(x)\doteq\gamma\bigl(\psi(x)-\psi(0)\bigr)-\frac{\gamma_c}{2}x^2,
\qquad x\in(-1,1).
\]
By Eq.~\eqref{eq:centeredObj} and substituting $\gamma_c=\lambda\xi_-/4$, for every $\bm{s}\in(-1,1)^m$ we have that,
\begin{equation}\label{eq:sepLower}
	\tobj(\bm{s})-\tobj(\bm{0})\ \ge\ \sum_e\phi_\gamma(s_e),
	\qquad\text{with equality iff } \bm{s}\in\mathcal{E}_- .
\end{equation}

The function $\phi_\gamma$ is even, $\phi_\gamma(0)=0$, and
$\phi_\gamma'(x)=\gamma\operatorname{atanh}(x)-\gamma_c x=\gamma x\,(\vartheta(x)-\gamma_c/\gamma)$, where we let
\[
\vartheta(x)\doteq\frac{\operatorname{atanh}(x)}{x}=\sum_{k\ge0}\frac{x^{2k}}{2k+1},
\qquad \vartheta(0)\doteq1 .
\]
By the power series, $\vartheta$ is continuous and strictly increasing on $[0,1)$, and $\vartheta(x)\to+\infty$ as $x\to1^-$.
Hence $\vartheta$ is a bijection from $[0,1)$ onto $[1,+\infty)$.
We now distinguish two cases as follows,
\begin{compactenum}[($a$)]
	\item If $\gamma\ge\gamma_c$, then $\vartheta(x)>1\ge\gamma_c/\gamma$ for $x\in(0,1)$, hence $\phi_\gamma'>0$ on $(0,1)$.
	Since $\phi_\gamma$ is even, $\phi_\gamma(x)>0$ for every $x\neq0$.
	\item If $0<\gamma<\gamma_c$, then consider $a_\gamma\doteq\vartheta^{-1}(\gamma_c/\gamma)$. 
	That is $a_\gamma$ corresponds to the unique solution in $(0,1)$ of
	$\operatorname{atanh}(a)=(\gamma_c/\gamma)a$, i.e., $a_\gamma$ is the solution to $a=\tanh(\gamma_c a/\gamma)$.
	Moreover, it holds that $\phi_\gamma'<0$ on $(0,a_\gamma)$ and $\phi_\gamma'>0$ on $(a_\gamma,1)$, i.e., $\phi_\gamma$ is strictly decreasing on $[0,a_\gamma]$ and strictly increasing on $[a_\gamma,1)$.
	Since $\phi_\gamma(0)=0$ and $\phi_\gamma$ is even, then we have that $\phi_\gamma(x)<0$ for $0<|x|\le a_\gamma$, and the minimum of $\phi_\gamma$ over $(-1,1)$ is attained exactly at the points $\pm a_\gamma$.
\end{compactenum}

\textbf{Proof of ($i$).}
Let $\gamma\ge\gamma_c$ and $\bm{s}\neq\bm{0}$.
From case~($a$), every summand in Eq.~\eqref{eq:sepLower} is non-negative and at least one is positive, thus we have that $\tobj(\bm{s})>\tobj(\bm{0})$, which proves ($i$).

\textbf{Proof of ($ii$).}
Let $0<\gamma<\gamma_c$ and $\bm{u}\in\mathcal{E}_-$ with $\|\bm{u}\|_\infty=1$.
For any $0<|t|\le a_\gamma$, the point $t\bm{u}$ lies in $\mathcal{E}_-\cap(-1,1)^m$, thus Eq.~\eqref{eq:sepLower} holds with equality since $\Mmat(t\bm{u})=-\xi_-t\bm{u}$.
Since $|tu_e|\le a_\gamma$ for every $e\in E$, by case~($b$) every summand $\phi_\gamma(tu_e)$ is non-positive, and it is negative when $u_e\neq0$.
Hence $\tobj(t\bm{u})<\tobj(\bm{0})$ for every $0<|t|\le a_\gamma$, and $\bm{0}$ is not a local minimizer of $\tobj$.

\textbf{Global minimizers for $0<\gamma<\gamma_c$.}
By Eq.~\eqref{eq:sepLower} and case~($b$), for every $\bm{s}\in(-1,1)^m$ it holds
\begin{equation}\label{eq:globalLower}
	\tobj(\bm{s})-\tobj(\bm{0})\ \ge\ \sum_e\phi_\gamma(s_e)\ \ge\ m\,\phi_\gamma(a_\gamma).
\end{equation}
When $\bm{s}=\pm a_\gamma\bm{q}$ both inequalities hold with equality: the first since $\pm a_\gamma\bm{q}\in\mathcal{E}_-$, the second since every coordinate of $\pm a_\gamma\bm{q}$ is $\pm a_\gamma$ and $\phi_\gamma$ is even.
Hence $\pm a_\gamma\bm{q}$ are global minimizers of $\tobj$, i.e., $\pv^{\pm}$ are global minimizers of $\obj$.

Now consider the case when $\mathcal{E}_-=\operatorname{span}(\bm{q})$ (that is, the eigenspace is one-dimensional), and let $\bm{s}^*$ be any global minimizer of $\tobj$.
Then both inequalities in Eq.~\eqref{eq:globalLower} hold with equality for $\bm{s}^*$.
By case~($b$), the second equality yields $s^*_e\in\{-a_\gamma,a_\gamma\}$ for every $e\in E$.
While the first equality yields $\bm{s}^*=c\bm{q}$ for some $c\in\R$.
Since $\bm{q}\in\{-1,1\}^m$, it follows that $|c|=a_\gamma$, i.e., $\bm{s}^*=\pm a_\gamma\bm{q}$, proving the remaining part of the claim.

\textbf{Limit of $a_\gamma$.}
Finally, consider the inverse $\vartheta^{-1}:[1,+\infty)\to[0,1)$, which is strictly increasing and onto $[0,1)$, hence $\vartheta^{-1}(c)\to1$ as $c\to+\infty$.
Since $a_\gamma=\vartheta^{-1}(\gamma_c/\gamma)$ and $\gamma_c/\gamma\to+\infty$ as $\gamma\to0^+$, we have $a_\gamma\to1$, concluding the proof.
\end{proof}

\begin{proof}[Proof of~\Cref{theo:accNL}]
First observe that \espressoNL computes {$L=\gamma+\lambda|\Wmax|/4\ge\gamma+\lambda\xi_+/4$, since $\xi_+\le|\Wmax|$ (by \Cref{lemma:GershEigen}). Thus, for $\gamma>\gamma_c$, by \Cref{lemma:Fprops}, $\obj$ is $L$-smooth and $\mu$-convex relative to $h$ for $\mu=\gamma-\gamma_c>0$.} We now use Theorem 3.1 from \citet{Lu2018NoLipsConvex}, which proved an analysis of the \nolips algorithm when $\obj$ is $\mu$-convex relative to $h$, for $\mu > 0$. 
That is, we get that at any iteration $k\ge 0$ of \espressoNL it holds that,
\begin{align}\label{eq:objBoundInner}
\obj(\pv^{k}) - \obj(\pv^{*}) \le L \left(1- \frac{\mu}{L}\right)^k D_h(\pv^*, \pv^{(0)}),
\end{align}
where $\pv^*$ is a minimizer of $\obj$.
We now bound Bregman divergence term, first recall that $\pv^{(0)} = 1/2 \cdot \bm{1}$, hence
\begin{align*}
h(\pv^{0}) &= \sum_{e\in E} \log \tfrac{1}{2} = - m \log 2 , \quad \text{ and}\\
\langle\nabla h(\pv^{0}), \pv^* - \pv^{0}\rangle &= \sum_e (p_e^*-1/2)\log \tfrac{1/2}{1/2}  = 0 .
\end{align*}
By noting that for any $e\in E$ and any $\pv \in (0,1)^m$  it holds that $p_e\log p_e + (1-p_e) \log (1-p_e) \le 0$ we have that
\begin{align*}
D_h(\pv^*, \pv^{0}) &= h(\pv^*) - h(\pv^{0}) - \langle \nabla h(\pv^{0}), \pv^* - \pv^{0}\rangle\\
&= h(\pv^*) + m \log 2 \le m \log 2 ,
\end{align*}
where the last step follows by the fact that $\pv^*\in(0,1)^m$. Finally, plugging in the above bound to Eq.~\eqref{eq:objBoundInner} we have that
\begin{align*}
L \left(1- \frac{\mu}{L}\right)^k D_h(\pv^*, \pv^{0}) & \le L \left(1- \frac{\mu}{L}\right)^k m \log 2 .
\end{align*}
Hence by the choice of $k=K$ we have that
\begin{align*}
\obj(\pv^{K}) - \obj(\pv^{*}) &\le L \left(1- \frac{\mu}{L}\right)^K m \log 2 \\
&\le L \exp\left(-K \frac{\mu}{L}\right) m \log 2 ,
\end{align*}
where the second inequality follows from the bound $1-x \le \exp(-x)$ for $x \in (0,1)$.
To guarantee that $\obj(\pv^{K}) - \obj(\pv^{*}) \le \errAcc$, we bound the above term with $\errAcc$, yielding
\[
\exp\left(-K \frac{\mu}{L}\right) \le \frac{\errAcc}{L m \log 2}.
\]
And by taking the logarithm of both sides
we have that
\[
K \ge \frac{L}{\mu} \log \left(\frac{L m \log 2}{\errAcc}\right).
\]
{For the choice of $\gamma=\lambda|\Wmax|/4+\delta$, we have that $\gamma_c\le\lambda|\Wmax|/4$ since $\xi_-\le|\Wmax|$ (by \Cref{lemma:GershEigen}), hence $\mu\ge\delta$, and $L/\mu\le L/\delta$. Thus any value of $K$ satisfying $K\ge\frac{L}{\delta}\log\left(\frac{Lm\log 2}{\errAcc}\right)$ with $L=\frac{\lambda|\Wmax|+2\delta}{2}$ corresponds to a sufficient choice for the algorithm to obtain its approximation guarantees from the claim}, concluding the proof.
\end{proof}

\begin{proof}[Proof of~\Cref{lemma:rtEx}]
First observe that
\begin{align*}
|\Wmax| &= \max_{e=(u,v)} \{|\Wset_e|\}
= \max_{e=(u,v)} \sum_{\wed=\{e,e'\}\in \Wset_e} 1\\
&\le \max_{e=(u,v)} (d_u+d_v-2) \le 2(\dMax-1).
\end{align*}
By \Cref{theo:accNL}, 
the total number of iterations $K$ of \espressoNL is bounded by $\bigO(\lambda \dMax \delta^{-1} \log (\lambda m/\errAcc))$. 
At each iteration $k\ge 0$, the computational cost is dominated by the evaluation of the full gradient $\nabla \obj(\pv^{k})$. Computing the gradient requires listing all triangles in $G$.
Using a classical result from Chiba and Nishizeki~\cite{Chiba1985Triangles}, we can enumerate all triangles in a graph in time $\bigO(\degen m)$.\footnote{The original result in~\cite{Chiba1985Triangles} is parameterized by the \emph{arboricity} of $G$, which is strictly bounded within a constant factor of the degeneracy $\degen$.} 
Multiplying the bound on the number of iterations $K$ by the $\bigO(\degen m)$ cost for each individual iteration, yields the claimed total time complexity, and concludes our proof.
\end{proof}

\subsection[appendix]{Proofs for \espressoNLAppx and \espressoR}\label{app:proofsAppx}

\begin{lemma}[Approximate three point descent property]\label{lemma:threePoint}
Let $h$ be a Legendre type function. At any iteration $i\ge 1$ it holds for points $\pv^{i}, \pv^{i-1}$ selected by~\espressoNLAppx and any $\pv\in \doma \obj = (0,1)^m$
\begin{align*}
L D_h(\pv,\pv^{i}) + L D_h(\pv^{i},\pv^{i-1}) &- L D_h(\pv,\pv^{i-1}) 
\le \langle\gradAppx f(\pv^{i-1}), \pv -\pv^{i}\rangle .
\end{align*}
\end{lemma}
\begin{proof}
For ease of notation we let $f=\obj$ in the rest of the proof.
First recall the three point property (e.g., see Lemma 3 in~\cite{Bauschke2017NoLips}), where for any $x\in \doma h$ and $y,z \in \inter \doma h$
\[
L D_h(x,z)-L D_h(x,y) - L D_h(y,z) = L \langle\nabla h(y) - \nabla h(z), x-y\rangle  .
\]
At any iteration $i\ge 1$ consider points $\pv^{i}, \pv^{i-1}$ explored by~\espressoNLAppx and any $\pv\in (0,1)^m$, then by substituting $z= \pv^{i-1}$ and $y= \pv^{i}$ and $x = \pv$ we get 
\begin{align}\label{eqproof:threedecomp}
L D_h(\pv,\pv^{i-1}) &- L D_h(\pv,\pv^{i})
- L D_h(\pv^{i},\pv^{i-1}) = L \langle\nabla h(\pv^{i}) - \nabla h(\pv^{i-1}), \pv-\pv^{i}\rangle  .
\end{align}
Note that at any iteration $i\ge 1$ the algorithm selects
\begin{align*}
\pv^{i} = \arg\min_{\pv\in (0,1)^m}\{f(\pv^{i-1}) &+ \langle \gradAppx f(\pv^{i-1}), \pv-\pv^{i-1}\rangle + L D_h(\pv,\pv^{i-1})\} ,
\end{align*}
by the optimization of $\obj$ according to the approximate gradient $\gradAppx$ in~\espressoNLAppx. 
Computing the gradient of the function within the $\arg\min$ above, with respect to $\pv$, we obtain
\[
\gradAppx f(\pv^{i-1}) + L (\nabla h(\pv) - \nabla h(\pv^{i-1})),
\]
by the form of the Bregman divergence $D_h$.
Hence by evaluating the above gradient at the optimal $\pv^i$ and for any $\pv \in (0,1)^m$ it holds that
$
\langle \gradAppx f(\pv^{i-1}) + L (\nabla h(\pv^{i}) - \nabla h(\pv^{i-1})),\pv-\pv^{i} \rangle \ge 0 ,
$
by the optimality of $\pv^i$. Therefore, by rearranging the terms we have that
\[
L\langle\nabla h(\pv^i) - \nabla h(\pv^{i-1}), \pv - \pv^i \rangle \ge  - \langle
\gradAppx f(\pv^{i-1}), \pv -\pv^{i}\rangle .
\]
By plugging in the above to Eq.~\eqref{eqproof:threedecomp} we get that
\begin{align*}
L D_h(\pv,\pv^{i-1}) - L D_h(\pv,\pv^{i})
- L D_h(\pv^{i},\pv^{i-1}) \ge  - \langle \gradAppx f(\pv^{i-1}), \pv -\pv^{i}\rangle ,
\end{align*}
and hence
\begin{align*}
L D_h(\pv,\pv^{i}) + L D_h(\pv^{i},\pv^{i-1}) - L D_h(\pv,\pv^{i-1}) 
\le  \langle\gradAppx f(\pv^{i-1}), \pv -\pv^{i}\rangle ,
\end{align*}
concluding the proof.
\end{proof}

\begin{lemma}\label{lemma:appxGradGuar}
Consider~\espressoNLAppx. Under the setting of parameters $\gamma \ge \lambda |\Wmax|/4 + \delta$ for $\alpha>0, \lambda>0, \delta>0, \errEst>0, \eta\in(0,1)$, for any $i\ge 1$ and $\pv\in (0,1)^m$ with probability at least $1-\eta$ over the execution of the algorithm it holds that
\begin{align*}
\obj(\pv^i) - \obj(\pv^*) \le L\left(1-\frac{\mu}{L}\right)^i D_h(\pv^*,\pv^0) + \lambda\errEst m i.
\end{align*}
\end{lemma}
\begin{proof}
Similar to the proof of \Cref{lemma:threePoint} we let $f=\obj$ for ease of notation.
Since $f$ is $L$-smooth and strictly $\mu$-convex relative to $h$ (see Def.~\ref{lemma:relativeSmoothandConv}) we have that for any $\pv,\qv\in\doma f$,
\begin{align*}
f(\bm{q})&\le f(\bm{p}) + \langle\nabla f(\bm{p}),\bm{q}-\bm{p} \rangle + L D_h(\bm{q},\bm{p}),\\
f(\bm{q})&\ge f(\bm{p}) + \langle\nabla f(\bm{p}),\bm{q}-\bm{p} \rangle + \mu D_h(\bm{q},\bm{p}).
\end{align*}
By evaluating the first inequality at $\qv=\pv^{i}$ and $\pv=\pv^{i-1}$ and the second inequality at $\pv = \pv^{i-1}$ and any $\qv\in (0,1)^m$
\begin{align*}
&f(\pv^i)\le f(\pv^{i-1}) + \langle\nabla f(\pv^{i-1}),\pv^i-\pv^{i-1}\rangle + L D_h(
\pv^i,\pv^{i-1}),\\
&f(\qv)\ge f(\pv^{i-1}) + \langle\nabla f(\pv^{i-1}),\qv-\pv^{i-1} \rangle + \mu D_h(\qv,\pv^{i-1}).
\end{align*}
Isolating the $f(\pv^{i-1})$ term in the second inequality and combining the two inequalities we get that
\begin{align*}
f(\pv^i)\le f(\qv) + \langle\grad f(\pv^{i-1}), \pv^i-\pv^{i-1} - \qv+ \pv^{i-1}\rangle + L D_h(\pv^i, \pv^{i-1}) - \mu D_h(\qv,\pv^{i-1}) ,
\end{align*}
that is 
\begin{align*}
f(\pv^i)\le f(\qv) + \langle\grad f(\pv^{i-1}), \pv^i - \qv\rangle + L D_h(\pv^i, \pv^{i-1}) - \mu D_h(\qv,\pv^{i-1}).
\end{align*}
Recall that from \Cref{lemma:threePoint} it holds that
\begin{align*}
L D_h(\pv^{i},\pv^{i-1}) 
\le 
\langle\gradAppx f(\pv^{i-1}), \pv -\pv^{i}\rangle+L D_h(\pv,\pv^{i-1}) - L D_h(\pv,\pv^{i}) .
\end{align*} 
Substituting the above with $\pv=\qv$ we obtain
\begin{align}
f(\pv^i)\le f(\qv) &+ \langle\grad f(\pv^{i-1}), \pv^i - \qv\rangle + \langle\gradAppx f(\pv^{i-1}), \qv -\pv^{i}\rangle+L D_h(\qv,\pv^{i-1}) - L D_h(\qv,\pv^{i}) - \mu D_h(\qv,\pv^{i-1}),\nonumber\\
\le f(\qv)  & + (L-\mu) D_h(\qv,\pv^{i-1}) - LD_h(\qv,\pv^{i})+ \langle\grad f(\pv^{i-1})-\gradAppx f(\pv^{i-1}), \pv^i - \qv\rangle\nonumber\\
\le f(\qv)  &+ (L-\mu) D_h(\qv,\pv^{i-1}) - LD_h(\qv,\pv^{i})+ ||\grad f(\pv^{i-1})-\gradAppx f(\pv^{i-1})||_1 ||\pv^i - \qv||_\infty\nonumber\\
\le f(\qv)  &+ (L-\mu) D_h(\qv,\pv^{i-1}) - LD_h(\qv,\pv^{i})+ \lambda \errEst m . \label{eq:ubAppx}
\end{align}
Where:
\begin{inparaenum}[($i$)]
\item in the second inequality we grouped the inner products and performed simple  algebraic manipulations;
\item in the third inequality we use the fact that for any $\bm{u}, \bm{v} \in \R^m$
\[
\langle\bm{u},\bm{v}\rangle \le ||\bm{v}|| \,||\bm{u}||_*,
\]
where $||\bm{v}||_* = \sup\{\bm{v}^T \bm{x} \,|\, ||\bm{x} || \le 1\}$ for any norm $|| \cdot||$ on $\R^m$. Where we set $\bm{u} = \grad f(\pv^{i-1})-\gradAppx f(\pv^{i-1})$ and $\bm{v} =\pv^i - \qv$ and $||\cdot||$ to be the $\ell_\infty$ norm, yielding as dual norm $||\cdot||_*$ the $\ell_1$ norm (see Appendix A.1.6 in~\cite{Boyd2004CO});
\item finally in the last inequality we used the facts that for any $\pv\in(0,1)^m$ and any $\pv^{i}\in(0,1)^m$ as processed by~\espressoNLAppx it holds that $||\pv^i - \qv||_\infty \le 1$, and that 
\begin{align*}
	||\grad f(\pv^{i-1})-\gradAppx f(\pv^{i-1})||_1 &\le \lambda||\countTriv(\pv^{i-1})-\appxTriv(\pv^{i-1})||_1\\
	& \le \lambda \errEst m .
\end{align*}
Where the first inequality follows by the definition of the approximate gradient $\gradAppx_e f(\pv)$ (see Eq.~\eqref{eq:appxGrad}) and the second inequality holds by the guarantees of our randomized algorithm (see \Cref{theo:appxGuarantees}), which hold
with probability at least $1-\eta$ for each iteration over the $K$ iterations of~\espressoNLAppx.
\end{inparaenum}

Consider an optimal vector $\pv^*$ then $f(\pv^i)-f(\pv^*) \ge 0$ for any $i\ge0$, hence by setting $\qv = \pv^*$ in Eq.~\eqref{eq:ubAppx} it holds
\[
D_h(\pv^*,\pv^i) \le (1-\mu/L) D_h(\pv^*, \pv^{i-1}) + \lambda/L\,\errEst m%
.
\]
By recursively applying the above inequality to the term $D_h(\pv^*, \pv^{i-1})$ we get that
\begin{align*}
D_h(\pv^*,\pv^i) \le &(1-\mu/L)^i D_h(\pv^*, \pv^{0})+ \frac{\lambda}{L}\sum_{j=1}^{i}\left(1-\tfrac{\mu}{L}\right)^{i-j} \errEst m .
\end{align*}
By considering Equation~\eqref{eq:ubAppx} for $\qv=\pv^*$ and noting that $-LD_h(\qv,\pv^i) \le 0$ since $L\ge 0$ and $D_h(\cdot,\cdot)\ge 0 $ we have that
\[
f(\pv^i) -f(\pv^*) \le (L-\mu )D_h(\pv^*,\pv^{i-1}) + \lambda\errEst m ,
\]
by dropping the $-LD_h(\qv,\pv^i)\le0$ term.
Substituting the recursion above we get that
\begin{align*}
f(\pv^i) -f(\pv^*) \le &(L-\mu)\Big((1-\tfrac{\mu}{L})^{i-1}D_h(\pv^*,\pv^0)+\frac{\lambda}{L}\sum_{j=1}^{i-1}\left(1-\tfrac{\mu}{L}\right)^{i-1-j}\errEst m \Big) + \lambda\errEst m.
\end{align*}
By noting that $L(1-\mu/L) = L-\mu$, and $L> 0$, we have that
\begin{align*}
f(\pv^i) -f(\pv^*) &\le L(1-\tfrac{\mu}{L})^{i}D_h(\pv^*,\pv^0)+\lambda\sum_{j=1}^{i-1}\left(1-\tfrac{\mu}{L}\right)^{i-j}\errEst m + \lambda\errEst m\\
&=L(1-\tfrac{\mu}{L})^{i}D_h(\pv^*,\pv^0) + \lambda\sum_{j=1}^{i}\left(1-\tfrac{\mu}{L}\right)^{i-j}\errEst m .
\end{align*}
Since we have that $\mu < L$ it holds that $(1-\mu/L)^{k}\le 1$ for any $k\ge 0$.
Hence, using the above inequality we have that
\[
f(\pv^i) - f(\pv^*) \le L (1-\frac{\mu}{L})^i D_h(\pv^*,\pv^0) + \lambda\errEst m i,
\]
which concludes the proof.
\end{proof}

\begin{proof}[Proof of~\Cref{theo:appxNL}]
Our proof follows the same structure as the proof of \Cref{theo:accNL}.
The key difference is that at iteration $K$, by \Cref{lemma:appxGradGuar} it holds with probability at least $1-\eta$ that
\begin{align*}\label{eq:objBoundInnerAppx}
\obj(\pv^{K}) - \obj(\pv^{*}) &\le L \left(1- \frac{\mu}{L}\right)^K D_h(\pv^*, \pv^{0}) + \lambda\sum_{j=1}^K \errEst m,
\end{align*}
where $\pv^*$ is a minimizer of $\obj$.

By an identical analysis to our~\Cref{theo:accNL}, under the choice of
\[
K \ge \frac{\lambda |\Wmax| + 2\delta}{2\delta} \log \left(\frac{(\lambda |\Wmax|+2\delta)m\log 2}{\errAcc}\right) ,
\]
it holds that
\begin{align}\label{eq:boundt1}
L \left(1- \frac{\mu}{L}\right)^K D_h(\pv^*, \pv^{(0)}) \le \frac{\errAcc}{2} .
\end{align}

While for the second term
\begin{align}\label{eq:boundt1b}
\lambda\sum_{j=1}^K \errEst m \le \lambda K\errEst m \le \errAcc / 2 ,
\end{align}
by the choice of $\errEst \le \frac{\errAcc}{2 \lambda K m}$.

Summing the two bounds of Eqs.~\eqref{eq:boundt1} and \eqref{eq:boundt1b} concludes the proof.
\end{proof}

\begin{proof}[Proof of~\Cref{lemma:rtAppx}]
First note that it follows that the time complexity of~\espressoNLAppx is bounded by 
\[
\bigO\left(\degen m + \sum_{e}s_e \frac{\lambda \dMax}{\delta} \log \frac{\lambda m}{\errAcc}\right) ,
\]
where we use $s_e$ to denote the sample size used for computing the estimate $\appxTri_e(\pv^k)$ at any iteration $k\ge1$ for each edge $e\in E$. 
The sampling step dominates the complexity of all iterations of~\espressoNLAppx. That is, all other steps of~\espressoNLAppx are bounded by $\bigO(m)$. The term $\degen m$ is due to the fact that we list all triangles in $G$ at the beginning of the algorithm (i.e., using Chiba and Nishizeki's approach~\cite{Chiba1985Triangles}).
by bounding $\sum_{e}s_e \le m s_{\max}$ the claim follows.
\end{proof}

\begin{proof}[Proof of~\Cref{cor:rtRistretto}]
The proof follows that of \Cref{lemma:rtAppx}. 
The only difference is that the initial $\bigO(\degen m)$ complexity of listing all triangles (i.e., using \cite{Chiba1985Triangles}) is now replaced by the $\bigO(\phi m)=\bigO(m)$ cost required to store the triangles of all edges $e=(u,v)\in E$ for which $\min\{d_u,d_v\} \le \phi$, for constant $\phi$. 
Setting the parameters via $B = 2(\dMax-1)$ has time complexity $\bigO(m)$ leaving our analysis unchanged since $|\Wmax|\le B$, and the claim follows.
\end{proof}

\subsection[appendix]{Proofs for the connections with hard STC and STC-densest subgraph}\label{app:proofsHardDense}

\begin{proof}[Proof of~\Cref{prop:rounding}]
Let $g(\bm z) \doteq \We(\bm z) + \V(\bm z)$. By linearity of expectation and the mutual independence of the $Z_e$ under $q(\cdot;\pv^*)$ (Eq.~\eqref{eq:factorization}), it holds that $\E_{\bm z\sim q(\cdot;\pv^*)}[g(\bm z)] = \We(\pv^*)+\V(\pv^*) = g(\pv^*)$.

Consider the following algorithm to round $\pv^* \in (0,1)^m$,
\begin{compactenum}
\item initialize $\bm z$ to $\bm{p}^*$;
\item for each edge $e_i$, $i\in[m]$, set 
\[
z_{e_i} \gets \arg\min_{b\in\{0,1\}} \E[ g(\bm{z}) \mid z_{e_1},\dots,z_{e_{i-1}}, z_{e_i}=b ] ,
\]
retaining $z_{e_j} = p^*_{e_j}$, $j>i$, when evaluating the expectation (ties broken arbitrarily);
\item for each wedge $\wed=(uvw) \in \Wset$ with $z_{uv}=1$ and $z_{vw}=1$ set one of such variables to 0.
\end{compactenum}	

It is immediate to see that the above algorithm runs in polynomial time. 
Step (1) takes $\bigO(m)$ time. 
In step (2), fixing $z_{e_i}$ requires evaluating uniquely the marginal difference $\E[ g(\bm{z}) \mid z_{e_1},\dots,z_{e_{i-1}}, z_{e_i}=1 ] - \E[ g(\bm{z}) \mid z_{e_1},\dots,z_{e_{i-1}}, z_{e_i}=0 ]$, 
which depends on the wedges 
$\Wset_{e_i}$ incident to $e_i$ and can be evaluated in  $\bigO(|\Wset_{e_i}|)$ time, yielding a total time complexity $\bigO(|\Wset|) = \bigO(m^2)$. Finally, step (3) requires $\bigO(|\Wset|) = \bigO(m^2)$ time. Hence the total time complexity is bounded by $\bigO(m^2)$.

Let $\hat{\bm z}$ be the solution at the end of step (2). Since 
$\E[g(\bm z)\mid z_{e_1},\dots,z_{e_{i-1}}]$ is a convex combination of its two conditional values $z_{e_i}\in\{0,1\}$, our choice never increases the conditional expectation; and chaining these yields
\[
g(\hat{\bm z}) \le \E[g(\bm z)] = \We(\pv^*)+\V(\pv^*).
\]

Let $\V(\hat{\bm z}) \ge 0$ be the STC violations of the solution $\hat{\bm z}$, and let $\bm{z'}$ be the final output of the above algorithm after step (3), it then holds
\[
\We(\bm z') \le \We(\hat{\bm z}) + \V(\hat{\bm z}) = g(\hat{\bm z}) \le \We(\pv^*)+\V(\pv^*),
\]
since each iteration in (3) makes one strong edge (with $\hat{z}_e=1$) weak, and the corresponding term $\We(\bm{\hat{z}})$ increases by exactly 1. 
Note that since an STC-violation requires \emph{both} edges of a wedge $\wed\in \Wset$ to be strong, making one edge weak ($\hat{z}_e=0$) cannot introduce new violations.
Hence, $\V(\hat{\bm z})$ strictly decreases at each iteration of step (3), which therefore performs at most $\V(\hat{\bm z})$ iterations. This yields a solution with $\We(\bm z') \le \We(\hat{\bm z}) + \V(\hat{\bm z})$ and $\V(\bm z')=0$, concluding the proof.
\end{proof}

\begin{proof}[Proof of~\Cref{lemma:expDense}]
The proof follows by~\Cref{eq:factorization}, since $z_e\sim\Bern(p_e)$ and therefore $\E_{\bm z\sim q(\cdot;\pv)}[z_e] = p_e, e\in E$, and by the linearity of expectation, it holds that,
\begin{align*}
\E_{\bm z\sim q(\cdot;\pv)}\left[\dens(U,\bm z;\lamD)\right]  = \frac{1}{|U|}\sum_{e\in E[U]} [p_e + (1-p_e)\lamD],
\end{align*}
the claim follows by a simple rewriting and the definition of $\hat{w}_e$.
\end{proof}

\spara{Rounding for \espressoDEN.}
Step ($iii$) of \espressoDEN solves \textsc{MaxSTC} on $G[U]$, which is $\NP$-hard.
We now discuss the polynomial-time approach mentioned in~\Cref{subsec:stcdensest}, which labels $E[U]$ with the procedure described in~\Cref{prop:rounding}, bounding the resulting objective in~\Cref{eq:objDense}.
Let $\V(\pv\mid U)$ and $\We(\pv \mid U)$ be the functions $\V(\pv)$ and $\We(\pv)$ restricted to the induced subgraph $G[U]$.

\begin{lemma}\label{lemma:stcdenCertificate}
Given $U\subseteq V$, we let $\bm z'$ be the STC-feasible labeling computed with the rounding procedure of~\Cref{prop:rounding} over $G[U]$ and a solution $\pv$, then it holds
\begin{equation}\label{eq:stcdenCertificate}
\dens(U,\bm z';\lamD) \ge \E_{\bm z\sim q(\cdot;\pv)}\left[\dens(U,\bm z;\lamD)\right] - (1-\lamD)\frac{\V(\pv \mid U)}{|U|}.
\end{equation}
\end{lemma}
\begin{proof}
Since every edge of $E[U]$ is either strong or weak under $\bm z'$, it holds that $m_s(U,\bm z') = |E[U]| - m_w(U,\bm z')$, hence
\begin{equation}\label{eq:densRewrite}
\dens(U,\bm z';\lamD) = \frac{|E[U]| - (1-\lamD)\, m_w(U,\bm z')}{|U|}.
\end{equation}
By using~\Cref{prop:rounding} applied to the induced subgraph $G[U]$ and to the values $\pv$ restricted to $E[U]$, the resulting labeling $\bm z'$ satisfies $\V(\bm z' \mid U) = 0$, and
\[
m_w(U,\bm z') = \We(\bm z' \mid U) \le \We(\pv \mid U) + \V(\pv \mid U).
\]
Since $\lamD\in[0,1]$, the value $(1-\lamD)$ in~\Cref{eq:densRewrite} is non-negative, hence we have
\[
\dens(U,\bm z';\lamD) \ge \frac{|E[U]| - (1-\lamD)\We(\pv \mid U)}{|U|} - (1-\lamD)\frac{\V(\pv \mid U)}{|U|}.
\]
Finally, since $\We(\pv \mid U) = \sum_{e\in E[U]}(1-p_e)$ by definition, from~\Cref{lemma:expDense}, we have
\begin{align*}
|E[U]| - (1-\lamD)&\We(\pv \mid U) =|E[U]| - (1-\lamD) \sum_{e\in E[U]}(1-p_e) \\
&=
|E[U]| - \sum_{e\in E[U]}(1-p_e) + \lamD \sum_{e\in E[U]}(1-p_e) \\
&=\sum_{e\in E[U]}p_e + \lamD \sum_{e\in E[U]}(1-p_e) \\
&=\sum_{e\in E[U]}
\left[\lamD + (1-\lamD)p_e\right] =\sum_{e\in E[U]} \hat{w}_e,
\end{align*}
and therefore $(|E[U]| - (1-\lamD)\We(\pv \mid U))/|U| = \E[\dens(U,\bm z;\lamD)]$ by~\Cref{eq:stcdenReduction}, concluding the proof.
\end{proof}

{The additive term $(1-\lamD)\V(\pv\mid U)/|U|$ in~\Cref{eq:stcdenCertificate} is zero when}
\begin{inparaenum}[($i$)]
\item $\V(\pv\mid U)=0$, i.e., $\pv$ has no expected STC violations over $G[U]$; or
\item $\lamD=1$, for which $\hat{w}_e=1$ for all $e\in E[U]$ and \stcden reduces to the densest subgraph problem.
\end{inparaenum}
In both cases it holds that $\dens(U,\bm z';\lamD)\ge \E_{\bm z\sim q(\cdot;\pv)}[\dens(U,\bm z;\lamD)]$.
Note that~\Cref{eq:stcdenCertificate} bounds the density of $\bm z'$ with respect to the \emph{expected STC-density} from~\Cref{lemma:expDense}.

\section[appendix]{Algorithms and implementation details}
\subsection[appendix]{\espressoNL}
Algorithm~\ref{alg:computeGradient} reports the pseudocode of the algorithm that implements the function $\mathtt{Comp\nabla}(G, \pv^{k},(\agg_u^{k})_{u\in V}, e)$ used in~\espressoNL.
More in detail, Algorithm~\ref{alg:computeGradient} first computes $\countTri_e(\pv^k)$ as from its definition (Line~\ref{algline2:computeTri}), for which we use the algorithm of \citet{Chiba1985Triangles} {to list all triangles in the graph}. 
That is, we can compute $(\bm{T}\pv^k)_e$ simultaneously for each edge $e\in E$, avoiding processing each triangle multiple times. 
We then compute the gradient $\grad_e\obj$ as from Eq.~\eqref{eq:gradient_entropy_open} (Line~\ref{algline2:computeGrad}) using Equation~\eqref{eq:gradAggSumTrick} for the computation of terms $\edPress_e(\pv^k)$. Finally the algorithm reports the resulting value of $\grad_e\obj(\pv^k)$ (Line~\ref{algline2:retVec}).

\begin{algorithm}[t]
\caption{$\mathtt{Comp\nabla}(G, \pv^{k},(\agg_u^{k})_{u\in V}, e)$}\label{alg:computeGradient}
\setcounter{AlgoLine}{0}
\PrintSemicolon
\KwIn{Graph $G$, probabilities $\pv^{k}$, aggregate sums $(\agg_u^{k})_{u\in V}$, edge $e=(u,v)$}
\KwOut{Gradient $\nabla_e\obj(\pv^{k})$ with respect to $p_e$}
$\countTri_e(\pv^k)
\gets \sum_{z \in \neigh(u)\cap\neigh(v)}
\!\bigl(p_{uz}^{k} + p_{vz}^{k}\bigr)$\label{algline2:computeTri}\;

$\nabla_e \obj(\pv^{k})
\gets -\alpha
+ \lambda\,\Bigl(
\agg_u^{k} + \agg_v^{k} - 2\,p_e^{k}
- \countTri_e(\pv^k)
\Bigr)
- \gamma\,\log\!\left(\dfrac{1-p_e^{k}}{p_e^{k}}\right)$\label{algline2:computeGrad}\;
\KwRet{$\nabla_e \obj(\pv^{k})$}\label{algline2:retVec}\;
\end{algorithm}

\subsection[appendix]{{Early stopping for \espressoNL}}\label{appsec:earlyStoppingNL}
Note that in practice when the magnitude of $|\obj|$ is extremely large, if $\errAcc$ is set as a small constant we may force~\espressoNL to perform an excessively large number of iterations $K$.
In such cases, it is desirable to obtain \emph{relative} error guarantees.
That is a guarantee of the form $\obj(\pv^{K}) - \obj(\pv^*) \le \varepsilon_{\text{rel}} |\obj(\pv^*)|$ for $\varepsilon_{\text{rel}} > 0$, corresponding to $\errAcc = \varepsilon_{\text{rel}} |\obj(\pv^*)|$.
Unfortunately, to make such bound practical for $K$ we require a tight lower bound on $ |\obj(\pv^*)|$,
which we leave as an open research direction.
In practice, we use the heuristic early-stopping condition
\begin{align*}
	\frac{|\obj(\pv^{k+1})-\obj(\pv^k)|}{|\obj(\pv^k)|+1}
	\le \varepsilon ,
\end{align*}
for $\varepsilon\in(0,1)$.
The denominator interpolates between absolute-difference (when $|\obj(\pv^k)|$ is small) and relative-difference stopping criteria (when $|\obj(\pv^k)|$ is large).

\subsection[appendix]{\espressoNLAppx}

\begin{algorithm}[!t]
\caption{\espressoNLAppx: \nolips for \espresso approximate}\label{alg:NoLipsAppx}
\setcounter{AlgoLine}{0}
\PrintSemicolon
\KwIn{%
Graph $G=(V,E)$,
parameters $\alpha>0$, $\lambda>0$, $\delta>0$,
accuracy $\errAcc>0$, max iterations $K$, absolute estimation error $\errEst>0$, error probability $\eta\in(0,1)$, sampling threshold $\phi\in \mathbb{N}_{\ge 1}$.%
}
\KwOut{Edge-strength probabilities $\pv^K\in(0,1)^m$.}
$\gamma \gets \dfrac{\lambda\,|\Wmax|}{4} + \delta$; $L \gets \dfrac{\lambda\,|\Wmax|}{4} + \gamma$; %
$\pv^{0} \gets (1/2,\dots,1/2)\in\R^m$\label{alglineNLA:p0}\;
\lFor{$u\in V$}{$\agg_u^{0} \gets d_u/2$\label{alglineNLA:sums}}
$\appxTriv(\pv^0) \gets \bm{0}\in \R^m$;
$\cntv \gets \bm{0}\in \R^m $\label{alglineNLA:initcounts}\;
\For{$\{u,v,w\} \in \triSet(G)$}{
$\appxTri_{uv}(\pv^0)  \gets \appxTri_{uv}(\pv^0)  + 1$;
$\appxTri_{vw}(\pv^0)  \gets \appxTri_{vw}(\pv^0)  + 1$;
$\appxTri_{wu}(\pv^0) \gets \appxTri_{wu}(\pv^0)  + 1$\label{alglineNLA:vec0}\;
$\cntv_{uv} \gets \cntv_{uv} + 1$;
$\cntv_{vw} \gets \cntv_{vw} + 1$;
$\cntv_{wu} \gets \cntv_{wu} + 1$\label{alglineNLA:count0};
}
$\triSet_e \gets \emptyset, e\in E$\label{alglineNLA:trisetEdge}\;
\For{$e=\{u,v\}\in E$\label{alglineNLA:forKeepSet}}{
\If{$\cntv_{e} \le \phi$\label{alglineNLA:ifSmallTriSet}}{
	$\triSet_e \gets \{\tau \in \triSet(G) : e\in \tau\}$\label{alglineNLA:keepMySet}\;
}
}
\For{$k \gets 0$ \KwTo $K-1$\label{alglineNLA:forIts}}{
\For{$e = (u,v) \in E$}{
	\If{$k\ge 1$\label{alglineNLA:estk1}}
	{
		$\appxTri_e(\pv^{k}) \gets 0$\label{alglineNLA:initEst}\;
		\If{$\cntv_e \le \phi$\label{alglineNLA:smallcntEx}}
		{
			$\appxTri_e(\pv^{k}) \gets \sum_{\tau=\{u,v,w\}\in \Delta_e} p_{uw} + p_{vw}$\label{alglineNLA:smallEx}\;
		}
		\Else
		{
			$w \gets \arg\min\{d_u, d_v\}$\label{alglineNLA:getMinDegNode}\;
			${p}_e^+ \gets {\max\{ p_{e'}: e'\in E,\ e'\cap e\neq\varnothing,\ e'\neq e\}}$\;
			$s \gets \frac{8d_w p^+_e\cntv_{e} + \frac{4}{3} d_w\errEst}{\errEst^2}  \log\left(\frac{2mK}{\eta}\right) $\label{alglineNLA:params}\;
			\For{$i\gets 1$ \KwTo $\lceil s\rceil$\label{alglineNLA:forSamples}}
			{
				$z\gets \mathtt{SampleNeighbor}(w)$\label{alglineNLA:getSamp}\;
				\If{$\{u,v,z\} \in \triSet(G)$\label{alglineNLA:ifTri}}
				{
					$ \appxTri_e(\pv^{k}) \gets \appxTri_e(\pv^{k}) +  p_{uz} + p_{vz}$\label{alglineNLA:updateEst}\;
				}
			}
			$ \appxTri_e(\pv^{k}) \gets \appxTri_e(\pv^{k}) \, d_w/s$\label{alglineNLA:finalEst}\;
		}
	}
	$\gradAppx_e\obj(\pv^k) \gets -\alpha + \lambda( \agg_u(\pv^k) + \agg_v(\pv^k) - 2p_e - \appxTri_e(\pv^k) )- \gamma\log \frac{1-p_e}{p_e}$\label{alglineNLA:appxGrad}\;
	$p_e^{k+1} \gets
	\sigma\left(
	\log\dfrac{p_e^{k}}{1-p_e^{k}}
	\;-\;
	\dfrac{\gradAppx_e \obj(\pv^{k})}{L}
	\right)$\label{alglineNLA:updateP}\;
}
\For{$u \in V$}{
	$\agg_u^{k+1} \gets
	\agg_u^{k}
	+ \sum_{e \in E : u\in e}
	(p_e^{k+1} - p_e^{k})$\label{alglineNLA:updatesums}\;
}
}
\KwRet{$\pv^{K}$\label{alglineNLA:return}}\;
\end{algorithm}

\textbf{Algorithm.} Algorithm~\ref{alg:NoLipsAppx} details our method \espressoNLAppx. 
The overall structure of the algorithm follows the one of Algorithm~\ref{alg:NoLips} (\espressoNL), with some key differences that we now detail below. That is, \espressoNLAppx:

\textbf{1.} computes the value 
$
\appxTriv(\pv^0) \gets (\bm{T}\pv^0)
$
by listing all triangles (Line~\ref{alglineNLA:initcounts}) in $G$. 
In addition, the algorithm also retains, for each edge $e\in E$ a count $\cntv_e = |\{\tau \in \triSet(G): e\in \tau\}|$ counting the number of triangles in $G$ that the edge $e\in E$ participates in (set to the correct value in Line~\ref{alglineNLA:count0}). 
The key idea is that, for any edge $e$ contained in a small number of triangles (i.e., less than a given value $\phi$) we can store the nodes closing a triangle with $e\in E$, improving the variance of the sampling approach and retaining a small time complexity, with a linear memory overhead $\bigO(\phi m)$. 
Similar techniques have been applied for the estimation of triadic coefficient in graphs (e.g., see~\cite{Sarpe2025Triad}), to reduce the variance of the estimates.
Therefore, for each edge $e\in E$ contained in a number of triangles $\cntv_e\le \phi$ we retain the triangles in memory (Line~\ref{alglineNLA:keepMySet}). Such triangles will be used for an exact computation of the term $\appxTri_e(\pv^i)$, at any iteration $i\ge 0$.

\textbf{2.} in the for loop starting at Line~\ref{alglineNLA:forIts}, for any iteration $k\ge1$, and for each edge $e=\{u,v\}$ the algorithm:
\begin{inparaenum}[($i$)]
\item computes exactly the value $\sum_{{uvw}\in\triSet_e} p_{uw} + p_{vw}$ for all edges $e\in E$ with $\cntv_e\le \phi$ (Line~\ref{alglineNLA:smallcntEx});
\item otherwise when $\cntv_e> \phi$, it estimates the value of $\countTri_e(p^k)$ by random sampling. 
The estimate $\appxTri_e(\pv^i)$ is computed such that for each edge $e=\{u,v\}$ \espressoNLAppx collects $s$ nodes $z_1,\dots,z_s$ among the neighborhood of $w=\arg\min\{d_u,d_v\}$ and if $z_i$ forms a triangle with $e$, i.e., $\{uvz_i\}\in\triSet(G)$, then   \espressoNLAppx updates its estimate $\appxTri_e$ by summing the term $(p^k_{uz}+p^k_{vz})d_w$, the output reports the average of the estimate over the $s$ samples. The entire procedure is detailed from Line~\ref{alglineNLA:getMinDegNode} to Line~\ref{alglineNLA:finalEst}.
For a sufficiently large choice of $s$ it holds that the estimates $\appxTri_e(\pv^k)$ are such that $|\countTri_e(\pv^k)-\appxTri_e(\pv^k)|\le \errEst$ simultaneously for all edges $e\in E$ over all iterations $k\ge0$ with probability at least $1-\eta$ for $\eta\in(0,1)$. Such result is formalized in our \Cref{theo:appxGuarantees}.
\end{inparaenum} 

\textbf{3.} computes the approximate gradient using $\appxTri_e$ as
\begin{align*}
\gradAppx_e\obj(\pv^k) \gets &-\alpha + \lambda( \agg_u(\pv^k) + \agg_v(\pv^k) - 2p_e - \appxTri_e(\pv^k) )- \gamma\log \frac{1-p_e}{p_e} ,
\end{align*}
in Line~\ref{alglineNLA:appxGrad}, by replacing the exact term $\countTri_e(\pv^k)$ with its estimate $\appxTri_e(\pv^k)$. The approximate gradient $\gradAppx_e\obj(\pv^k)$ is then used in Line~\ref{alglineNLA:updateP} to find the vector $\pv^k$ optimizing the mirror descent step as for \espressoNL (see Section~\ref{subsec:randomNespresso}).

We now prove the approximation guarantees required by~\espressoNLAppx, that is we show that each iteration $k\ge 0$ it holds $||\countTriv(\pv^k) - \appxTriv(\pv^k)||_1 \le \errEst m $ with probability at least $1-\eta$.
Note that by Equations~\eqref{eq:appxGrad} and~\eqref{eq:gradient_entropy_open} this immediately implies that $||\grad \obj - \gradAppx \obj ||_1 \le \lambda \errEst m$ as desired.
\begin{theorem}\label{theo:appxGuarantees}
For an execution of~\espressoNLAppx under the setting of \Cref{theo:appxNL} and $\phi\ge1$, {with probability at least $1-\eta$ it holds simultaneously for every edge $e\in E$ and every iteration $k\in[0,K-1]$ that}
\[
|\countTri_e(\pv^k)-\appxTri_e(\pv^k)|\le \errEst,
\]
that is $||\countTriv(\pv^k)-\appxTriv(\pv^k)||_1\le \errEst m$ {for every iteration $k\in[0,K-1]$}.
\end{theorem}
\begin{proof}
Fix an iteration $k\ge 1$ as the claim trivially follows for $k=0$ since $\appxTriv(\pv^0) = \countTriv(\pv^0)$. 
Since the samples at each iteration $k$ are independent of the previous iterations, our analysis considers $\pv^k$ as fixed (i.e., determined by the $s$ samples).
Furthermore, it trivially follows that $|\countTri_e(\pv^k)-\appxTri_e(\pv^k)| = 0$ for all edges such that $\cntv_e\le \phi$. 
Now, let $\appxTri_e(\pv^k)$ be the estimate computed by \espressoNLAppx for an edge $e=\{u,v\}\in E$ such that $\cntv_e> \phi$. First we show that $\E[\appxTri_e(\pv^k)] = \countTri_e(\pv^k)$. 
Note that the algorithm samples nodes in the neighborhood of the node with minimum degree among $u$ and $v$. Let such node be $u$ without loss of generality.
The estimate $\appxTri_e(\pv^k)$ considered by the algorithm corresponds thus to 
\[
\appxTri_e(\pv^k) = \frac{1}{s}\sum_{i=1}^s\sum_{\tau=\{uvz\} \in \triSet_e} (p_{uz}^k+p_{vz}^k) d_u X_\tau\enspace,
\]
where $X_\tau$ is a Bernoulli random variable denoting if triangle $\tau$ is sampled. Clearly, $\E[X_\tau] = 1/d_u$. Hence, by the linearity of the expectation it holds
\[
\E[\appxTri_e(\pv^k)] =\frac{1}{s}\sum_{i=1}^s\sum_{\tau=\{uvz\} \in \triSet_e} (p_{uz}^k+p_{vz}^k) d_u \E[X_\tau] = \countTri_e(\pv^k).
\]
We now compute the variance of $\appxTri_e(\pv^k)$, note that
\[
\Var[X] = \E[X^2] - \E[X]^2.
\]
We now upper bound $\E[\appxTri_e(\pv^k)^2]$. We first obtain the bound for $s=1$ and then we will consider the average over $s$ such terms. 
For a fixed $e=\{u,v\}\in E$ and $\tau=\{uvz\}\in \triSet_e$, we let $p_\tau = p_{uz}+p_{vz}$.\footnote{Note that since $\pv^k$ is fixed we therefore drop the superscript $k$ for ease of notation.}
\begin{align*}
\E[\appxTri_e(\pv^k)^2] &= \E\left(\sum_{\tau\in \triSet_e} p_{\tau} d_uX_\tau\right)^2\\
=&\,\E\big(\sum_{\tau\in \triSet_e}p_\tau^2 d_u^2X_\tau^2 + \sum_{\tau_1\in \triSet_e}\sum_{\tau_2\in \triSet_e\setminus\{\tau_1\}}p_{\tau_1}p_{\tau_2} d_u^2 X_{\tau_1} X_{\tau_2}\big)\\
= &\sum_{\tau\in \triSet_e}d_u^2 p_{\tau}^2  \E[X_\tau^2] + \sum_{\tau_1\in \triSet_e}p_{\tau_1}\sum_{\tau_2\in \triSet_e\setminus\{\tau_1\}} p_{\tau_2} d_u^2 \E[X_{\tau_1}  X_{\tau_2}]\\
\le &2d_u\countTri_e(\pv^k).
\end{align*}
Where the first equality is by definition of $\appxTri_e(\pv^k)$, the second equality follows by a re-writing of the squared term.
The third equality follows by the linearity of expectation. 
Finally the last inequality follows by the facts that $\E[X_\tau^2] = 1/d_u$, $\E[X_{\tau_1}  X_{\tau_2}] = 0$ for $\tau_1\neq\tau_2$ (since a fixed sample collects at most one triangle) and that $p_\tau^2 = p_\tau \, p_\tau \le 2 p_\tau$.
Since $\Var[1/s \sum_{i=1}^s X_i] = 1/s\Var[X_1]$ for i.i.d.\ random variables $X_1,\dots,X_s$, it holds that
\begin{align*}
	\Var[\appxTri_e(\pv^k)] \le \frac{(2d_u-\countTri_e(\pv^k)) \countTri_e(\pv^k)}{s} \le \frac{2d_u \countTri_e(\pv^k)}{s} \le \frac{4d_u{p}_e^+\cntv_e}{s},
\end{align*}
since $\countTri_e(\pv^k)\le 2{p}_e^+\cntv_{e}$, where for an edge $e=\{u,v\}$ we let ${p}_e^+ \doteq \max\{ p^k_{e'}: e'\in E,\ e'\cap e\neq\varnothing,\ e'\neq e\}$ be the largest probability of an edge sharing a node with $e$.
Furthermore, note that $\appxTri_e(\pv^i)= \frac{1}{s} \sum_{i}^s  Z_i $ where $Z_i = \sum_{\tau=\{uvz\} \in \triSet_e} (p_{uz}^i+p_{vz}^i) d_u X_\tau \le 2d_u$.

We now apply Theorem \ref{thm:bernstein} to show that the choice of $s$ as from the statement yields the claim. Let $X_i = \sum_{\tau=\{uvz\} \in \triSet_e} (p_{uz}^k+p_{vz}^k) d_u X_\tau$. Note that
\[
X_i - \mathbb{E}[X_i]  \le 2d_u-\countTri_e(\pv^k)\le 2d_u \quad\text{and}\quad \mathbb{E}[X_i]-X_i\le \countTri_e(\pv^k)\le 2d_u,
\]
almost surely since $p_{uz}^k+p_{vz}^k \le 2$, $X_i\ge0$, and $X_\tau$ is 1 for a single triangle in $\triSet_e$ at most.
Thus, $|X_i-\mathbb{E}[X_i]|\le b$ with $b = 2d_u$.

Since the variables $X_i$ are identically distributed, and $\sigma^2\doteq  4d_u{p}_e^+\cntv_e \ge \text{Var}(X_1)$, then the sum of variances is bounded by $s \sigma^2$.

Let $\varepsilon = \errEst $.
Substituting $q = s\varepsilon$, $v = s\sigma^2$, and $b$ into Eq.~\eqref{eq:bersteinBound} we have
\begin{align*}
	\mathbb{P}\left(\left|\frac{1}{s}\sum_{i=1}^s X_i - \countTri_e(\pv^k)\right| \ge \varepsilon\right)
	&\le 2\exp\left( - \frac{(s\varepsilon)^2}{2(s\sigma^2 + \frac{b s \varepsilon}{3})} \right) \\
	&= 2\exp\left( - \frac{s^2 \varepsilon^2}{2s(\sigma^2 + \frac{b \varepsilon}{3})} \right) \\
	&= 2\exp\left( - \frac{s \varepsilon^2}{2\sigma^2 + \frac{2}{3} b \varepsilon} \right).
\end{align*}

To guarantee that the above probability is at most $\eta$, for all edges over the $K$ iterations and for all the $m$ edges we upper-bound the right-hand side by $\eta/(mK)$, that is
\begin{equation}\label{eq:ssProb}
	2\exp\left( - \frac{s \varepsilon^2}{2\sigma^2 + \frac{2}{3} b \varepsilon} \right) \le \eta/(mK).
\end{equation}

Dividing by 2 and taking the natural logarithm yields
\[
- \frac{s \varepsilon^2}{2\sigma^2 + \frac{2}{3} b \varepsilon} \le \log\left(\frac{\eta}{2mK}\right) = -\log\left(\frac{2mK}{\eta}\right).
\]

Which yields
\begin{equation}\label{eq:ubSS}
	s \ge \frac{8d_u p^+_e\cntv_e + \frac{4}{3} d_u\errEst}{\errEst^2} \log\left(\frac{2mK}{\eta}\right),
\end{equation}
where we substituted $\sigma^2=4d_u{p}_e^+\cntv_e$ and $b=2d_u$, and performed simple algebraic manipulations. 
That is, by the above choice of $s$
it holds that $|\appxTri_e(\pv^k)-\countTri_e(\pv^k)|> \errEst$ with probability at most $\eta/(mK)$. We now bound the error probability of~\espressoNLAppx. 
For any $k\ge 1$, 
let $F_k$ be the event  ``there exists an edge $e\in E$ such that $|\appxTri_e(\pv^k)-\countTri_e(\pv^k)|> \errEst$ at iteration $k$'' and let $F_k(e)$ be the event  
``the estimate of edge $e\in E$ is such that $|\appxTri_e(\pv^k)-\countTri_e(\pv^k)|> \errEst$'', 
hence by the union bound on $k$ and on $E$ we have that $\Prob[\cup_{k=1}^{K-1} F_k] \le \sum_{k=1}^{K-1} \Prob[\cup_{e\in E} F_k(e)] \le \sum_{k=1}^{K-1} \sum_{e\in E}\Prob[F_k(e)] \le \sum_{k=1}^{K-1} \sum_{e\in E} \frac{\eta}{mK} \le \eta$ where the last inequality follows by Eq.~\eqref{eq:ssProb}, establishing the claim and concluding the proof.
\end{proof}

{\textbf{Guarantees of \espressoR.}
The proof of~\Cref{theo:appxGuarantees} holds also for \espressoR with the following three remarks.
\begin{compactenum}[($i$)]
	\item We use $\min\{d_u,d_v\}\le\phi$ instead of $\cntv_e\le\phi$ to determine the edges for which triangles are computed deterministically.
	\item Eq.~\eqref{eq:ubSS} holds with $\cntv_e$ replaced by its upper bound $\min\{d_u,d_v\}-1$ since we do not list the triangles of the graph $G$.
	\item \espressoR also estimates $\appxTriv(\pv^0)$ by sampling, thus the union bound accounts for $K$ iterations $k\in[0,K-1]$. Our result from Eq.~\eqref{eq:ssProb} already accounts for $K$ iterations and $m$ edges, and its correctness remains thus unaltered.
\end{compactenum}}

\begin{theorem}[Bernstein's inequality, Eq. (2.10) from \cite{Boucheron2016Concentr}]\label{thm:bernstein}
Let $X_1, \dots, X_s$ be independent real-valued random variables. 
Assume that there exists a constant $b > 0$ such that $X_i - \E[X_i] \le b$ almost surely for all $i = 1, \dots, s$. Let $S_s = \sum_{i=1}^s (X_i - \mathbb{E}[X_i])$ and let $\sum_{i=1}^n \Var(X_i)\le v$ be an upper bound on the variance of the sum. Then, for any $q > 0$,
\[
\mathbb{P}(S_s \ge q) \le \exp\left( - \frac{q^2}{2(v + \frac{bq}{3})} \right) .
\]
By applying the same argument to $-X_i$, it holds that
\begin{equation}\label{eq:bersteinBound}
\mathbb{P}(|S_s| \ge q) \le 2\exp\left( - \frac{q^2}{2(v + \frac{bq}{3})} \right) .
\end{equation}
\end{theorem}

\begin{theorem}[Hoeffding Bound, Theorem 4.12 from~\cite{Mitzenmacher2017}]\label{theo:Hoeffding}
Let $X_1,\dots, X_s$ be independent random variables such that for all $i=1,\dots, s$, $\E [X_i] = \theta $ and $\Prob[X_i \in [a,b]] =1$. Then
\[
\Prob\left(\left|\frac{1}{s}\sum_{i=1}^s X_i - \theta \right|\ge \varepsilon\right) \le 2 \exp\left(\frac{-2s\varepsilon^2}{(b-a)^2}\right).
\]
\end{theorem}

\subsection[appendix]{Practical optimizations}\label{appsec:earlyStopping}
Our method \espressoNLAppx does not necessarily
retain the monotonicity of the sequence $(\obj(\pv^k))_{k\ge0}$: due to
the randomness introduced in the gradient computation, it is possible
that
\[
\obj(\pv^{k+1})-\obj(\pv^k)>0.
\]

As for \espressoNL, we now discuss a practical early-stopping condition.
Computing the exact objective $\obj$ at every iteration would require
listing all triangles in the graph, i.e., the procedure that we want to avoid. 
We therefore consider an approximate objective by using the values of
$\appxTriv(\pv^k)$. That is we evaluate
\[
\objAppx(\pv^k)
=
\alpha\We(\pv^k)
+
\frac{\lambda}{2}
(\pv^k)^\top\tilde{\edPress}(\pv^k)
-
\gamma\Rf(\pv^k),
\]
where, for $e=\{u,v\}$,
\[
\tilde{\edPress}_e(\pv^k)
=
\agg_u(\pv^k)+\agg_v(\pv^k)-2p_e^k-\appxTri_e(\pv^k).
\]
In fact, since
$\Mmat\pv=\edPress(\pv)$ and
$\grad\V(\pv)=\Mmat\pv$, we have
\[
\V(\pv)
=
\frac{1}{2}\pv^\top\Mmat\pv
=
\frac{1}{2}\pv^\top\edPress(\pv).
\]
Hence $\edPress(\pv^k)$ can be replaced by its estimate
$\tilde{\edPress}(\pv^k)$ without any additional asymptotic overhead in the estimation of the objective $\obj$.

With probability at least $1-\eta$, simultaneously for every iteration of the algorihtm
we thus have,
\begin{align*}
|\objAppx(\pv^k)-\obj(\pv^k)|
&=
\frac{\lambda}{2}
\left|
(\pv^k)^\top
\bigl(
\tilde{\edPress}(\pv^k)-\edPress(\pv^k)
\bigr)
\right|\\
&=
\frac{\lambda}{2}
\left|
\left\langle
\pv^k,
\appxTriv(\pv^k)-\countTriv(\pv^k)
\right\rangle
\right|\\
&\le
\frac{\lambda}{2}
\|\pv^k\|_\infty
\|\appxTriv(\pv^k)-\countTriv(\pv^k)\|_1\\
&\le
\frac{\lambda}{2}\errEst m.
\end{align*}

Hence for a fixed parameter
$\ell\in[K-1]$, we consider $\objAppx(\pv^k)\ge\objAppx(\pv^{k-j})$
for every $j=1,\dots,\ell$ as a practical stopping condition at any iteration $k\ge\ell$.
Upon this condition, let
\[
j^*
\in
\arg\min_{j\in\{k-\ell,\dots,k\}}
\objAppx(\pv^j)
\qquad\text{and consider}\qquad
\pv_{\mathrm{appx}}\doteq\pv^{j^*}.
\]
Then it holds that
\[
|\obj(\pv_{\mathrm{appx}})
-\objAppx(\pv_{\mathrm{appx}})|
\le
\frac{\lambda}{2}\errEst m
\]
with probability at least $1-\eta$.

\subsection[appendix]{A data-dependent method to select $\phi$}\label{appsec:selectingPhi}

Recall that the value of $\phi$ controls the threshold for which if $\cntv_e \le \phi$ we retain deterministically the triangles of $\triSet_e$ for $e=(u,v) \in E$, in~\espressoNLAppx and according to the upper bound $\cntv_e\le \min\{d_u,d_v\}\le \phi$ for~\espressoR (see optimization $(ii)$ in~\Cref{subsec:ristretto}).

This mechanism, assuming a fixed sample size $s$,\footnote{In our experiments we set $s$ to a small constant.} implies
\begin{compactenum}
\item if an edge $e$ has its $\triSet_e$ cached then we can process its gradient update in $\bigO(\cntv_e)$ time, while using $\bigO(\cntv_e)$ memory;
\item if an edge $e$ has $\cntv_e > \phi$ then performing its gradient update requires $\bigO(s)$ time complexity at the expense of $\bigO(1)$ extra memory.
\end{compactenum}

Hence an efficient approach for any edge $e$ is to store $\triSet_e$ only if $\cntv_e \le s$, since under this condition processing $\triSet_e$ is both more efficient than sampling and it yields no error. 
While when $\cntv_e> s$, sampling is clearly the most efficient strategy.
That is, fixing $\phi = s$ makes our caching mechanism never more expensive than sampling, for each gradient update.

In~\espressoR we avoid the explicit listing of all triangles in $\triSet(G)$, hence we do not know the values of $\cntv_e, e\in E$ to apply the rule detailed above. 
Instead we rely on the upper bound $\cntv_e \le \min\{d_u,d_v\}, e=(u,v)\in E$ as it is efficiently computable. 
Now note that when $s$ is small (i.e., often in practice), then setting $\phi = s$ may be extremely inefficient for~\espressoR, due to the use of the upper bound $\cntv_e \le \min\{d_u,d_v\}$. 
To show why this holds, let the \emph{local edge clustering} of an $e=(u,v)\in E$ be
\begin{equation}\label{eq:edgeClustering}
c_e \doteq \frac{\cntv_e}{\min\{d_u,d_v\}} \le \frac{\min\{d_u,d_v\}-1}{\min\{d_u,d_v\}}\in[0,1),
\end{equation}
i.e., the fraction of closed triangles around $e\in E$ among all possible triangles. 
Note that the ratio is well defined since $\min\{d_u,d_v\}\ge 1$, and $\cntv_e=0$ whenever $\min\{d_u,d_v\}=1$. 
Let $c \doteq \tfrac{1}{|E|}\sum_{e\in E} c_e$ be the average local edge clustering of $G$.
By Eq.~\eqref{eq:edgeClustering}, it holds $\cntv_e = c_e\min\{d_u,d_v\}$, hence $\cntv_e \le s$ is equivalent to $\min\{d_u,d_v\} \le s/c_e$. 
By using $c$ as a proxy for $c_e$ for each $e\in E$ we have that our condition $\cntv_e\le s$ can be alternatively evaluated as
\begin{equation}\label{eq:phiRuleOfThumb}
\min\{d_u,d_v\}\le \frac{s}{c} = \phi \quad \text{ for each } e=(u,v)\in E,
\end{equation}
where the left-hand side term does not depend on $\cntv_e$.
The intuition behind the above formula is as follows:
\begin{inparaenum}[($i$)]
\item networks that are very sparse globally force a large value of $\phi$, since it holds that $\cntv_e$ is low for all edges of $G$;
\item dense networks instead have a large value of $c$ forcing a low value of $\phi$. That is, caching $\cntv_e$ is extremely expensive compared to the randomized approximation.
\end{inparaenum}
While~\Cref{eq:phiRuleOfThumb} provides us a data-dependent approach to set the value of $\phi$, it still depends on $c$ which in turn depends on the values of $\cntv_e$ for each edge $e\in E$. We now show how to obtain estimate $\hat{c}$ such that $\Prob[|\hat{c} -c| \ge \varepsilon]\le \delta$ with an extremely efficient approach, for $(\varepsilon,\delta)\in (0,1)^2$. 

\textbf{Estimating $c$.}
The average local edge clustering $c$ can be estimated through the following randomized algorithm.
\begin{compactenum}
\item draw $S$ edges $e_1,\dots,e_S$ independently and uniformly from $E$;
\item for each $e_i=(u,v)$ with $i\in [S]$ let $w$ be its lower-degree node and $q$ the other remaining node. 
\item sample a single neighbour $z_i\in\neigh(w)$ uniformly at random and set $X_i \doteq \bm{1}[z_i\in\neigh(q)]$;
\item Output $\hat c \doteq \tfrac{1}{S}\sum_{i=1}^S X_i$.
\end{compactenum} 
Since each $z_i$ is drawn uniformly over the neighbours of $w$, it is simple to see that $\E[X_i\mid e_i=(u,v)] = \Prob[X_i=1\mid e_i=(u,v)] = \countTri_{e_i}/\min\{d_u,d_v\} = c_{e_i}$. 
By the linearity of expectation 
$\hat c$ is an unbiased estimator of $c$. 
Each sample can be processed in $\bigO(1)$,\footnote{Assuming that edge lookups and the random neighbour draw can be both performed in constant time} so $\hat c$ is computed in $\bigO(S)$ time.

Since $X_i\in\{0,1\}$ are i.i.d.\ with expectation $c$, by Hoeffding's bound (\Cref{theo:Hoeffding}) it holds $\Prob(|\hat c-c|\ge\varepsilon)\le 2e^{-2S\varepsilon^2}$. 
Hence, by setting $S=\bigO(\varepsilon^{-2}\log(1/\delta))$ it holds  $\Prob[|\hat{c} -c| \ge \varepsilon]\le \delta$ for desired $(\varepsilon,\delta)\in (0,1)^2$, on any $G$. 
Note that a similar estimator has been used to estimate the average \emph{node} clustering coefficient~\cite{Seshadhri2014CC}.

We use the estimate $\hat c$ to set 
\begin{align}\label{eq:phiSelect}
\phi = \big\lceil s/\max(\hat c,c_0)\big\rceil,
\end{align}
where we use $c_0>0$ to avoid division by zero, which can happen on bipartite graphs or trees, for example.

It is important to note that while in some applications an accurate \emph{relative} approximation to $c$ could be of interest, in our setting an estimate of $c$ with $\varepsilon=0.05$ and $\delta=0.02$ yields highly reliable values, and it requires only $S\approx 922$ samples.\footnote{We set $S$ to 1000 in practice.} This is mainly because
\begin{inparaenum}[($i$)]
\item if $c$ is very small (e.g., $<0.05$), then $\phi = s/\max(\hat{c}, c_0)$ for $\hat{c} \in [c-\varepsilon, c+\varepsilon]\approx 0$ corresponds to a very large value of $\phi$. That is, on sparse graph we can set $\phi$ extremely large;
\item while when $c$ is large then $\phi = s/\max(\hat{c}, c_0)$ is almost $s$, and recall that for our purposes the value of $s$ is often constant. Reconciling the fact that on dense graphs, retaining $\triSet_e$ for the various edges is extremely expensive.
\end{inparaenum}

\begin{table}[t]
	\centering
	\caption{Selection of $\phi$ according to~\Cref{eq:phiSelect}. $\phi$: is large on sparse graphs (\pokData) and small on dense graphs (\hollData). Runtimes are averaged over two runs and are reported in seconds.}
	\label{tab:phiAblation}
	\begin{tabular}{lccccc}
		\toprule
		Dataset & $\hat{c}$ & $\phi$ & Setup& Optimize & Total \\
		\midrule
		\pokData & 0.082 & 122 & 271.9 & 106.9 & 378.8 \\
		\hollData & 0.604 & 17 & 13.6 & 2\,879.0 & 2\,892.6 \\
		\bottomrule
	\end{tabular}
\end{table}

\textbf{Empirical validation.} 
To further support our design we show empirically how our rule for selecting the value of $\phi$ aligns with its desired behavior. 
Results are reported in~\Cref{tab:phiAblation} where we validate the selection of $\phi$ according to the rule of~\Cref{eq:phiSelect}, on two datasets with contrasting properties. 
That is, \pokData is a large but extremely sparse dataset, where exact triangle listing can be performed extremely efficiently. 
In contrast, \hollData is an extremely dense graph with a large number of triangles, where exact triangle listing is extremely demanding.
To produce the results in \Cref{tab:phiAblation} we estimated $\hat{c}$ on a \emph{single} run and then averaged the runtime of~\espressoR under the threshold of $\phi$ with the fixed  $\hat{c}$, for two independent runs. 
We set the value of $s=10$, $c_0=0.01$, and the number of iterations to $K=10$ and report under columns
\begin{inparaenum}[($i$)]
\item ``Setup'': the time required to estimate $\hat{c}$ and to cache the triangles for all those edges such that~\Cref{eq:phiRuleOfThumb} holds, with the value of $\phi$ as from~\Cref{eq:phiSelect}.
\item ``Optimize'': the runtime to perform the iterations of the gradient updates;
\item ``Total'': the sum of Setup and Optimize.
\end{inparaenum}

We observe that our rule from~\Cref{eq:phiSelect} well aligns with the desired behavior on the two graphs. 
A high value of $\phi$ on \pokData yields an extremely fast optimization (each of the $K$ iterations require $\approx$ 10 seconds), at the expense of a costly ``Setup'' phase (dominated by the caching of a large number of triangles in $\Delta(G)$). 
While on \hollData a small value of $\phi$ forces~\espressoR to sample the value of $\countTri_e(\pv)$ at each iteration for most edges, which is the desired behavior on dense graphs.

\section[appendix]{Reproducibility}\label{appsec:reprod}

\subsection[appendix]{Environment}\label{appsubsec:envCode}
Experimental results where performances were not the key measurement on small datasets have been performed on a MacOS M2 system with 16GB of RAM memory. 
Larger experiments have been performed on an HPC machine with 4x Intel Xeon Gold 6252N CPU @ 2.30/3.60GHz and 3TB RAM, under an image running Ubuntu 22.04 and generated with Apptainer.\footnote{\url{https://apptainer.org/}.}
All our code is written in Python 3.10. The scripts to reproduce our results are in~\CODE.

\subsection[appendix]{Methods and parameters}\label{appsubsec:baselines}

We now describe all methods we used in our experimental evaluation, and their parameter choice.

\espressoNL: unless otherwise stated we set $\errAcc= 0.001$.

\espressoGD: 
Minimizes $\obj(\bm{p})$ using projected gradient descent~\cite{bertsekas1999nonlinear} (more specifically Adam~\cite{kingma2014adam}). The update step at iteration $k$ is
\begin{equation*}
\bm{p}^{k+1} 
\;=\;
\operatorname{Clip}_{[0,1]}
\bigl(\,
\bm{p}^{k} - \eta_k\,\nabla\obj(\bm{p}^{k})
\bigr),
\end{equation*}
where $\eta_k$ is the learning rate and $\operatorname{Clip}_{[0,1]}$ is applied component-wise to ensure that all $p_e \in [0,1]$.
We fix the learning rate to $0.005$ by default. 
Such method has no convergence
guarantee. 

\espressoNL: Our exact mirror-descent algorithm
(\Cref{sec:optimization}). Unless otherwise stated, we consider an early stopping condition as from~\Cref{sec:optimization}, where we set $\varepsilon$ (the threshold for early stopping) to $0.0001$, and a maximum number of iterations set to $500$. Such algorithm has an automatic selection of the learning rate $1/L$ (i.e., as from~\Cref{lemma:Fprops}).

\espressoR: Unless otherwise stated we set its number of samples to $s=4$, the values of $c_0=0.01$, and $S=1000$ (used to estimate $\phi$). We set its number of iterations to be fixed to $K$, which we specify according to the setting of interest, all other parameters are set as for~\espressoNL. We fixed a random seed while performing its execution to enforce reproducibility.

\spara{Baselines for \textsc{MaxSTC}}

{\blGreedy}: 
Such method, sorts the edges by decreasing values of either $w(e)$ or
$p_e$, 
and labels edges as strong, one at a time, following the given order.
In doing so it avoids introducing STC violations. 
This produces a valid STC solution by
construction. 
The quality of the final solution (for the STC optimization problem) depends on the initial ordering of the edges.

{\blVertexCover}:
Edges are ranked using either weights $w(e)$ or probabilities $p_e$.
The method marks every edge in $G$ as strong, then it enumerates all open wedges (i.e., STC violations) in $\Wset$.
Each violation $\wed=\{e_1,e_2\}\in \Wset$ is sorted by decreasing maximum rank of the two edges $e_1$ or $e_2$. 
The method, then processes the violations by their rank: for each $\wed \in \Wset$ whose two edges are both marked strong, it marks the edge with lowest rank as weak, resolving the conflict.
Hence, the highest-priority conflicts are resolved first. 
The resulting weak edge set is a vertex cover of the graph
defined by the wedge structure.
Since
marking an edge weak can only decrease the number of violations and never introduce new ones,
the result is a valid STC assignment by construction.

\spara{Baselines for \stcden}

We consider the two methods of~\citet{wickrama2024dense}, which we denote with \blGreedyKDD and \blPeel (Algorithm~4 and Algorithm~1 in~\citep{wickrama2024dense}, respectively).
Both methods use the \emph{wedge graph} $\Gw$ of $G$, which has a vertex for each edge of $E$, and an edge between $e$ and $e'$ if $\{e,e'\}\in\Wset$.
Labeling as weak the edges of a vertex cover of $\Gw$ yields an STC-feasible labeling of $G$~\cite{sintos2014using}.
Both methods compute such a cover with the standard $2$-approximation based on a maximal matching.
Building $\Gw$ requires $\bigO(\sum_{v\in V} d_v^2)$ time and $\bigO(|\Wset|)$ space.
In contrast, \espressoDEN does not build $\Gw$: by~\Cref{lemma:expDense}, its step ($ii$) only requires the edge weights $\hat w_e$.

Both methods are peeling techniques that remove one vertex at a time, and differ in how the labeling is handled while $G$ is peeled. \blGreedyKDD updates the vertex cover after each deletion, avoiding its recomputation, 
while \blPeel fixes the labeling during the peeling and recomputes the cover only on the densest subgraph identified by the peeling.
Both methods return a subgraph $U$ and an STC-feasible labeling of $E[U]$, and require no parameter other than $\lamD$.
We use the code provided by the authors\footnote{\url{https://version.helsinki.fi/dacs/dense-stc-subgraph}.}
where we removed an unused import to Gurobi. The code is also available in our codebase.

\espressoDEN:
The algorithm is as detailed in~\Cref{subsec:stcdensest}.
We recall that
step ($iii$) requires solving a \textsc{MaxSTC} problem on $G[U]$, which is $\NP$-hard. We therefore rely on our method \blGreedy by using the values of $p_e$ to rank the edges of $E$, i.e., the same setup of~{\Cref{appsubsec:informedSTC}}.
As an ablation, we also consider the method proposed in~\Cref{lemma:stcdenCertificate}, which runs in polynomial time, denoted with \roundExp.
We compare the two approaches in~\Cref{appsubsec:stcdenRounding}.

\spara{Parameters for~\Cref{RQ:stcden}.}
Every method was run with a wall-clock budget of $900$ seconds per configuration (dataset, $\lamD$); methods that do not terminate within the time limit are marked with \xmark.

Regarding \espressoDEN.
The probabilities $\pv$ are obtained from a single \espressoNL run per dataset in the non-convex regime, with $\alpha=1$, $\lambda=0.05$, $\gamma=0.1$, $\errAcc=10^{-4}$, $p^0_e = 0.05$, and at most $500$ iterations.
The node set $U$ is obtained by a weighted peeling approach (providing a 2-approximation to the weighted densest subgraph). 

\subsubsection[appendix]{{LP relaxation and Chebyshev center}}\label{appsubsec:lpChebyshev}
{For~\Cref{RQ:informedSTC} we use the second LP relaxation of~\citet{adriaens2020relaxing} with $d=1$ (LP2), which has one variable $x_e\ge 0$ per edge, maximizes $\sum_{e\in E}x_e$, and has one constraint for each wedge of $\Wset$ and three constraints for each triangle in $G$.
LP2 can be formulated as $\max\{\bm 1^\top\bm x : A\bm x\le\bm b\}$, where each row $\bm a_i^\top\bm x\le b_i$ corresponds to one constraint, and we denote with $\mathrm{OPT}$ its optimal value.
The set of optimal solutions of LP2 corresponds to a face $F=\{\bm x : A\bm x\le\bm b,\ \bm 1^\top\bm x=\mathrm{OPT}\}$ of the polytope of LP2.
On Zachary's Karate Club, $70$ of the $78$ edges have $x_e=1/2$ in every optimal solution, while the values of the remaining eight edges vary over $F$.
Hence, the optimal solution of LP2 is not unique, and an LP solver returns one of the vertices of $F$, depending on the solver.
Following~\citet[Section~5.3.3]{adriaens2020relaxing}, we instead report the Chebyshev center of $F$.
The Chebyshev center of a polytope corresponds to the center of the largest ball contained in it, i.e., the point that is farthest from all of its facets.
For a fixed $F$, we consider the distance of a point $\bm y$ from the hyperplane of the $i$-th constraint to be $(b_i-\bm a_i^\top\bm y)/\|\bm a_i\|_2$, and as in~\citet{adriaens2020relaxing}, $\|\bm a_i\|_2$ corresponds to the norm of the $i$-th row of $A$.
Hence, the Chebyshev center can be computed by the LP that maximizes the smallest such distance $r$, i.e., $\max\{r : \bm 1^\top\bm y=\mathrm{OPT},\ \bm a_i^\top\bm y=b_i \text{ for } i\in I,\ \bm a_i^\top\bm y + r\|\bm a_i\|_2\le b_i \text{ for } i\notin I\}$, where $I$ is the set of rows of $A$ that hold with equality on all of $F$. We solve all LPs with HiGHS through SciPy.}

\subsection[appendix]{Metrics}\label{appsubsec:metrics}
In~\Cref{appsec:weightsVsStructure} we compare $\pv$ with a hard labeling $\bm z\in\{0,1\}^m$ using two standard ranking metrics.

AUC-ROC is the area under the curve of the true positive rate of $\{e: p_e\ge\tau\}$ as a function of its false positive rate, for varying $\tau\in[0,1]$. Such metric corresponds to the probability that a randomly chosen edge with $z_e=1$ has a higher $p_e$ than a randomly chosen edge with $z_e=0$, hence $0.5$ corresponds to a random ranking and $1$ to a perfect ranking.

AUC-PR is the area under the curve of the precision as a function of the recall, for varying $\tau\in[0,1]$. 
Since the metric is computed only with respect to the edges with $z_e=1$, it accounts for class imbalance, and a random ranking attains $|\{e: z_e=1\}|/m$.

\subsection[appendix]{Datasets} \label{appsubsec:data}

All our small datasets but \egofbData consist of directed timestamped interactions.
We aggregate events into a single weighted undirected graph.
For each unordered pair $e= \{u,v\}$, we assign a weight $w(e)$ defined as the mean of the
available directed weights.

\textbf{Bitcoin OTC Trust (\otcData).}
The Bitcoin OTC trust network from SNAP\footnote{\url{https://snap.stanford.edu/data/soc-sign-bitcoin-otc.html}}
contains signed ratings between users in the range $[-10,10]$.
We average multiple ratings per ordered pair and map the result to $[0,1]$ via
$w = (r+10)/20$.
The final undirected graph contains 5,881 nodes and 21,492 edges.

\textbf{CollegeMsg (\cmData).}
The CollegeMsg dataset\footnote{\url{https://snap.stanford.edu/data/CollegeMsg.html}}
records private messages between university users.
We count messages per ordered pair and normalise by the global maximum,
yielding $w \in (0,1]$.

\textbf{EU Email (\euData).}
The EU Email dataset\footnote{\url{https://snap.stanford.edu/data/email-Eu-core-temporal.html}}
contains timestamped internal emails within a European institution.
Edge weights correspond to normalised email counts.

\textbf{Ego-Facebook (\egofbData).}
The \egofbData dataset\footnote{\url{https://snap.stanford.edu/data/ego-Facebook.html}}
merges $10$ Facebook ego-networks into a single static undirected graph.
We use such dataset only in~\Cref{RQ:stcden}, as its high edge
clustering coefficient (\Cref{tab:datastats}) makes it the densest of our
small graphs.

We also consider larger datasets: \pokData, \twitchData, \gemsecData, and \wikiData are
downloaded from SNAP\footnote{\url{https://snap.stanford.edu/data/}}, while
\hollData and \bmgData are downloaded from Network
Repository\footnote{\url{https://networkrepository.com/}}. 
\hollData and
\bmgData are used exclusively for scalability experiments
in~\Cref{RQ:scalability} (\Cref{tab:scalability}), as their size and density is such that
{our methods based on an exact-gradient computation exceed their time and memory limits}.

The scripts to download and process each dataset are in our code repository: \CODE.

\section[appendix]{Additional experiments}\label{appsec:additExps}

\subsection[appendix]{Additional results for~\Cref{RQ:params}}
We report additional results for the experiments answering~\Cref{RQ:params} in~\Cref{fig:collapse-vs-spread-app} and~\Cref{fig:gamma-free-bimodal-app}. Overall the experiments match our theoretical characterization of~\Cref{lemma:Fprops} and our practical guidance for~\Cref{RQ:params} detailed in~\Cref{subsec:params}.

\begin{figure*}[ht]
\centering
\includegraphics[width=0.6\linewidth]{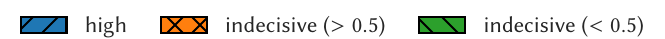}
\begin{subfigure}[b]{\linewidth}
\centering
\includegraphics[width=0.8\linewidth]{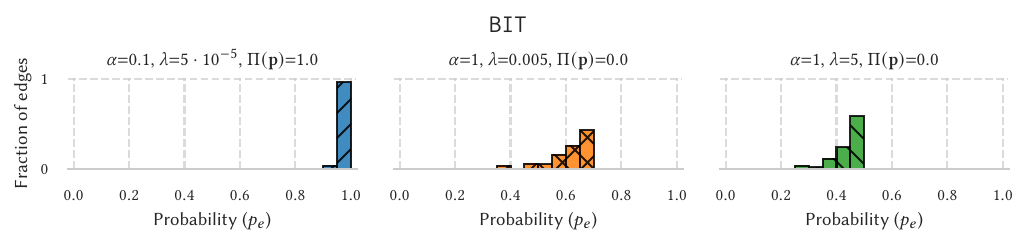}
\end{subfigure}
\begin{subfigure}[b]{\linewidth}
\centering
\includegraphics[width=0.8\linewidth]{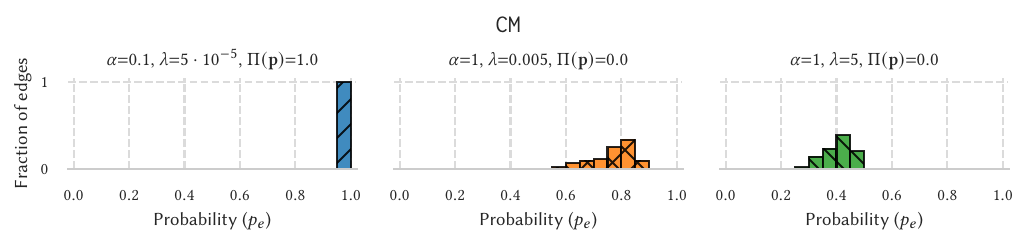}
\end{subfigure}
\begin{subfigure}[b]{\linewidth}
\centering
\includegraphics[width=0.8\linewidth]{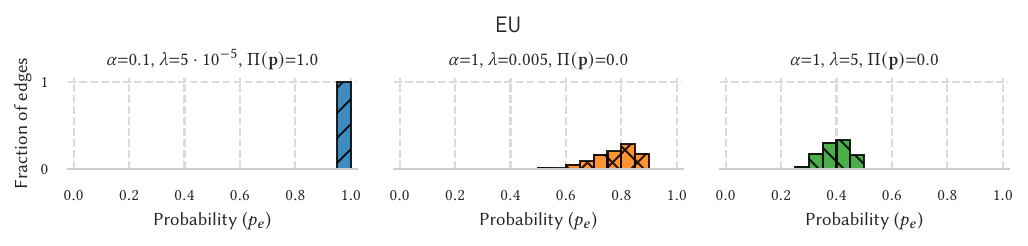}
\end{subfigure}
\caption{Convex regime ($\gamma$ from~\Cref{lemma:Fprops}). 
Distribution of the values of $p_e$ as a fraction of the total edges{; $\Pi(\pv)$ is the fraction of edges with $p_e<0.1$ or $p_e\ge0.9$}.
Left: {all edges have $p_e\ge0.9$}. Center: a representative solution for which most values of $p_e$ are widely spread $>0.5$. Right: a solution for which the majority of $p_e$ is widely spread with values $\le0.5$.
}
\label{fig:collapse-vs-spread-app}
\end{figure*}

\begin{figure}[t]
\centering
\includegraphics[width=0.4\linewidth]{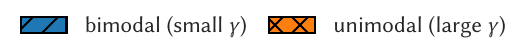}
\begin{tabular}{cc}
\begin{subfigure}[b]{0.5\linewidth}
	\centering
	\includegraphics[width=\linewidth]{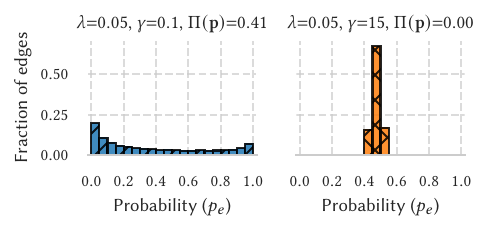}
	\caption{\cmData}
\end{subfigure} & 
\begin{subfigure}[b]{0.5\linewidth}
	\centering
	\includegraphics[width=\linewidth]{figures/gamma_free_bimodal_eu_email.pdf}
	\caption{\euData}
\end{subfigure}\\
\end{tabular}
\caption{Non-convex regime on \cmData (Left) and \euData (Right), with
$\alpha=1$. Each panel shows the distribution of edge probabilities
$p_e$ as a fraction of all edges. A bimodal distribution has two
distinct probability modes, as observed for some smaller values of
$\gamma$; larger $\gamma$ yields a single dominant mode.}
\label{fig:gamma-free-bimodal-app}
\end{figure}

\subsection[appendix]{Spectral scaling across synthetic and real graphs}
\label{app:synth_graph}
{We detail the experiment of~\Cref{subsec:params} that compares the threshold $\gamma_c$ of~\Cref{lemma:Fprops} with the observed onset $\widehat{\gamma}_{\Psi}$.
We compare $\widehat{\gamma}_{\Psi}$ with the threshold $\gamma_c^-\doteq\gamma_c=-\lambda\lambda_{\min}(\Mmat)/4$, the value $\gamma_c^+\doteq\lambda\lambda_{\max}(\Mmat)/4$, and its upper bound $\gamma_{\mathrm{Gersh}}\doteq\lambda|\Wmax|/4$ (\Cref{lemma:GershEigen}).
The last two values appear in the value of $L$ of \espressoNL, while the value of $\gamma_c^-$ determines the convexity of $\obj$ (\Cref{lemma:Fprops}).}

{\textbf{Graphs and parameters.}
We generate $80$ graphs with $n=200$ from four families. We consider, Erd\H{o}s--R\'enyi (\famER), four-block stochastic block models (\famSBM), Barab\'asi--Albert (\famBA), and toroidal random geometric graphs (\famRGG), with a mean degree $\bar{k}\in\{8,16\}$. 
We generate datasets using $10$ seeds per setting.
We also consider \otcData, \cmData, and \euData.
We set $\lambda=1$ on the synthetic graphs, $\lambda=0.05$ on the real datasets, and $\alpha=\lambda|\Wset|/m$.}

{\textbf{Optimization.}
For every graph we optimize $\obj$ for $81$ logarithmically spaced values of $\gamma\in[0.05,16]$.
For each value of $\gamma$ we use three different initializations for \espressoNL.
That is, we consider two signed perturbations of $\tfrac12\bm 1$ along an eigenvector of $\lambda_{\min}(\Mmat)$, and one seeded random initialization.
We terminate a run when the early-stopping condition of~\Cref{sec:optimization} holds with $\varepsilon=10^{-8}$, limiting the iterations to  $1\,000$. If no convergence is reached, then we repeat each setting increasing the limit on the iterations to $10\, 000$.
Among the runs that converge, we retain the solution with smallest objective value.}

{\textbf{Onset.}
For each retained solution we compute the value of $\Psi(\pv)$ from~\Cref{subsec:params}, which is $0$ for $\pv=\tfrac12\bm 1$ and approaches $1$ as the probabilities approach $0$ or $1$.
As $\gamma$ decreases, the value of $\Psi$ of a solution to $\obj$ increases from
values close to $0$ (uncertain solutions) to values close to $1$ (nearly
binary solutions).
We define the \emph{onset} $\widehat{\gamma}_{\Psi}$ of a graph as the
value of $\gamma$ at which $\Psi$ changes fastest as a function of
$\log\gamma$, i.e., the maximizer of $|d\Psi/d\log\gamma|$.
We compute $\widehat{\gamma}_{\Psi}$ from the values of $\Psi$ as follows.
Let $\gamma_1<\dots<\gamma_{81}$ be the grid of values we considered and let $x_i=\log\gamma_i$.
The points $x_i$ are equally spaced, i.e., $x_{i+1}-x_i=c$ with
$c=\log(16/0.05)/80$.
Let $y_i=\Psi(\pv^{(i)})$, where $\pv^{(i)}$ corresponds to the solution computed with 
$\gamma_i$.}

{\emph{Step 1.} To reduce the noise of the individual executions, we replace each $y_i$ by
\[
\bar y_i=\tfrac{1}{21}\left(-2y_{i-3}+3y_{i-2}+6y_{i-1}+7y_i
+6y_{i+1}+3y_{i+2}-2y_{i+3}\right),
\qquad 4\le i\le 78 .
\]
This is the value for $x_i$ of the parabola fitted by least squares to the
seven points $(x_{i-3},y_{i-3}),\dots,(x_{i+3},y_{i+3})$. 
For $i\le 3$ and $i\ge 79$, $\bar y_i$ corresponds to the value for $x_i$ of the
parabola fitted by least squares to the first (resp.\ last) seven points.}

{\emph{Step 2.} For each value of the grid, except the first and the last, we
estimate $|d\Psi/d\log\gamma|$ by the central difference
\[
s_i=\frac{|\bar y_{i+1}-\bar y_{i-1}|}{2c},
\qquad 2\le i\le 80,
\]
and let $j$ be the index with the largest such value, i.e.,
$j=\arg\max_{2\le i\le 80}s_i$.}

{\emph{Step 3.} The largest value of $s_i$ usually
lies between two grid points. 
To locate it, we consider the parabola through
the three points $(x_{j-1},s_{j-1})$, $(x_j,s_j)$, $(x_{j+1},s_{j+1})$ and
use the position of its maximum as
\[
\widehat{\gamma}_{\Psi}=\exp\!\left(x_j+\frac{c}{2}\cdot
\frac{s_{j-1}-s_{j+1}}{s_{j-1}-2s_j+s_{j+1}}\right).
\]
If $s_{j-1}=s_j=s_{j+1}$, we set
$\widehat{\gamma}_{\Psi}=\gamma_j$.}

{\textbf{Statistics.}
In our experiments over the $80$ synthetic graphs we considered, $\widehat{\gamma}_{\Psi}$ and $\gamma_c^-$ have Pearson correlation $r=0.946$ and Spearman correlation $\rho=0.922$.
For comparison, $\gamma_c^+$ yields $r=0.418$ and $\rho=0.338$, and $\gamma_{\mathrm{Gersh}}$ yields $r=0.454$ and $\rho=0.490$.
On every synthetic graph, the error $|\log(\widehat{\gamma}_{\Psi}/\gamma')|$ is smaller for $\gamma'=\gamma_c^-$ than for $\gamma'=\gamma_c^+$ and $\gamma'=\gamma_{\mathrm{Gersh}}$.
The experiments for the real datasets are reported in~\Cref{fig:spectral-predicted}.
\Cref{fig:spectral-collapse} reports $\Psi$ as a function of $\gamma/\gamma_c^-$ for all graphs.
For $\gamma>\gamma_c^-$, $\obj$ has a unique minimizer (\Cref{lemma:Fprops}).
Such a minimizer is $\tfrac12\bm 1$ only when every edge belongs to the same number of wedges.
Hence, for $\gamma/\gamma_c^->1$ the value of $\Psi$ is positive on all graphs of~\Cref{fig:spectral-collapse}, and it approaches $0$ as $\gamma/\gamma_c^-$ increases.
Therefore, the convexity threshold $\gamma_c^-$ of~\Cref{lemma:Fprops} predicts the onset across different graph families, as captured by our analysis.}

\begin{figure}[t]
	\centering
	\includegraphics[width=.42\linewidth]
	{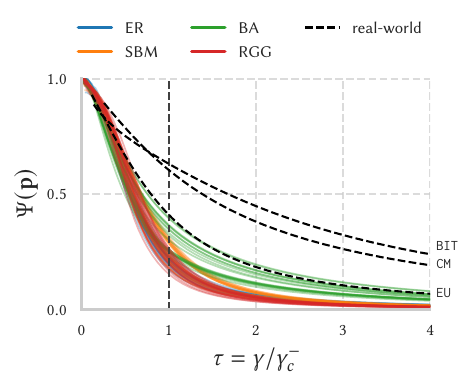}
	\caption{
		\textbf{Normalized entropy transition.}
		Value of ${\Psi}$ as a function of
		$\gamma/\gamma_c^-$ for the four synthetic graph families and
		the three smallest real-world datasets.
	}
	\label{fig:spectral-collapse}
\end{figure}

{\textbf{Gallai irregularity.}
When every edge belongs to the same number of wedges, the solution for $\gamma=\gamma_c^-$ corresponds to $\pv=\tfrac12\bm 1$ (\Cref{prop:balancedTransition}), hence $\Psi=0$.
We study whether $\Psi$ for $\gamma=\gamma_c^-$ increases when the numbers of wedges of the edges vary.
We measure this difference through the Gallai irregularity $R_\Gamma=2\operatorname{std}(\bm d_\Gamma)/\xi_-$ defined in~\Cref{subsec:params}, where $(\bm d_\Gamma)_e=|\Wset_e|$ corresponds to the number of wedges that contain $e$, and $R_\Gamma=0$ when all edges belong to the same number of wedges.
For each synthetic graph, we compute $\Psi$ for $\gamma=\gamma_c^-$ by linear interpolation between the two closest grid values of $\gamma$.
\Cref{fig:synth-residual} reports $\Psi$ at $\gamma=\gamma_c^-$ as a function of $R_\Gamma$.
Within each family ($20$ graphs), the Pearson correlation $r$ and the Spearman correlation $\rho$ between the two quantities are high for \famBA ($r=0.97$, $\rho=0.90$) and \famRGG ($r=0.87$, $\rho=0.88$), moderate for \famER ($r=0.68$, $\rho=0.59$), and low for \famSBM ($r=0.17$, $\rho=0.24$).
Hence, $R_\Gamma$ does not necessarily determine $\Psi$ when $\gamma=\gamma_c^-$.}

\begin{figure*}[t]
	\centering
	\includegraphics[width=\linewidth]
	{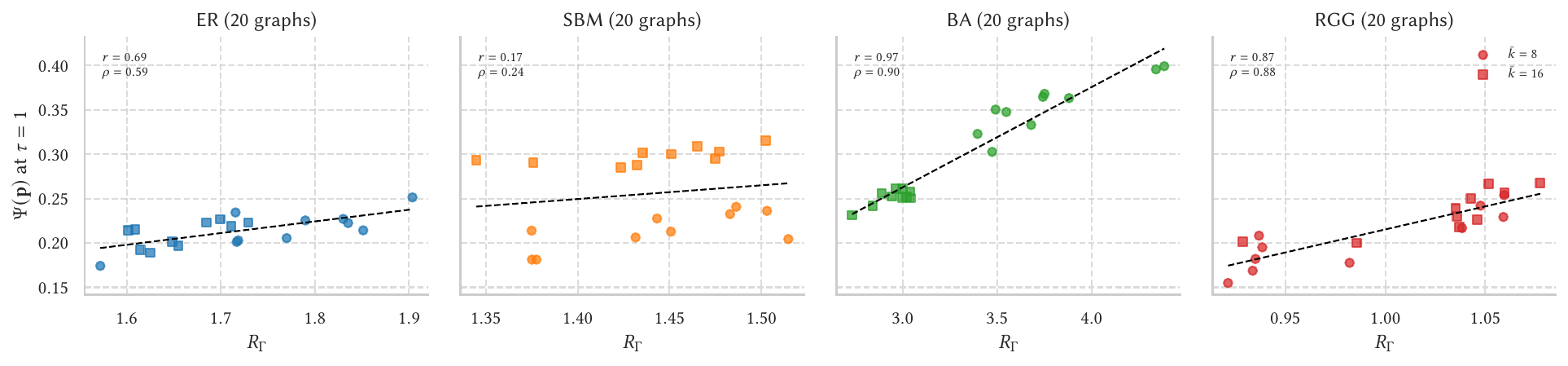}
	\caption{
		\textbf{{Gallai irregularity and entropy deficit when $\gamma=\gamma_c^-$.}}
		{$\Psi(\pv)$ corresponds to the value of  the solution for $\gamma=\gamma_c$ ($=\gamma_c^-$) as a function of the Gallai irregularity $R_\Gamma$, for the $20$ graphs of each family.
		Circles and squares corresponds to graphs with target mean degree $\bar{k}=8$ and $\bar k=16$, respectively.
		Dashed lines are least-squares lines fitted to the graphs of each family, and $r$ and $\rho$ correspond to the Pearson and Spearman correlations.}
	}
	\label{fig:synth-residual}
\end{figure*}

\subsection[appendix]{Exact validation against optimal hard STC}
\label{appsec:hardOptimaValidation}

We complement the parameter experiments of~\Cref{subsec:params} by
studying whether the probabilities inferred by \espresso reflect
optimal hard-STC labelings.
Since a instance to the hard STC problem may admit several equally optimal labelings, we
compare against the complete optimal set of solutions rather than selecting an arbitrary
single optimum.

For a graph $G$, let
\[
\mathcal Z^*
\doteq
\operatorname{argmin}_{\bm z\in\{0,1\}^m}
\hardCost(\bm z;\alpha,\lambda)
\]
denote the set of globally optimal labelings for the penalized hard-STC
objective. %
For every edge we define its hard-optimum marginal
\begin{equation}\label{eq:averageHard}
r_e
\doteq
\frac{1}{|\mathcal Z^*|}
\sum_{\bm z\in\mathcal Z^*} z_e.
\end{equation}
Hence, for a given edge $e\in E$ it holds $r_e=1$ if the edge is strong in every hard optimum of $\mathcal Z^*$, $r_e=0$ if
it is weak in every optimum, and $r_e \approx 0.5$ denotes disagreement
among optimal hard labelings.
We remark that $r_e$ is used only as an empirical measure of the ambiguity of solutions over $\mathcal Z^*$.

\textbf{Setup.} We generate $100$ small graphs, $25$ each from Erd\H{o}s--R\'enyi,
stochastic block model, planted-structure, and Watts--Strogatz families.
The graphs have $7$--$9$ vertices and between $5$ and $17$ edges, allowing
us to enumerate all $2^m$ binary labelings exactly.
We fix $\alpha=0.7, \lambda=1$, and  $\gamma=0.5$.
For each graph we enumerate $\mathcal Z^*$ exactly and independently
optimize the objective $\obj$ for \espresso {with \espressoNL.
In the convex regime (over $39$ graphs) we run \espressoNL once with $\pv^0=\tfrac12\bm 1$.
When the objective $\obj$ is in the non-convex regime (over $61$ graphs) we run \espressoNL using $20$ initializations ($\tfrac12\bm 1$, $0.1\cdot\bm 1$, $0.9\cdot\bm 1$, and $17$ random ones) retaining the solution with smallest objective value.
We set $\varepsilon=10^{-11}$ and at bound the number of iterations to $20\, K$, such that each execution converges.
We denote by $\pv^*$ the solution of \espressoNL.}
{Since $\hardCost(\bm z;\alpha,\lambda)$ penalizes STC violations instead of forbidding them, a
labeling in $\mathcal Z^*$ may contain violations when $\lambda\le\alpha$.
For $\lambda>\alpha$ no violation occurs.
In fact consider a solution $\bm z$ with a violation when $\lambda>\alpha$, then by labeling as weak one edge of a wedge that yields a violation, increases the first term of $\hardCost(\bm z;\alpha,\lambda)$ by $\alpha$ and
decreases the second term of by at least $\lambda$, i.e., $\bm z$ is not optimal.
Hence, under our setting of $\alpha=0.7$ and $\lambda=1$, $\mathcal Z^*$ corresponds to the set of
optimal solutions of \textsc{MaxSTC}~\cite{sintos2014using}.}

We compare $\pv^*$ with the hard-optimum marginals $\bm r$ using Pearson
and Spearman correlation and the mean absolute error
\[
\operatorname{MAE}(\pv^*,\bm r)
\doteq
\frac{1}{m}\sum_{e\in E}|p^*_e-r_e|.
\]
We also test two deterministic {roundings} of $\pv^*$.
The first assigns $z_e=1$ iff $p^*_e>1/2$.
The second assigns $z_e=1$ to the $k$ edges with largest $p^*_e$, where $k$
is the number of strong edges in an optimal hard labeling of $\mathcal Z^*$, i.e., $k=|\{z_e=1, \bm z\in \mathcal Z^*\}|$.
We will report the fraction of graphs for which each {rounding} is itself a
globally optimal hard-STC labeling.

\begin{table}[t]
\centering
\caption{
Comparison of the probabilities {for the objective $\obj$} with exact optimal hard-STC
labelings on $100$ small graphs.
Rows group the graphs by their number of optimal hard-STC labelings,
and ``Graphs'' is the number of graphs in each group.
``Thresh.'' is the fraction of graphs for which labeling as strong the
edges with $p^*_e>1/2$ gives an optimal hard-STC labeling.
``Top-$k$'' is the same fraction when labeling as strong the $k$ edges
with largest $p^*_e$, where $k$ is the number of strong edges in an
optimal hard-STC labeling.
$\rho_P$ and $\rho_S$ denote the Pearson and Spearman correlations between
$p^*_e$ and the hard-optimum marginal $r_e$ {(see \Cref{eq:averageHard})}.
Correlations are omitted when $\bm r$ is constant.
}
\label{tab:hardOptimaValidation}
\begin{tabular}{lrrrrrr}
\toprule
Optima & Graphs & Thresh. & Top-$k$ & $\rho_P$ & $\rho_S$ & MAE \\
\midrule
All & 100 & .480 & .820 & .928 & .874 & .152 \\
Unique & 30 & .900 & .900 & .940 & .817 & .224 \\
Multiple & 70 & .300 & .786 & .922 & .900 & .121 \\
$2$--$4$ & 43 & .442 & .814 & .915 & .890 & .146 \\
$\geq5$ & 27 & .074 & .741 & .933 & .915 & .082 \\
\bottomrule
\end{tabular}
\end{table}

\textbf{Results.}
\Cref{tab:hardOptimaValidation} shows two complementary behaviors.
When the hard-STC optimum is unique, thresholding the 
probabilities of $\pv^*$ recovers the optimum exactly on $90\%$ of the $30$ graphs with such a
unique optimum.
Thus, when the discrete problem provides an unique solution (i.e., there is no ambiguity),
our probabilistic model and its objective $\obj$ often captures the same structural information.

Similarly, when the hard problem admits several optima, $\pv^*$ is also strongly aligned with the hard-optimum marginals.
In fact, among the $70$ instances where $|\mathcal{Z}^*|>1$, the MAE value is $0.121$,
while Pearson and Spearman correlations are $0.922$ and $0.900$.
The alignment between $\pv^*$ and the solutions of $\mathcal{Z}^*$ is further supported by the $27$ graphs with at least five optimal
labelings, which have a MAE of $0.082$ and a high Pearson correlation
of $0.933$.
That is, our results support our probabilistic view of the values of $\pv^*$ for \espresso:
such probabilities
capture the ambiguity in networks that yields to several equally optimal (and often contrasting) hard
solutions.

We further observe that the ranking induced by $\pv^*$ is also consistent with hard STC.
Rounding a solution $\pv^*$ by selecting as strong its $k$ edges with largest probabilities yields an exact hard optimum over
$82\%$ of all graphs, including $78\%$ of the graphs with multiple
optima.
These results further support the alignment of a solution $\pv^*$ for $\obj$ with the set of optimal hard solutions $\mathcal{Z}^*$.

\textbf{Recovering hard solutions.}
{For the value of $\gamma=0.5$ we considered above, the two roundings we discussed do not always yield an optimal
hard labeling (\Cref{tab:hardOptimaValidation}).
The entropy term in the objective encourages a solution $\pv^*$ to assign values far from $0$ or $1$ for those edges that have ambiguous structural information.
Note that when $\gamma$ is small, the entropy term is negligible, and $\obj$ has a solution close to a
hard-STC assignment.
We therefore consider the previous setting using \espressoNL on $30$ of the $100$ graphs
varying $\gamma$ between $0.010\,\gamma_c$ and $0.017\,\gamma_c$, and using the same setting of
$\alpha=0.7$ and $\lambda=1$.
Our choice forces the probabilities of a solution $\pv^*$ to be essentially $0$ or $1$.
Interestingly, both the above
roundings applied to $\pv^*$, they yield an optimal hard labeling on all such $30$ graphs.
In contrast, when $\gamma=0.5$, the first rounding yields a hard optimal solution only on $14$ out of 30 graph, and the rounding based on the top-$k$ edges yields an optimal solution on $24$ of the $30$ graphs.
Hence, increasing the value of $\gamma$ encourages more uncertain solutions, as captured by our analysis.}

\subsection[appendix]{Exact Gibbs-marginal validation}
\label{appsec:gibbsMarginals}

{By our discussion in \Cref{app:gibbs-mean-field}, $q(\cdot;\pv^*)$ corresponds to the factorized
distribution minimizing $\mathrm{KL}(q\,\|\,\pi_\gamma)$, where
$\pi_\gamma(\bm z)\propto\exp(-\hardCost(\bm z)/\gamma)$ is the Gibbs
distribution of the penalized STC cost.
Since $q$ does not model dependencies among edge labels, $p_e^*$ differs in
general from the Gibbs marginal $\mu_e\doteq\Prob_{\pi_\gamma}(Z_e=1)$
(see also \Cref{sec:espresso_model}).
We now study the difference between $\pv^*$ and $\pi_\gamma$, on graphs where $\pi_\gamma$ can be computed
exactly.}

\textbf{Setup.}
{We consider $92$ random graphs, $23$ from each of the families \famER,
\famSBM, \famBA, and \famRGG, and in addition the $8$ even cycles with a number of nodes ranging from $6$
to $20$.
Every graph has at most $m=20$ edges, hence we compute
$\mu_e=\sum_{\bm z\in\{0,1\}^m}z_e\,\pi_\gamma(\bm z)$ by enumerating all
$2^m$ labelings.
For each graph we consider ten values of $\tau=\gamma/\gamma_c$, including values
smaller and larger than $1$ (i.e., we optimize $F$ in both the convex and non-convex regimes).
For each value we execute \espressoNL using ten deterministic initializations and
retain the solution with smallest objective value.
For a graph $G$ and a value of $\gamma$ we compute
\[
\mathrm{TV}(G,\gamma)=\frac{1}{m}\sum_{e\in E}|p_e^*-\mu_e|,
\qquad
\Delta(G,\gamma)=\frac{1}{m}\sum_{e\in E}
\big(|p_e^*-1/2|-|\mu_e-1/2|\big).
\]
The value $|p_e^*-\mu_e|$ corresponds to the total-variation (TV) distance between the
Bernoulli distributions with parameters $p_e^*$ and $\mu_e$.
Thus, $\mathrm{TV}(G,\gamma)$ corresponds to the average distance between $\pv^*$ and $\bm\mu$.
Furthermore, when $\Delta(G,\gamma)>0$, the values $p_e^*$ are, on average, further from $1/2$
than $\mu_e$, i.e., $\pv^*$ is more confident than $\bm\mu$.
For each family of graphs and regime of $\obj$ ($\tau\geq1$ or $\tau<1$), we will report the
averages of $\mathrm{TV}(G,\gamma)$ and $\Delta(G,\gamma)$ over the graphs
and values of $\gamma$, the largest absolute difference $|p_e^*-\mu_e|$ over all $e\in E$, and the Spearman
correlation $\rho$ between $p_e^*$ and $\mu_e$.}

\begin{table}[t]
\centering
\caption{
Distance between the \espresso probabilities $p_e^*$ and the exact
Gibbs marginals $\mu_e$ on graphs with at most $20$ edges.
For each edge, $|p_e^*-\mu_e|$ is the total-variation (TV) distance
between the Bernoulli distributions with parameters $p_e^*$ and $\mu_e$.
``Mean'' and $\Delta$ are the averages of $\mathrm{TV}(G,\gamma)$ and
$\Delta(G,\gamma)$ over the graphs and values of $\gamma$;
``Max'' is the largest $|p_e^*-\mu_e|$.
$\Delta>0$ means that $p_e^*$ is on average further from $1/2$ than
$\mu_e$. $\rho$ is the Spearman correlation
between $p_e^*$ and $\mu_e$, undefined for even cycles since
$\mu_e=1/2$ for every edge.
}
\label{tab:gibbsMarginals}
\small
\setlength{\tabcolsep}{4pt}
\begin{tabular}{lrrrrrrrr}
\toprule
& \multicolumn{4}{c}{$\tau\geq1$ (convex)} & \multicolumn{4}{c}{$\tau<1$ (non-convex)} \\
\cmidrule(lr){2-5}\cmidrule(lr){6-9}
& \multicolumn{2}{c}{TV} & & & \multicolumn{2}{c}{TV} & & \\
\cmidrule(lr){2-3}\cmidrule(lr){6-7}
Family & Mean & Max & $\Delta$ & $\rho$ & Mean & Max & $\Delta$ & $\rho$ \\
\midrule
\famER & .034 & .186 & .029 & .986 & .126 & .586 & .115 & .930 \\
\famSBM & .035 & .170 & .032 & .990 & .091 & .500 & .087 & .963 \\
\famBA & .027 & .157 & .022 & .990 & .113 & .667 & .099 & .938 \\
\famRGG & .026 & .165 & .025 & .991 & .056 & .498 & .054 & .966 \\
Even cycle & .000 & .000 & .000 & -- & .407 & .500 & .407 & -- \\
\bottomrule
\end{tabular}
\end{table}

\textbf{Results.}
{\Cref{tab:gibbsMarginals} reports our results.
For $\tau\geq1$, $\pv^*$ is close to $\bm\mu$ on the four random families.
That is, the mean TV distance is at most $0.035$, and the maximum distance is at most $0.186$,
and $\rho\geq0.986$.
We note that the TV distance decreases as $\tau$ increases, and for the value of $\tau=4$ the mean TV
distance, averaged over the four random families, is $0.002$.
When $\tau<1$, the mean TV distance is between $0.056$ (\famRGG) and $0.126$
(\famER), and the maximum attains $0.667$ (\famBA), showing that a solution to $\obj$ in the non-convex regime may not be well-aligned with $\bm \mu$ .
The correlation between the rankings induced by $\pv^*$ and $\bm \mu$ follow the same trends, with $\rho$ being in the range of $0.930$
and $0.966$.
In both regimes $\Delta(G,\gamma)$ is close to the mean TV distance, hence $\pv^*$
differs from $\bm\mu$ by being more confident in the values of $\pv^*$, which may be due to the fact that our model does not capture dependencies among edges.}

\subsection[appendix]{Probabilistic strength vs.\ pairwise weights}\label{appsec:weightsVsStructure}

We  briefly quantify the relationship between the tie strength $p_e$ estimated
by \espressoNL and the aggregated dataset weights $w(e)$ (trust rating,
message volume, or email count) on \otcData, \euData, \cmData. See \Cref{appsubsec:data} for a detailed description on how such weights are computed. 
We emphasize that $w(e)$ is not a
ground truth for tie strength: it is a pairwise quantity, whereas $p_e$ is a relational quantity that depends on
the topology surrounding each edge. 
The two therefore measure different
properties. 
We executed~\espressoNL with $\varepsilon =\errAcc=0.01$ under the setting of $\alpha=1, \lambda=0.05, \gamma=0.1$ which makes the distribution of a solution to~\espresso bimodal (see our answer to~\Cref{RQ:params} explaining our parameter choice). 
For the resulting $\pv$ we computed ranking metrics according to the ranking induced by the weights $w(e)$. 
In detail, we report the Spearman rank correlation
between $p_e$ and $w(e)$.
In addition, we assigned a strong label $z_e=1$ to the edges within the top 10\% of the weights $w(e)$, and $z_e=0$ otherwise, we then evaluated classification metrics: AUC-ROC, AUC-PR.

The results are reported in~\Cref{tab:espresso_vs_weights}.
The three datasets exhibit different behavior, indicating that
tie strength does not purely depend on pairwise interactions.
On \otcData, structural tie strength is essentially uncorrelated with ratings ($\rho \approx 0$), and $p_e$ ranks the top-10\% edges almost randomly (AUC-ROC $0.497$).
On \cmData, we observe a negative correlation:
a high value of  $w(e)$ (corresponding to a large volume of messages exchanged between nodes), negatively correlates with tie strength as captured by our model; consistently, its AUC-ROC is below $0.5$.
In contrast, \euData shows a mildly positive correlation
($\rho=0.280$, AUC-ROC $0.743$).
The AUC-PR values are consistent with our observations: since the strong labels are the top 10\% of the weights, a random score attains AUC-PR $0.10$.
These behaviors show that pairwise interactions alone cannot capture the
relational structure that STC enforces, motivating our probabilistic
formulation.

\begin{table}[t]
	\centering
	\caption{Comparison between a solution to \espresso and pairwise weights $w(e)$.}
	\resizebox{0.5\columnwidth}{!}{%
	\begin{tabular}{lccc}
		\toprule
		Dataset & Spearman $\rho$ & AUC-ROC (top-10\%) & AUC-PR (top-10\%) \\
		\midrule
		\otcData & $-0.028$ & 0.497 & 0.120 \\
		\cmData & $-0.176$ & 0.355 & 0.071 \\
		\euData & $+0.280$ & 0.743 & 0.238 \\
		\bottomrule
	\end{tabular}}
	\label{tab:espresso_vs_weights}
\end{table}

\subsection[appendix]{Application: informed hard STC}
\label{appsubsec:informedSTC}

A natural question is if $\pv$ can be used to compute a hard STC-assignment. 
We consider two settings.
Recall that for \otcData, \cmData, \euData and \pokData we obtain a weight $w(e)$ from the data, while \twitchData, \gemsecData and \wikiData have no edge metadata, hence we set $w(e)=1$.
We then executed our two informed baselines for hard STC (\blGreedy and \blVertexCover) once with the weights $w(e)$ and once with the values of $\pv$ provided by~\espressoNL under the convex
regime ($\gamma$ selected as from~\Cref{lemma:Fprops}).
We use the convex regime
since under such setting \espressoNL is optimal and converges extremely fast
(see~\Cref{appsec:solutionComparison}).
We then compared the quality of the two hard STC solutions denoted with $S_w=\{e : y_e^{w}=1\}$ (when using weights $w(e)$) and
$S_p=\{e : y_e^{p}=1\}$ (when using $\pv$ by \espressoNL). 
Let $|S_w| = m_w$ and $|S_p| = m_p$.
We computed:
$m^+ = (m_p - m_w)/m_w$, the relative increase (in percentage) in strong edges obtained using $\pv$ over $w(e)$,
and $ m_p / m$, the fraction of strong edges in the hard solution obtained using \espressoNL as informed ranking.

\begin{table}[t]
	\centering
	\caption{\espressoNL for hard-STC as informed ranking. Missing configurations did not terminate for high resource usage (>1hr of runtime or >100GB of RAM) of the hard STC baseline. See~\Cref{appsubsec:informedSTC} for a detailed description.}
	\resizebox{0.4\columnwidth}{!}{%
	\begin{tabular}{llcc}
		\toprule
		Dataset & Method & $m^+$ & $m_{p}/m$ \\
		\midrule
		\multirow{2}{*}{\otcData} & \blGreedy & +20.9\% & 0.086 \\
		 & \blVertexCover & +31.8\% & 0.086 \\
		\midrule
		\multirow{2}{*}{\cmData} & \blGreedy & +27.5\% & 0.061 \\
		 & \blVertexCover & +31.5\% & 0.061 \\
		\midrule
		\multirow{2}{*}{\euData} & \blGreedy & +17.9\% & 0.113 \\
		 & \blVertexCover & +18.9\% & 0.113 \\
		\midrule
		\multirow{1}{*}{\pokData} & \blGreedy & +22.5\% & 0.053 \\
		\midrule
		\multirow{1}{*}{\twitchData} & \blGreedy & +36.5\% & 0.022 \\
		\midrule
		\multirow{2}{*}{\gemsecData} & \blGreedy & +26.1\% & 0.046 \\
		 & \blVertexCover & +303.3\% & 0.046 \\
		\midrule
		\multirow{1}{*}{\wikiData} & \blGreedy & +14.6\% & 0.013 \\
		\bottomrule
	\end{tabular}}
	\label{tab:informed_stc}
\end{table}

\spara{Results.}
Results are reported in~\Cref{tab:informed_stc}. We observe that across all datasets, a solution obtained from the rankings provided by~\espressoNL produces more strong edges than using the weights $w(e)$.
The increase ranges from $17.9$ to $31.8\%$ on the datasets where $w(e)$ is available, and from $14.6$ to $36.5\%$ on those without edge metadata, with a large value of $+303.3\%$ for \blVertexCover on \gemsecData, where the solution $S_w$ has an extremely small number of strong edges.
Overall, this shows that \espressoNL is effective in providing an ordering of the edges for computing hard STC-solutions.

\subsection[appendix]{Ablation: rounding for \espressoDEN}
\label{appsubsec:stcdenRounding}

We now compare the two different rounding approaches for \espressoDEN, as discussed in~\Cref{subsec:stcdensest}.
We recall that both approaches are applied at step $(iii)$ to compute an STC-feasible labeling on the \emph{same} subgraph $U$ (identified over step $(ii)$).
We denote with ``\roundExp'' the first rounding based on~\Cref{prop:rounding}, and with ``\blGreedy'' the one used in~\Cref{subsec:stcden}, which ranks the edges of $G[U]$ by their values $p_e$ (see~\Cref{appsubsec:baselines}).

\Cref{tab:stcden_rounding} compares the two rounding approaches.
The \roundExp approach optimizes $\We(\bm z)+\V(\bm z)$: on a dense $G[U]$, as is often the case, each edge belongs to many wedges and the procedure labels almost the whole $G[U]$ as weak.
That is, its fraction of weak edges ranges from $0.91$ to $0.99$ in all configurations.
Instead \blGreedy, since it focuses on~\Cref{eq:densFixedU}, provides solutions with high $\dens(\cdot)$ for all configurations, improving the solutions of \roundExp up to $61.7\%$ on \egofbData at $\lamD=0.2$.
Finally, by definition both methods report an STC-feasible solution,
hence every value of $\dens(\cdot)$ we report is that of a feasible solution for the \stcden problem.

We therefore observe, that for this specific application, our approach based on \blGreedy is the most effective choice for step $(iii)$, motivating our choice for the setup of~\Cref{subsec:stcden}.

\begin{table}[t]
	\centering
	\caption{Objective of the rounded solution $q(U,L;\lamD)$ by the two methods reported in~\Cref{subsec:stcdensest}. The experiments are as in~\Cref{subsec:stcden}. We also report the fraction of weak edges in each hard labeling.}
	\label{tab:stcden_rounding}
	\begin{tabular}{llrrrr}
		\toprule
		& & \multicolumn{2}{c}{$q(U,L;\lamD)$} & \multicolumn{2}{c}{frac.\ weak of $E[U]$} \\
		\cmidrule(lr){3-4}\cmidrule(lr){5-6}
		Dataset & $\lamD$ & \roundExp & \blGreedy & \roundExp & \blGreedy \\
		\midrule
		\otcData & 0.2 & 3.105 & \textbf{3.905} & 0.910 & 0.823 \\
		 & 0.4 & 5.855 & \textbf{6.345} & 0.912 & 0.849 \\
		 & 0.6 & 10.683 & \textbf{10.905} & 0.933 & 0.901 \\
		 & 0.8 & 13.929 & \textbf{14.016} & 0.929 & 0.904 \\
		\midrule
		\cmData & 0.2 & 3.064 & \textbf{3.731} & 0.994 & 0.938 \\
		 & 0.4 & 6.411 & \textbf{6.924} & 0.993 & 0.939 \\
		 & 0.6 & 9.937 & \textbf{10.263} & 0.989 & 0.939 \\
		 & 0.8 & 13.330 & \textbf{13.479} & 0.987 & 0.943 \\
		\midrule
		\euData & 0.2 & 6.032 & \textbf{8.138} & 0.912 & 0.794 \\
		 & 0.4 & 11.563 & \textbf{12.893} & 0.927 & 0.842 \\
		 & 0.6 & 16.778 & \textbf{17.613} & 0.936 & 0.859 \\
		 & 0.8 & 22.297 & \textbf{22.671} & 0.946 & 0.878 \\
		\midrule
		\egofbData & 0.2 & 17.455 & \textbf{28.218} & 0.909 & 0.698 \\
		 & 0.4 & 29.561 & \textbf{37.674} & 0.924 & 0.720 \\
		 & 0.6 & 47.055 & \textbf{51.662} & 0.953 & 0.802 \\
		 & 0.8 & 62.443 & \textbf{64.598} & 0.957 & 0.818 \\
		\bottomrule
	\end{tabular}
\end{table}

\subsection[appendix]{Accuracy of the randomized gradient}
\label{appsec:gradientAccuracy}

We additionally evaluate the accuracy of the sampling approach used by
\espressoR for estimating the gradient.
Since this validation requires computing the exact gradient for comparison,
we use \wikiData and \pokData, for which exact triangle enumeration is
feasible within one hour.

\begin{figure}[t]
\centering
\includegraphics[width=0.45\linewidth]
{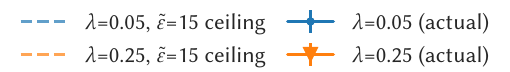}\\
\begin{tabular}{lr}
\includegraphics[width=0.35\linewidth]
{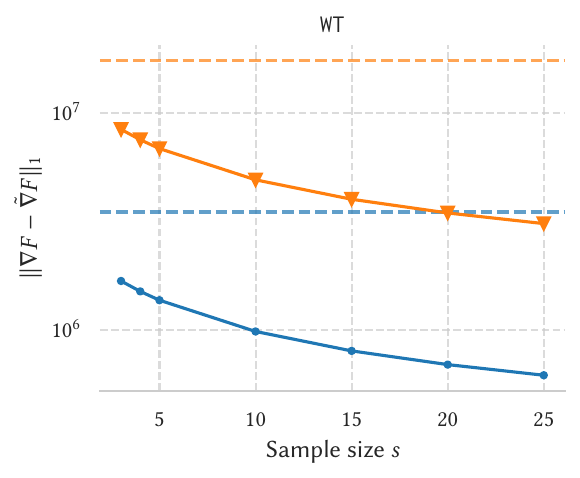} &
\includegraphics[width=0.35\linewidth]
{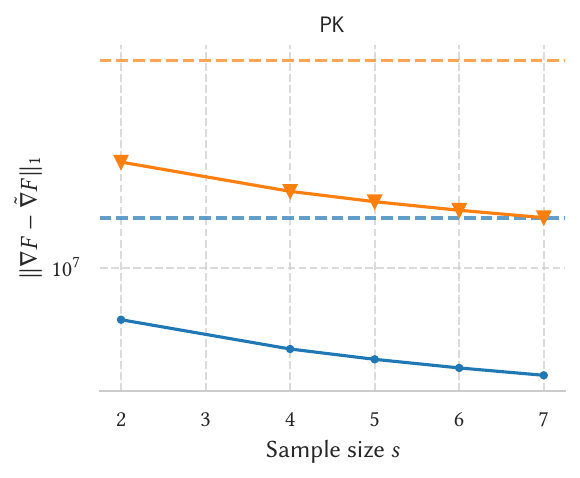}
\end{tabular}
\caption{Approximation quality of~\espressoR on \wikiData and
\pokData for increasing $s$.
Solid lines show the actual $\ell_1$ distance
$\lambda\cdot\|\countTriv(\pv)-\appxTriv(\pv)\|_1$
between the exact and approximate gradient, for two values of
$\lambda$.
Dashed lines show the theoretical upper bound
$\lambda\errEst m$ guaranteed by
\Cref{theo:appxGuarantees}, for $\errEst=15$.}
\label{fig:gradient_l1_ceiling}
\end{figure}

\Cref{fig:gradient_l1_ceiling} shows that the sampling approach of
\espressoR accurately estimates the gradient in terms of the $\ell_1$
error required by~\Cref{theo:appxGuarantees}.
The error decreases substantially as $s$ increases and is below
the {bound} from~\Cref{theo:appxGuarantees} {for every value of $s$ and both values of $\lambda$}.
Moreover, $\lambda\in\{0.05,0.25\}$ lies within the range of values used
throughout our experiments.

\Cref{fig:gradient_l1_ceiling} validates the bound for $\errEst=15$.
When $\errEst$ is set according to a relative accuracy of
$|\obj(\pv^*)|$ (see~\Cref{theo:appxNL}), its value scales with the
magnitude of the objective, which is large in our experiments
(see~\Cref{tab:solution_comparison}).
For example, in our experiments, $\errEst=15$ corresponds to an output
$\pv$ whose objective $\obj(\pv)$ has approximately $10\%$ relative
error with respect to $|\obj(\pv^*)|$.

We perform this validation on \wikiData and \pokData because computing
the exact gradient requires listing all triangles, which is feasible
within one hour on these networks.
In contrast, the main advantage of~\espressoR arises on large and dense
graphs, where the number of triangles is much larger than $m$ and exact
triangle enumeration {exceeds the limits of our implementations}, as {shown} by the
scalability results in~\Cref{tab:scalability}.

\subsection[appendix]{Comparing \espressoNL and  \espressoGD}\label{appsec:solutionComparison}

\begin{table}[t]
	\centering
	\caption{Comparison between \espressoNL and \espressoGD under the same number of iterations (bounded by 50). $\alpha=1$, $\lambda=0.05$, convex ($\gamma$ by~\Cref{lemma:Fprops}) and non-convex ($\gamma=0.1$) regimes. \espressoGD is run with the same number of iterations of \espressoNL. Lower objectives are marked in bold, together with the faster runtimes.}
	\resizebox{0.8\columnwidth}{!}{%
	\begin{tabular}{llrrrrr}
		\toprule
		\multirow{2}{*}{Dataset} & \multirow{2}{*}{Regime} & \multirow{2}{*}{$K$} & \multicolumn{2}{c}{Objective~(\obj)}  & \multicolumn{2}{c}{Runtime (s)} \\
		\cmidrule(lr){4-5}\cmidrule(lr){6-7}
		 & & & \espressoNL & \espressoGD & \espressoNL & \espressoGD \\
		\midrule
		\multirow{2}{*}{\gemsecData} & convex & 9 & $\mathbf{-1.74 \cdot 10^{7}}$ & $-7.83 \cdot 10^{6}$ & \textbf{2.1} & 3.6 \\
		 & non-convex & 50 & $7.34 \cdot 10^{5}$ & $\mathbf{7.15 \cdot 10^{5}}$ & \textbf{10.6} & 19.6 \\
		\midrule
		\multirow{2}{*}{\pokData} & convex & 9 & $\mathbf{-3.08 \cdot 10^{9}}$ & $-1.39 \cdot 10^{9}$ & \textbf{68.7} & 108.4 \\
		 & non-convex & 50 & $2.07 \cdot 10^{7}$ & $\mathbf{1.89 \cdot 10^{7}}$ & \textbf{285.6} & 528.3 \\
		\midrule
		\multirow{2}{*}{\twitchData} & convex & 9 & $\mathbf{-2.50 \cdot 10^{9}}$ & $-1.16 \cdot 10^{9}$ & \textbf{45.7} & 80.4 \\
		 & non-convex & 50 & $6.68 \cdot 10^{6}$ & $\mathbf{6.39 \cdot 10^{6}}$ & \textbf{236.5} & 490.4 \\
		\midrule
		\multirow{2}{*}{\wikiData} & convex & 9 & $\mathbf{-4.24 \cdot 10^{9}}$ & $-1.96 \cdot 10^{9}$ & \textbf{11.3} & 19.4 \\
		 & non-convex & 50 & $4.80 \cdot 10^{6}$ & $\mathbf{4.45 \cdot 10^{6}}$ & \textbf{58.7} & 104.9 \\
		\bottomrule
	\end{tabular}}
	\label{tab:solution_comparison}
\end{table}

In this setting we compared~\espressoNL with~\espressoGD under the same parameter setting.
We use this setup to assess the practical advantage of \espressoNL, based on mirror descent, which converges with guarantees in both the convex and the non-convex regime and, unlike \espressoGD, requires no tuning of the learning rate and no clipping.

We ran~\espressoNL with $\errAcc=0.001$, measured its number of iterations $K$ (limited at 50), and then performed $K$ iterations for~\espressoGD (with learning rate to $0.05$).
All algorithms are deterministic, therefore we do not perform multiple runs, and we report the runtime of a single execution.
The results are shown in~\Cref{tab:solution_comparison}.
In the convex regime, \espressoNL terminates after $K=9$ iterations, consistently with its linear convergence from~\Cref{theo:accNL}, and attains an objective at least $2.1\times$ lower than \espressoGD run for the same $K$, while being from $1.6$ to $1.8\times$ faster.
In the low-temperature regime ($\gamma=0.1$), we observe the opposite behavior:
\espressoNL performs $K=50$ iterations and reports a higher objective value than \espressoGD on all four datasets. We observe that solutions between \espressoGD and \espressoNL in the non-convex regime do not differ for more than $9.5\%$ in quality, and remaining \espressoNL is from $1.8$ to $2.1\times$ faster than \espressoGD.
This is captured in our discussion of the non-convex regime in~\Cref{sec:optimization}.
When $\obj$ is no
longer $\mu$-convex relative to $h$, the step size $1/L$ that \espressoNL sets as from~\Cref{theo:accNL} is conservative, and the (local) convergence
guarantee drops from a linear to a sublinear
$\bigO(1/k)$ rate~\cite{Bolte2018Lips}, retaining the monotone descent property.
By contrast, \espressoGD's step size is effective in the non-convex regime, but it comes with no convergence guarantees. 
Furthermore, we recall that \espressoGD requires clipping its gradient updates, since the gradient of $\obj$ diverges over the boundary of $(0,1)^m$, while the updates of \espressoNL lie in $\doma \obj$ by definition (see~\Cref{sec:optimization}).
Overall, \espressoNL converges with guarantees in both regimes and without tuning, showing its theoretical and practical applicability.

\fi

\end{document}